\documentclass[a4paper,fleqn,usenatbib]{mnras}

\usepackage[T1]{fontenc}
\usepackage{ae,aecompl}

\usepackage{graphicx}	
\usepackage{amsmath}	
\usepackage{amssymb}	
\usepackage{makecell}
\usepackage{multirow}
\usepackage{ulem}
\usepackage{xcolor}

\newcommand{\Msold}{M$_{\odot}$\,yr$^{-1}$}
\newcommand{\kms}{km\,s$^{-1}$}
\newcommand{\dego}{$^{\circ}$}

\title[New features of the millimetre emission of Mira Ceti]{Revealing new features of the millimetre emission of the circumbinary envelope of Mira Ceti}
\author[D.T. Hoai et al.]{
{D. T. Hoai\thanks{E-mail: dthoai@vnsc.org.vn}, P. Tuan-Anh, P.T. Nhung, P. Darriulat, P.N. Diep, N.T. Phuong}
\newauthor{   and T.T. Thai}
\\
Department of Astrophysics, Vietnam National Space Center, VAST, 18 Hoang Quoc Viet, Hanoi, Vietnam\\
}
\date{Accepted XXX. Received YYY; in original form ZZZ}

\pubyear{2020}

\begin{document}
\label{firstpage}
\pagerange{\pageref{firstpage}--\pageref{lastpage}}
\maketitle

\begin{abstract}
We study the morpho-kinematics of the circumbinary envelope of Mira Ceti between $\sim$100 and $\sim$350 au from the stars using ALMA observations of the SiO ($\nu$=0, $J$=5-4) and CO ($\nu$=0, $J$=3-2) emissions with the aim of presenting an accurate and reliable picture of what cannot be ignored when modelling the dynamics at stake. A critical study of the uncertainties attached to imaging is presented. The line emissions are shown to be composed of a few separated fragments. They are described in detail and plausible interpretations of their genesis are discussed. Evidence for a focusing effect of the Mira A wind by Mira B over the past century is presented; it accounts for only a small fraction of the overall observed emission but its accumulation over several orbital periods may have produced an enhancement of CO emission in the orbital plane of Mira B. We identify a South-western outflow and give arguments for the anti-correlation observed between CO and SiO emissions being the result of a recent mass ejection accompanied by a shock wave. We discuss the failure of simple scenarios that have been proposed earlier to explain some of the observed features and comment on the apparent lack of continuity between the present observations and those obtained in the close environment of the stars. Evidence is obtained for the presence of large Doppler velocity components near the line of sight aiming to the star, possibly revealing the presence of important turbulence at $\sim$5 to 10 au away from Mira A. 
\end{abstract}

\begin{keywords}
stars: AGB and post-AGB -- circumstellar matter -- stars: individual: Mira AB -- radio lines: stars.
\end{keywords}



\section{Introduction} \label{sec1}

Mira Ceti is one of the most observed AGB stars, remarkable for the large amplitude of its visual variability, $\sim$8 mag, and known to be accompanied by a distant companion ($\sim$80 au away). Table \ref{tab1} lists some basic parameters and related references.

Recent radio observations at millimetre wavelengths have explored the close environment of the star below $\sim$20 au \citep{Vlemmings2015, Matthews2015, Planesas2016, Wong2016, Khouri2018, Kaminski2016a,Kaminski2016b}. They reveal a complex dynamics and grain chemistry, influenced by variability, displaying significant inhomogeneity and suggesting the presence of shocks related to star pulsations. A similar picture is obtained when observing the infrared emission of this inner region of the star envelope \citep{Khouri2018,Kaminski2016a,Kaminski2016b}.

At larger distances from the star, recent millimetre observations have revealed the extreme inhomogeneity of the wind, made almost exclusively of separated fragments moving away radially at typical velocities of $\sim$5 \kms\ with respect to the systemic velocity \citep{Ramstedt2014, Nhung2016} and contrasting with the picture that could have been drawn by earlier observations of much lower angular resolution \citep{Planesas1990, Josselin2000, Fong2006}. Detection of CI emission has been recently reported, consistent with the expected dissociation of CO molecules by the interstellar UV radiation \citep{Saberi2018}.

The companion is now detected separately from the AGB star and its presence is seen to influence the morpho-kinematics of the wind, in particular by focusing the wind of Mira A \citep{Nhung2016} as qualitatively expected from Wind Roche Lobe Overflow models \citep{Mohamed2007,Mohamed2012}. The possible reduction of emission in the accretion wake of Mira B and the presence of a spiral arc have also been speculated \citep{Ramstedt2014}. Infrared observations detect the presence of dust forming an arc between the two stars \citep{Khouri2018}, they have been tentatively interpreted as suggesting accretion of the Mira A wind by a Main Sequence Mira B \citep{Ireland2007}.

Observations in the visible and at shorter wavelengths, UV and X-rays, have revealed a rather different picture. The first image showing a clear separation between Mira A and its companion was obtained in 1997 by the Hubble Space Telescope \citep{Karovska1997} but over a decade earlier a standard picture of the dynamics as seen in the UV had been drawn \citep{Reimers1985}. According to this picture, Mira B is a White Dwarf surrounded by a thin accretion disc rotating at very high velocity, between 100 and 1000 \kms\ depending on radius, and its UV emission dominates over that of Mira A. As Mira is moving at very high velocity through the interstellar medium ($\sim$130 \kms), it leaves a turbulent wake behind it that has been detected in the UV \citep{Martin2007} with evidence for a bow shock; the astropause has been detected in the FIR \citep{Ueta2008}. Closer to the star, observations in the FUV and H$\alpha$ have revealed the possible presence of a fast bipolar outflow \citep{Meaburn2009} with a velocity of 160$\pm$10 \kms\ and probably inclined by some 70\dego\ with respect to the plane of the sky. While this outflow would be emitted by Mira A, variations of the UV emission of the companion \citep{Wood2006, Sokoloski2010} have been interpreted as evidence for Mira B to be a White Dwarf surrounded by a disc accreting the wind of the central star, as had been postulated by \citet{Reimers1985}. Such a disc would be two orders of magnitude smaller than that postulated by \citet{Ireland2007} around a Main Sequence star rather than a White Dwarf. The mass loss rate of Mira B would then be 2.5 10$^{-12}$ \Msold, the wind velocity $\sim$250 \kms\ and one might expect that bipolar jets would be emitted along the axis of the accretion disc. However, a recent survey \citep{Montez2017} suggests that the intensity of the UV emission of AGB stars is essentially related to their distance, in apparent contradiction with the earlier accepted picture of UV emission dominated by the companion.

The importance of binarity in the dynamics of the symbiotic Mira pair remains a matter of debate. A clear bridge linking the two stars on a straight path has been observed in X-ray by Chandra \citep{Karovska2005} and at millimetre wavelength by ALMA \citep{Nhung2016}. Mira B may simply focus the wind of Mira A flowing in its direction and accrete part of it; or it may have its own wind interacting with the Mira A envelope and producing shocks playing an important role in shaping the circumbinary envelope in its environment. The presence of arcs in Herschel IR observations has been interpreted in such terms \citep{Mayer2011}.

The complex morpho-kinematics of the gas and dust envelope at distances beyond some 20 au has been studied using ALMA observations of the CO(3-2) line emission by \citet{Ramstedt2014} and \citet{Nhung2016} but has not yet been the object of a reliable interpretation. The observed inhomogeneity may be the result of recent episodes of enhanced mass loss rate, such as suggested by the occurrence of an X-ray outburst in December 2003 \citep{Karovska2005}; at variance with such a scenario, it may also, in principle, have been produced by progressive condensation of the wind material.

Among the fragments that make up the circumbinary envelope is a blue-shifted south-eastern arc, referred to as a bubble by \citet{Ramstedt2014} and which can be described as a ring inclined with respect to the plane of the sky and born some 2000 years ago \citep{Nhung2016}. \citet{Ramstedt2014} have suggested that it may have been created by the wind of Mira B blowing a bubble in the expanding envelope of Mira A. Beyond some 250 au from the stars and red-shifted from them by some 5 \kms\ a few arcs are detected. \citet{Ramstedt2014} suggest that some of these form a spiral confined to the orbital plane of Mira B, fed by the Mira A wind leaking through the inner Lagrangian point of the system. At shorter distances from Mira A, within some 250 au, two broad outflows are seen in the south-western and north-eastern quadrants. They display important inhomogeneity, the north-eastern outflow being strongly depleted at distances between $\sim$100 au and $\sim$200 au, which \citet{Ramstedt2014} interpret as the result of an accretion wake behind Mira B. The present work explores this central region with the study of the emission of the SiO($\nu$=0, $J$=5-4) line, which happens to be confined to the south-western outflow for distances from the star in excess of some 50 au. At the same time we revisit the analysis of the CO(3-2) observations that had been presented in \citet{Nhung2016} in the light of the recently published studies of the close environment of the star. The aim is to obtain new information that might help with a better understanding of the fragmented emission.

\begin{table*}
  \caption{The Mira AB pair: some basic parameters and related references.}
  \label{tab1}
  \begin{tabular}{ccc}
    \hline
    Quantity&Value&Reference\\
    \hline
    \multirow{2}{*}{Distance(a)}&92$\pm$10 pc&\citet{vanLeeuwen2007}\\
    &110$\pm$9 pc&\citet{Haniff1995}\\
    \hline
    Right ascension (J2000)(b) & 02h 19m 20.79210 s &\multirow{3}{*}{\citet{vanLeeuwen2007}}\\
    Declination (J2000)(b)& $-$02$^\circ$58$'$ 39.4956$''$&\\
    Proper motion(b)&(9, $-$237) mas yr$^{-1}$&\\
    \hline
    Space velocity (in LSR frame)&123 \kms &\multirow{1}{*}{(c)}\\
    \hline
    Spectral type &M5-9IIIe&\citet{Skiff2014}\\
    \hline
    Pulsation period&333 d&\citet{Templeton2009}\\
    \hline
    Apparent magnitude (visible) & 2.0 to 10.1&\makecell{\citet{Kukarkin1971}\\See \citet{Reid2002}}\\
    \hline
    Mass & 1.2 M$_\odot$&\citet{Wyatt1983}\\
    \hline
    Temperature &2900 to 3200 K&\citet{Woodruff2004}\\
    \hline
    Mass-loss rate & A few 10$^{-7}$ \Msold&\citet{Ryde2001}\\
    \hline
    Orbit: inclination & $\sim$112\dego&\multirow{3}{*}{\makecell{\citet{Prieur2002}\\See also \citet{Planesas2016}}}\\
    Orbit: period&$\sim$500 yr&\\
    Orbit: Semi-major axis &$\sim$80 au&\\
    \hline
  \end{tabular}\\
  Notes: (a) We adopt a distance of 100 pc, average between the two values in the table. \\(b) We use coordinates corrected for proper motion using the values given in the table.\\
  (c): calculated for the proper motion listed in the table and for the radial velocity of 47.7 \kms\ used in the present paper. 
\end{table*}

\section{Observations and data reduction} \label{sec2}

\subsection{Generalities}\label{sec2.1}

The data used here are retrieved from ALMA archives and correspond to two different sets of observations.

Project ADS/JAO.ALMA\#2011.0.00014.SV is used for the analysis of the SiO($\nu$=0, $J$=5-4) line emission at distances from the star exceeding some 100 au; the emission of the same line, together with that of other lines has been studied by \citet{Wong2016} for distances from the star not exceeding some 100 au. The Mira AB system was observed with ALMA on October 29th and November 1st as part of the 2014 ALMA Long Baseline Science Verification Campaign with the longest baseline reaching 15.24 km \citep{ALMA2015}. The very detailed description of observations and data reduction given by \citet{Wong2016} does not need to be repeated. In the region explored in this article, beyond some 50 au from Mira A, the contribution of continuum emission is negligible and does not need to be subtracted. Therefore, we recalibrated and reduced the data using the provided ALMA scripts and CASA version 4.2.2 omitting continuum subtraction. The only significant emission of the SiO($\nu$=0, $J$=5-4) line occurs in the western hemisphere. The other lines covered by the observations, SiO($\nu$=2, $J$=5-4), $^{29}$SiO($\nu$=0, $J$=5-4) and H$_2$O(5$_{5,0}-6_{4,3}$) display no significant emission in the region explored here.

Project ADS/JAO.ALMA\#2013.1.00047.S is used for the analysis of the CO($\nu$=0, $J$=3-2) line emission; the same observations have been analysed earlier by \citet{Planesas2016} to study continuum emission and by \citet{Nhung2016} to study line emission. Here again we do not need to repeat the detailed descriptions given by these authors. However, a very complex and fragmented morphology is observed in the region of interest, as has been described earlier by \citet{Ramstedt2014} and \citet{Nhung2016}. It prompted us to revisit the imaging process with the aim of assessing as rigorously as possible the reliability of the results obtained. This is discussed in detail below (Section \ref{sec2.2}). The value of the recoverable scale associated with the CO observations implies that a large fraction of the extended flux is missing. We checked that this had no significant impact on the results presented in the present article, which deal with distances not exceeding 3 arcsec from the stars.

In what follows we refer to the two sets of observations as simply SiO and CO. A summary of relevant parameters is given in Table \ref{tab2}. We use orthonormal coordinates with the $x$ axis pointing east and the $y$ axis pointing north. The origin is taken to match the centre of continuum emission of Mira A as observed by \citet{Planesas2016} and \citet{Wong2016}. The projected angular distance of pixel ($x,y$) to the origin is $R=\sqrt{x^2+y^2}$ and its position angle, measured counter-clockwise from north, is $\varphi=\tan^{-1}(x/y)$. The $z$ axis is parallel to the line of sight pointing away from us. The measured brightness is $f(x,y,V_z)$ and its integral over the measured Doppler velocity $V_z$ is the intensity $F(x,y)=\int{f(x,y,V_z)\mbox{d}V_z}$. As origin of velocity coordinate, we use a systemic velocity of 47.7 \kms\ with respect to the local standard of rest (LSR) as obtained by \citet{Khouri2018} from a fit of the CO($\nu$=1, $J$=3-2) line emission in the close environment of the star. This is 1.0 \kms\ larger than obtained by \citet{Wong2016} from a fit of the SiO($\nu$=0, $J$=5-4) line over a large field of view: the uncertainty on this number may be as large as 1 \kms.    

\begin{table*}
  \caption{Parameters of relevance to observations and data reduction.}
  \label{tab2}
  \begin{tabular}{ccc}
    \hline
    &SiO &CO \\
    \hline
    ALMA project (ADS/JAO.ALMA\#)&2011.0.00014.SV &2013.1.00047.S\\
    \hline
    PI&ALMA&P. Planesas\\
    \hline
    Date of observation& 29 October and 1st November 2014 &12-15 June 2014 \\
    \hline
    Mira A phase &0.47 &0.09\\
    \hline
    Line&SiO($\nu$=0, $J$=5-4)&CO($\nu$=0, $J$=3-2)\\
    \hline
    Frequency (GHz)&217.1&345.8\\
    \hline
    Number of antennas&35/36&34/36\\
    \hline
    Maximal baseline (km)&15.24&0.65\\
    \hline
    Maximal recoverable scale (arcsec)&11.3 &$\sim$4\\
    \hline
    Beam size (mas$^2$)&60$\times$30&390$\times$360\\
    \hline
    Beam PA&30\dego&83\dego\\
    \hline
    Velocity channel width (\kms)&0.4&0.4\\
    \hline
    Pixel size (arcsec$^2$)&0.01$\times$0.01&0.06$\times$0.06\\
    \hline
    Noise level (mJy\,beam$^{-1}$)&0.66&32\\
    \hline
  \end{tabular}
\end{table*}

\subsection{De-convolution of the measured CO visibilities}\label{sec2.2}

\begin{figure*}
  \includegraphics[width=0.3\textwidth,trim=-1cm -.5cm 0.cm 0.cm,clip]{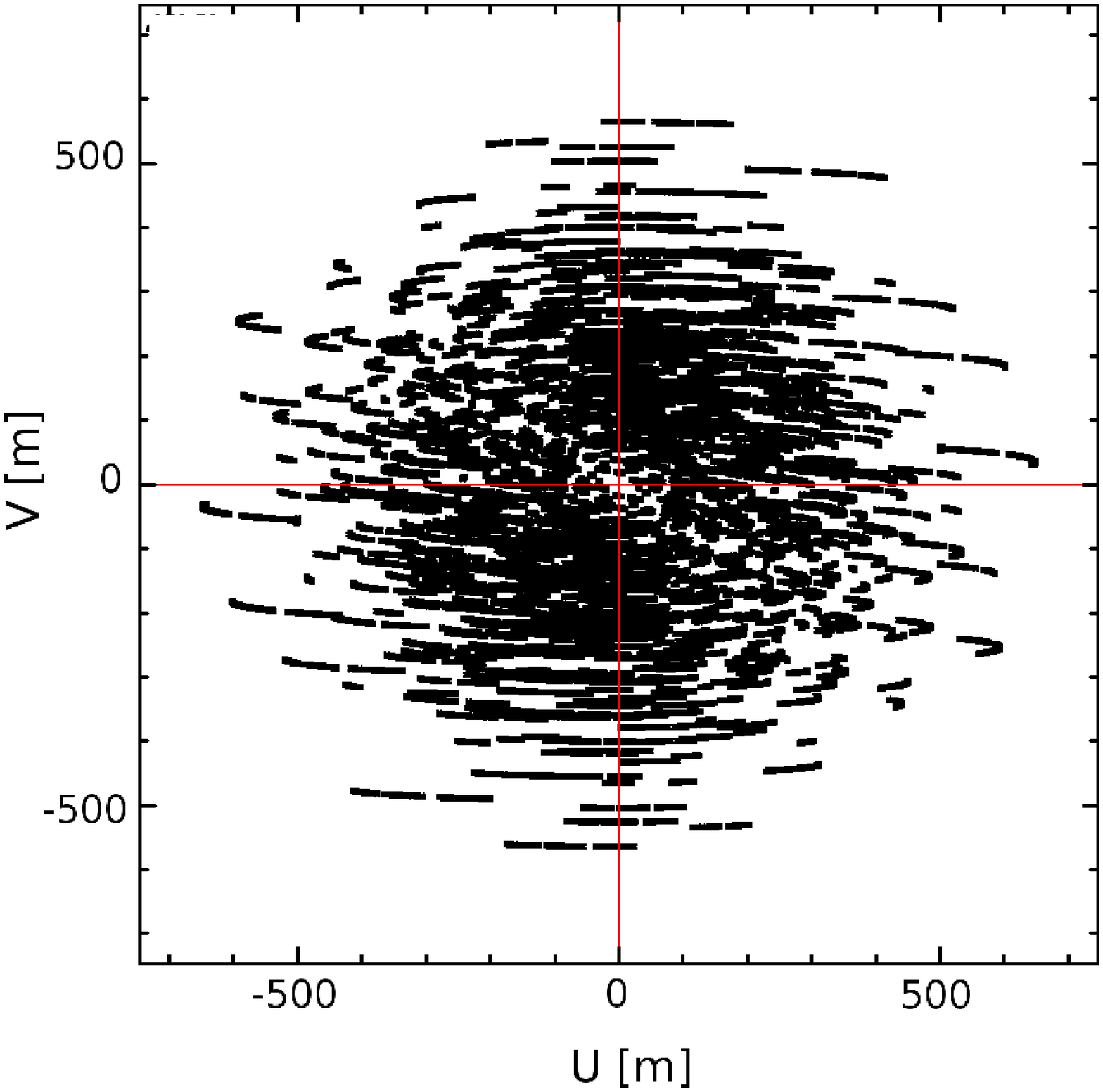}
  \includegraphics[width=0.3\textwidth,trim=-1cm -.5cm 0.cm 0.cm,clip]{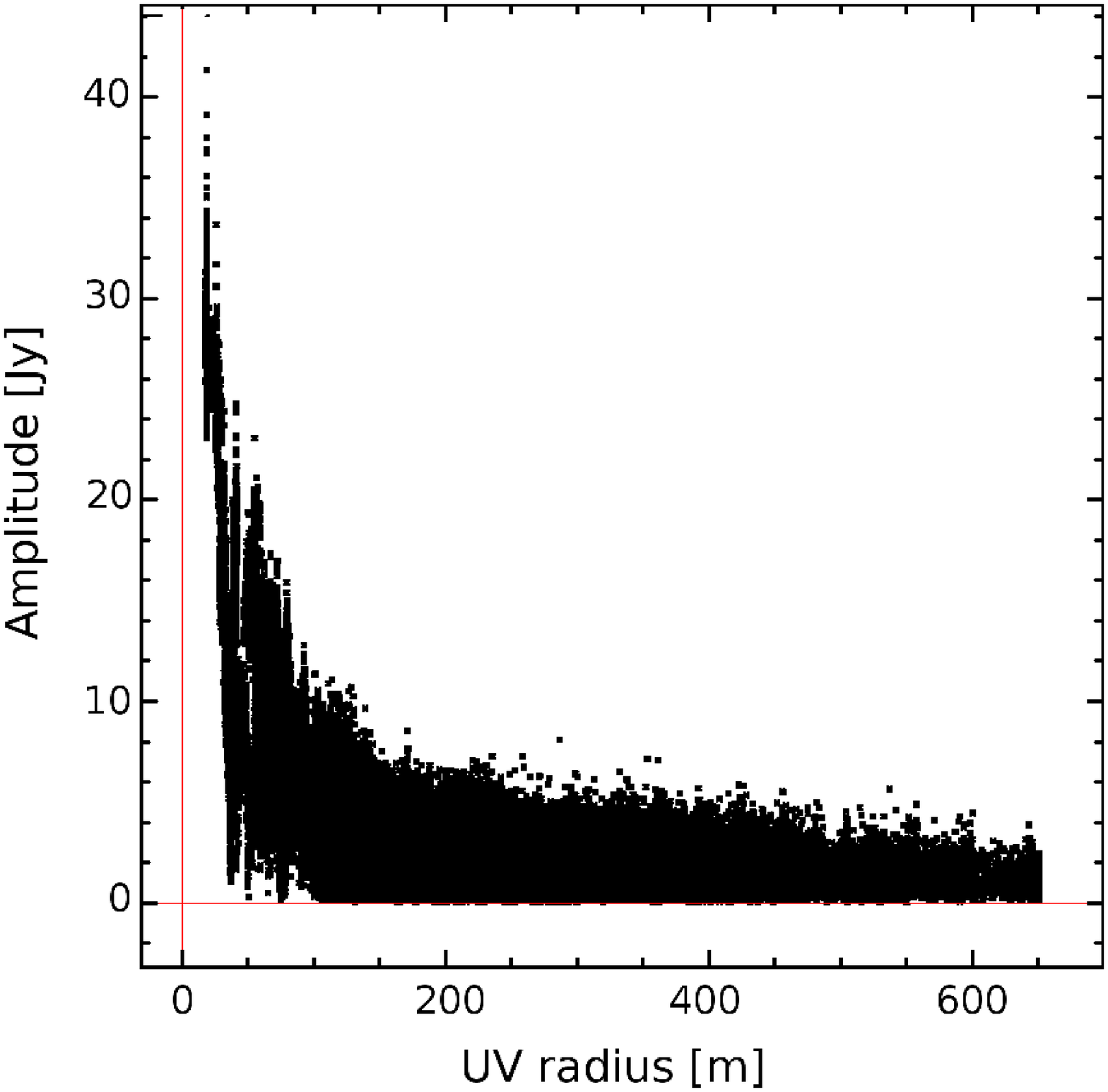}
  \includegraphics[width=0.355\textwidth]{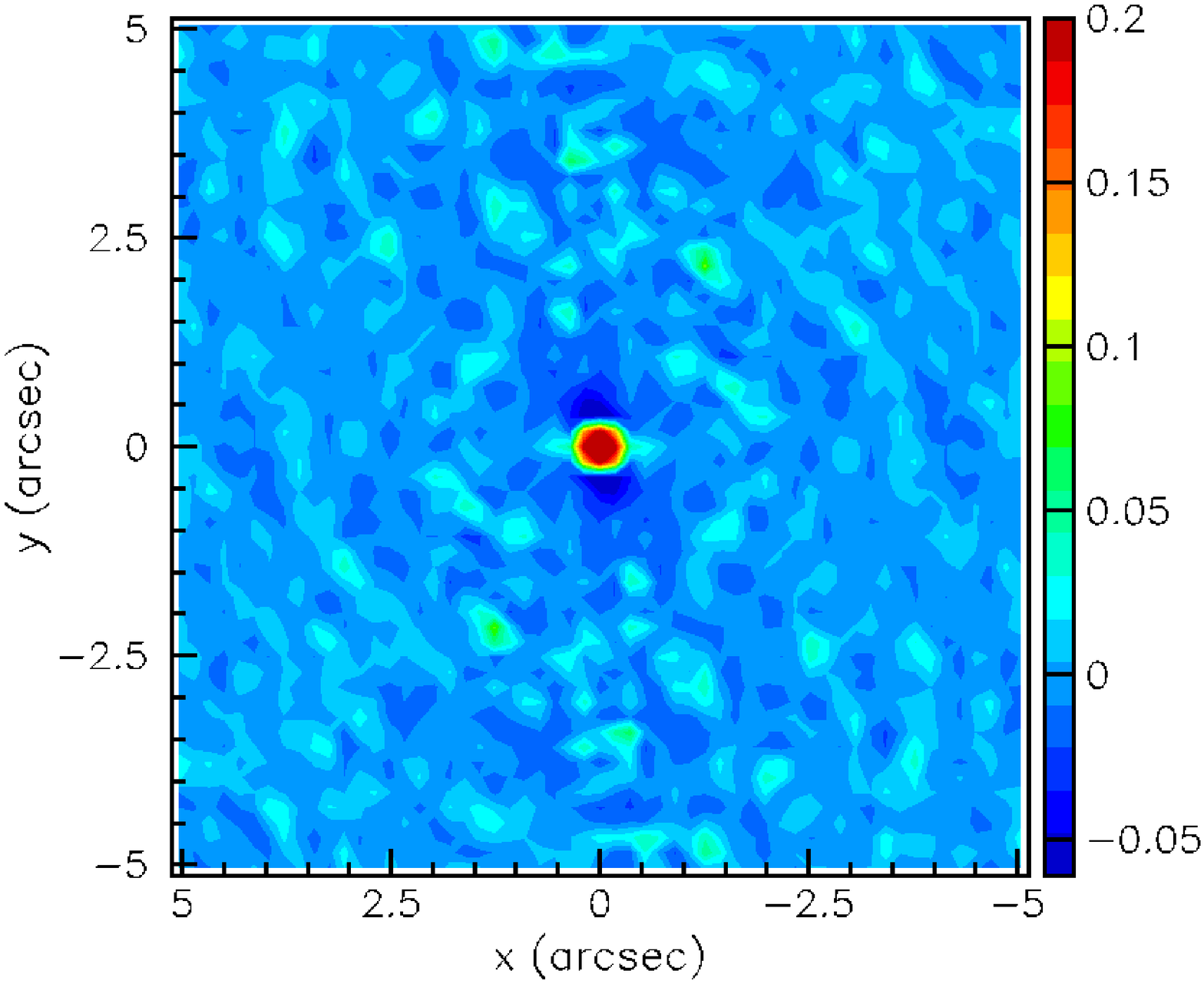}
  \caption{CO observations. Left: $uv$ coverage. Centre: visibility amplitude vs baseline for the Doppler velocity interval of 2.2 to 2.6 \kms. Right: map of the dirty beam.}
  \label{fig1}
\end{figure*}

\begin{figure*}
  \includegraphics[height=4.9cm]{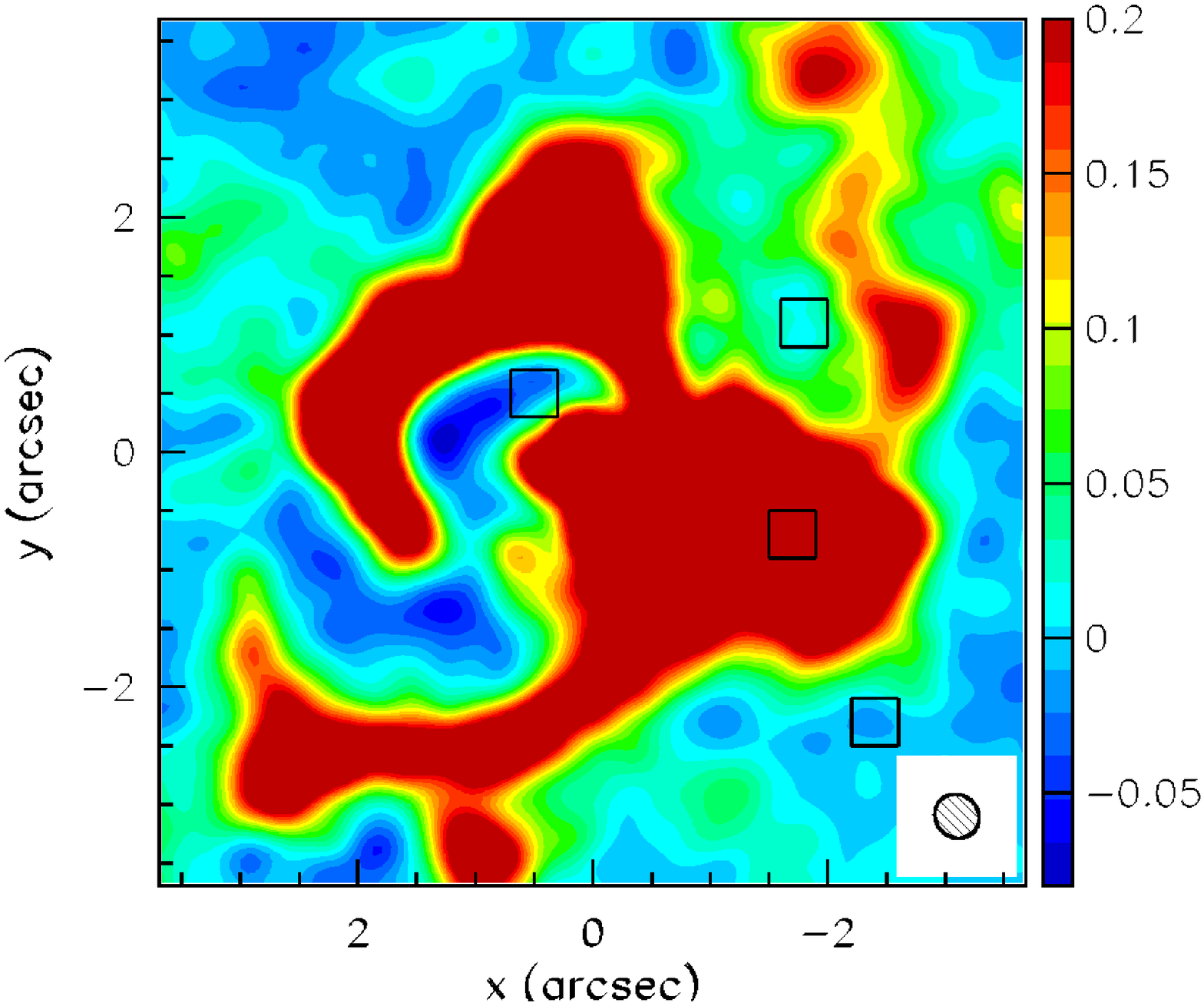}
  \includegraphics[height=4.9cm,trim=0.cm 0.cm 3.cm 0.cm,clip]{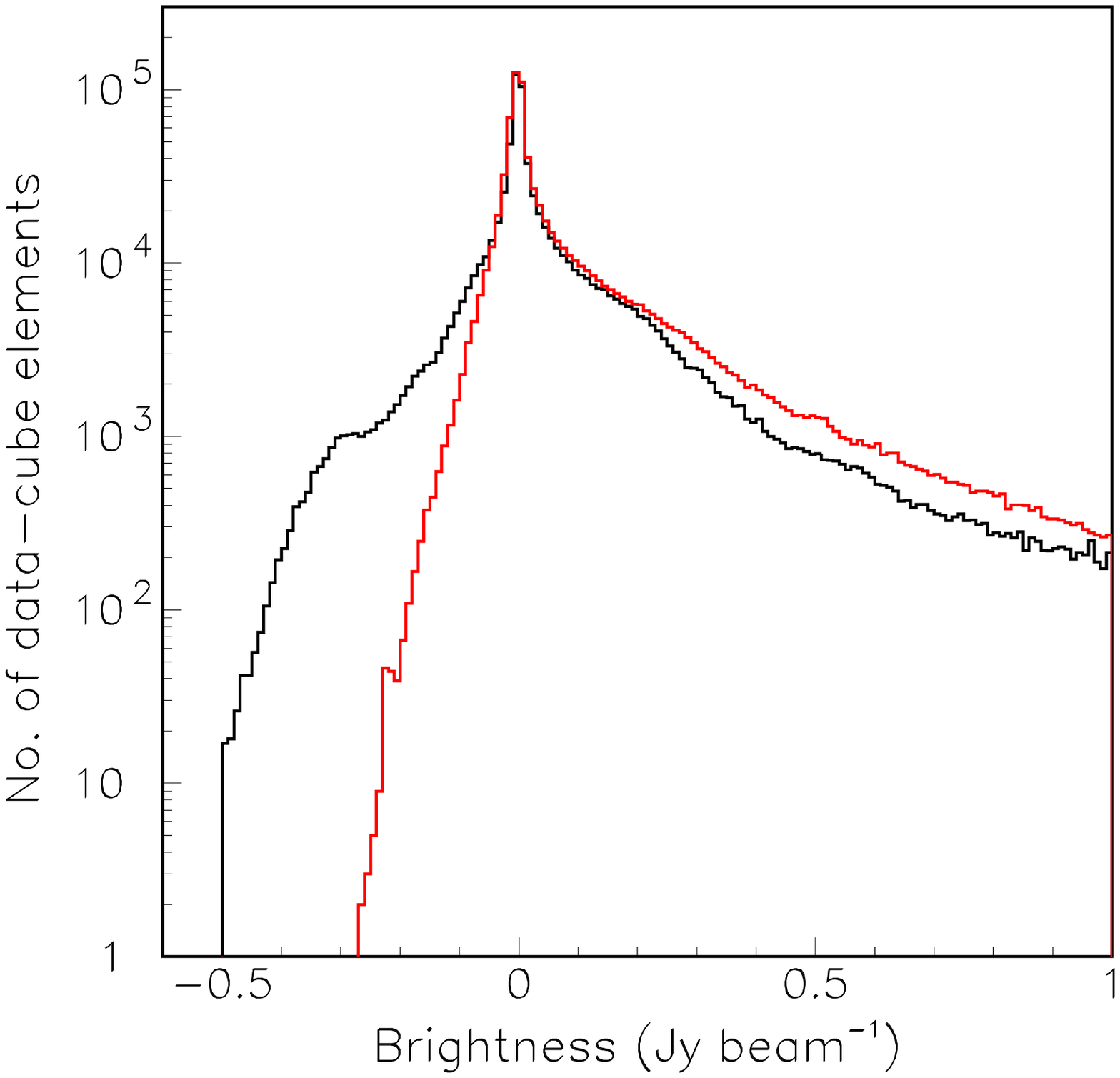}
  \includegraphics[height=4.9cm,trim=0.cm 0.cm 2.cm 0.cm,clip]{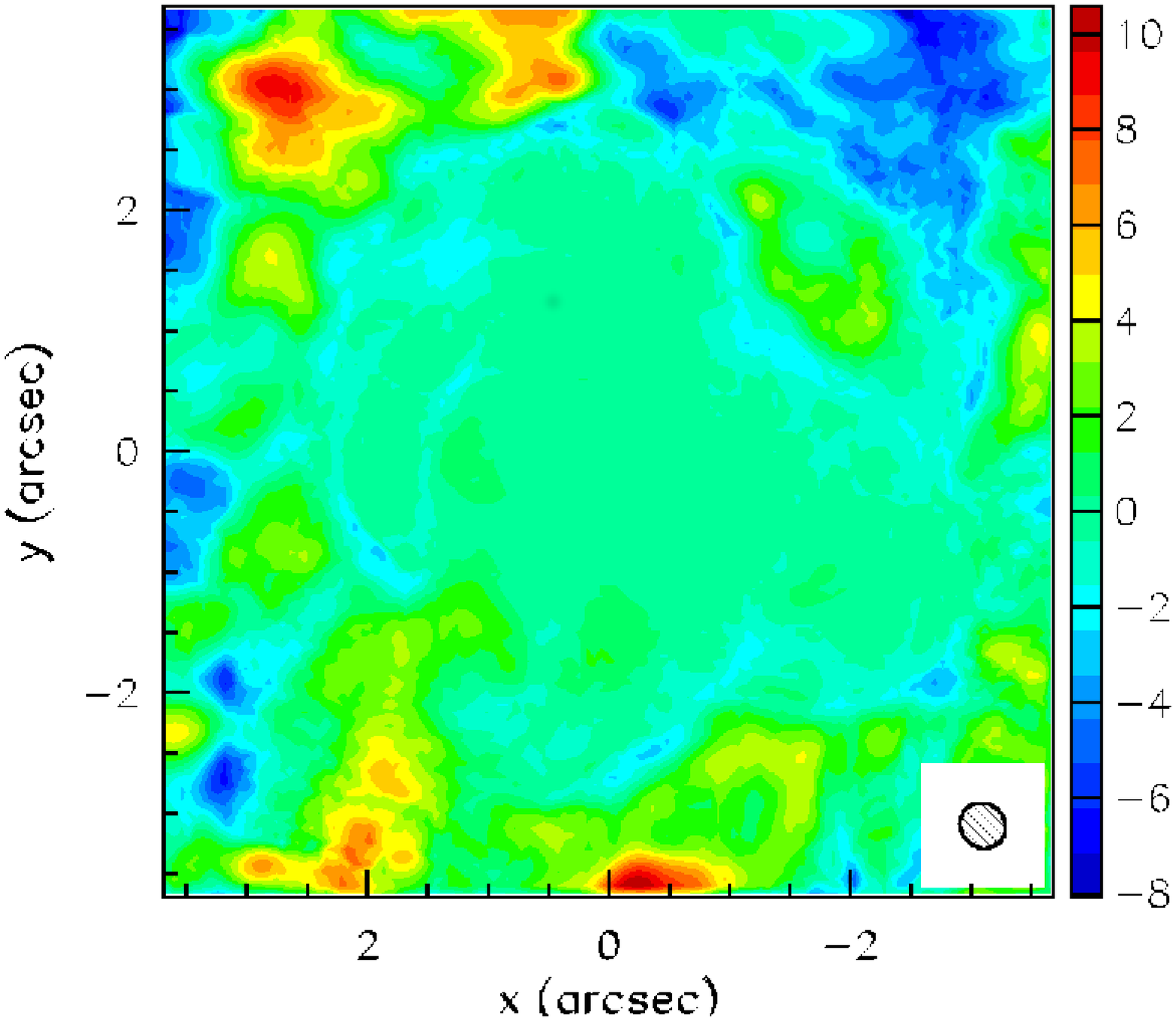}
 \caption{Left: clean map of CO emission (Jy\,beam$^{-1}$) for the frequency channel corresponding to a Doppler velocity interval of 2.2 to 2.6 \kms\ with squares showing the regions selected to fill Table \ref{tab3}.  Centre: brightness distribution obtained for $|x|$ and $|y|$ smaller than 3.7 arcsec and $|V_z|<10$ \kms\ with natural weighting and using a single circular mask of 6 arcsec radius (black) or optimized masks for each frequency channel (red). Right: map (integrated over $|V_z|<5$ \kms) of the difference, multiplied by $R$, between the intensity obtained with and without applying a 0.08 Jy\,beam$^{-1}$ brightness cut for the optimal data cube listed in the third line of Table \ref{tab3} (Jy\,arcsec$^{-1}$\kms).}
 \label{fig2}
\end{figure*}

In the CO data, image de-convolution is complicated by the concurrence of significant side-lobes of the dirty beam with sharp-edged emissivity fragments and by the large extension of the emission. In such cases, in the imaging process, the Fourier transform of the $uv$ map associated with a sharp-edged source produces a negative component that faces it. The task of cleaning the dirty map, namely of producing a clean map free of unphysical negative brightness regions, is then made difficult. In order to understand the impact of this problem on the reliability of the result and to evaluate the uncertainties attached to the brightness measured in each data cube element, we have studied its effect on the present data in great detail. In particular we have compared the performance of two software packages, GILDAS\footnote{https://www.iram.fr/IRAMFR/GILDAS} and CASA\footnote{https://casa.nrao.edu} and explored the relevant parameter space: number of iterations, weighting scheme, masking, threshold, etc. The details of the study go beyond the scope of the present article; it should be sufficient to present the main results. The map and profile of the dirty beam (obtained with natural weighting) are displayed in Figure \ref{fig1} together with the $uv$ coverage. To illustrate the effects of cleaning, we select a single frequency channel, corresponding to $2.2<V_z<2.6$ \kms, representative of the data sample; Figure \ref{fig2} displays a clean map obtained from GILDAS after 20,000 iterations. We limit the illustration to a square defined as $|x|<3.7$ arcsec and $|y|<3.7$ arcsec, within which the main part of the present work is confined. We define four small regions in which we measure the value of the brightness, $f_{i}$, $i=1,4$, three of which have a negative dirty map brightness and the third one being associated with a clear positive signal. Table \ref{fig3} summarizes some of the main results. They show that GILDAS gives generally better results than CASA and illustrate the importance of using masks well suited to the task: a very significant improvement is obtained by using different masks for each frequency channel, well adapted to the specific morphology. In practice, we explored several masking strategies to minimize the remaining negative brightness contribution to the clean map. The data used in the subsequent sections of the article were processed by GILDAS using optimal masks and natural weighting; they correspond to the third line of Table \ref{fig3} and are illustrated in the central panel of Figure \ref{fig2}. However, we have also produced data cubes obtained with other sets of cleaning parameters in order to evaluate the uncertainties attached to our results and to assess their reliability.  We studied the effect of applying a brightness cut on the optimal data cube, namely to only retain in the analysis brightness values exceeding such a cut. We remarked that with or without applying a cut at 0.08 Jy beam$^{-1}$, very similar results are produced in the region of the sky plane of relevance to the present study. Namely, as illustrated in the right panel of Figure \ref{fig2}, significant differences occur only at larger distances from the stars. Consequently, we present in the remaining of the article results obtained by applying a default cut of 0.08 Jy\,beam$^{-1}$ and we systematically check that the relevant results are consistent with what is obtained without cut.  In practice this means that although the brightness distribution shown in Figure \ref{fig2} does not display a Gaussian noise peak at the origin, the cut at 0.08 Jy\,beam$^{-1}$ corresponds to an effective 2.5$\sigma$ cut (see next section). Systematically, we made sure that all the statements made in the remaining part of the article are robust with respect to this brightness cut and to de-convolution and cleaning processes.

\section{Global description of the observed morpho-kinematics}\label{sec3}

\begin{figure*}
  \includegraphics[height=4.3cm,trim=0.2cm -.2cm 3.9cm 0.cm,clip]{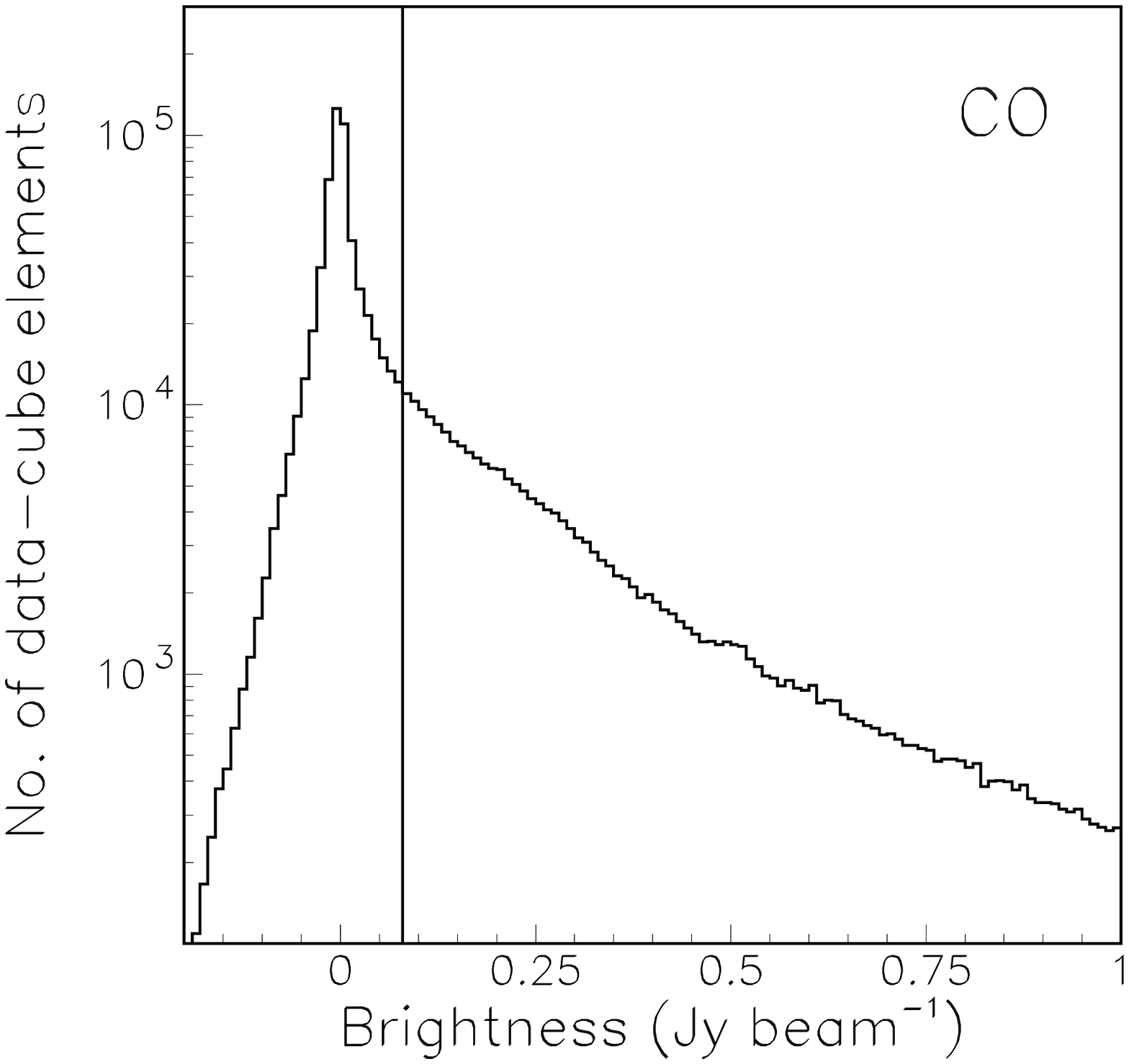}
  \includegraphics[height=4.3cm,trim=1.5cm 0.4cm 2.5cm 0.cm,clip]{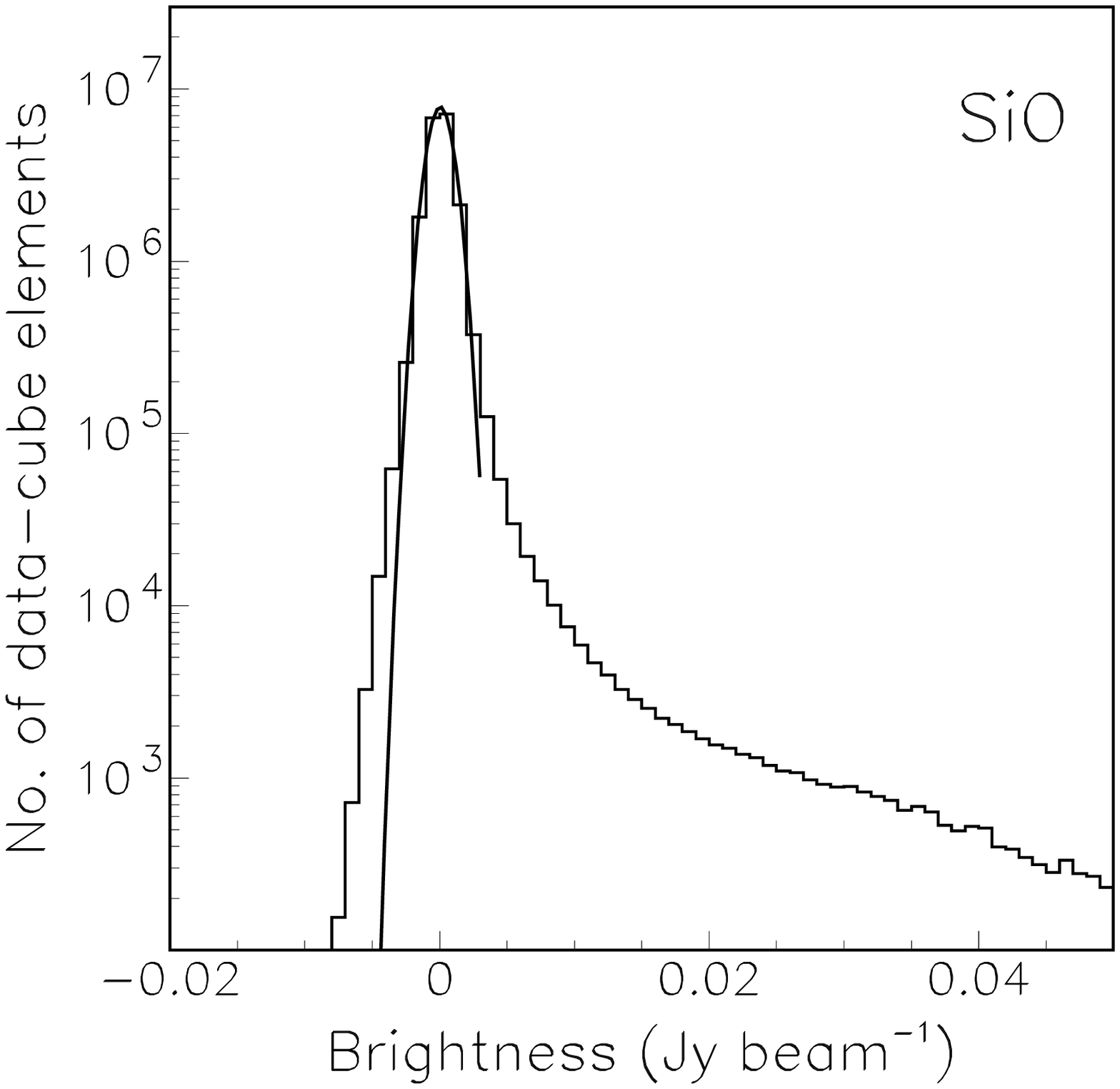}
  \includegraphics[height=4.3cm,trim=0.cm 0.4cm 3.9cm 0.cm,clip]{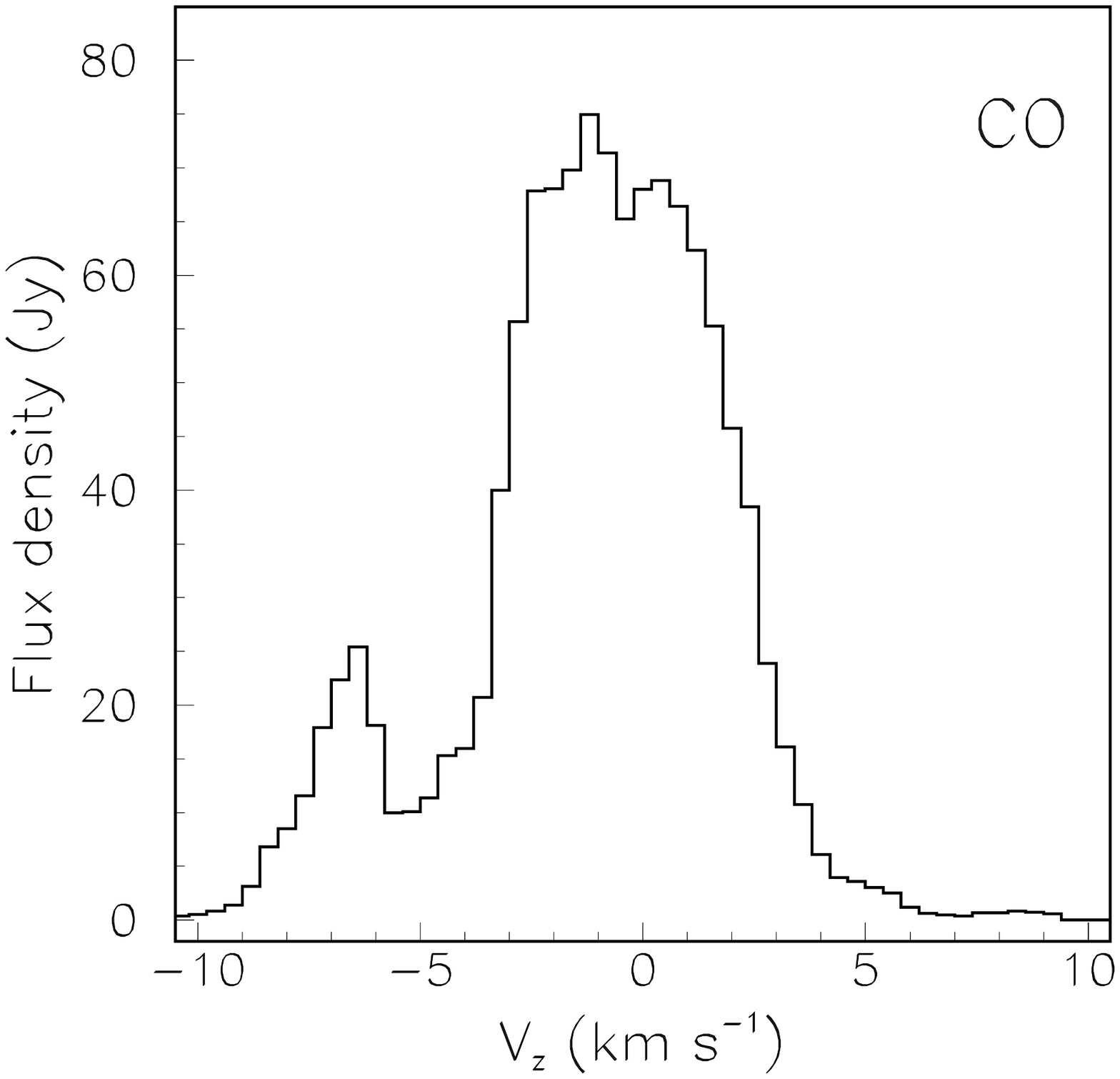}
  \includegraphics[height=4.3cm,trim=1.5cm 0.4cm 3.9cm 0.cm,clip]{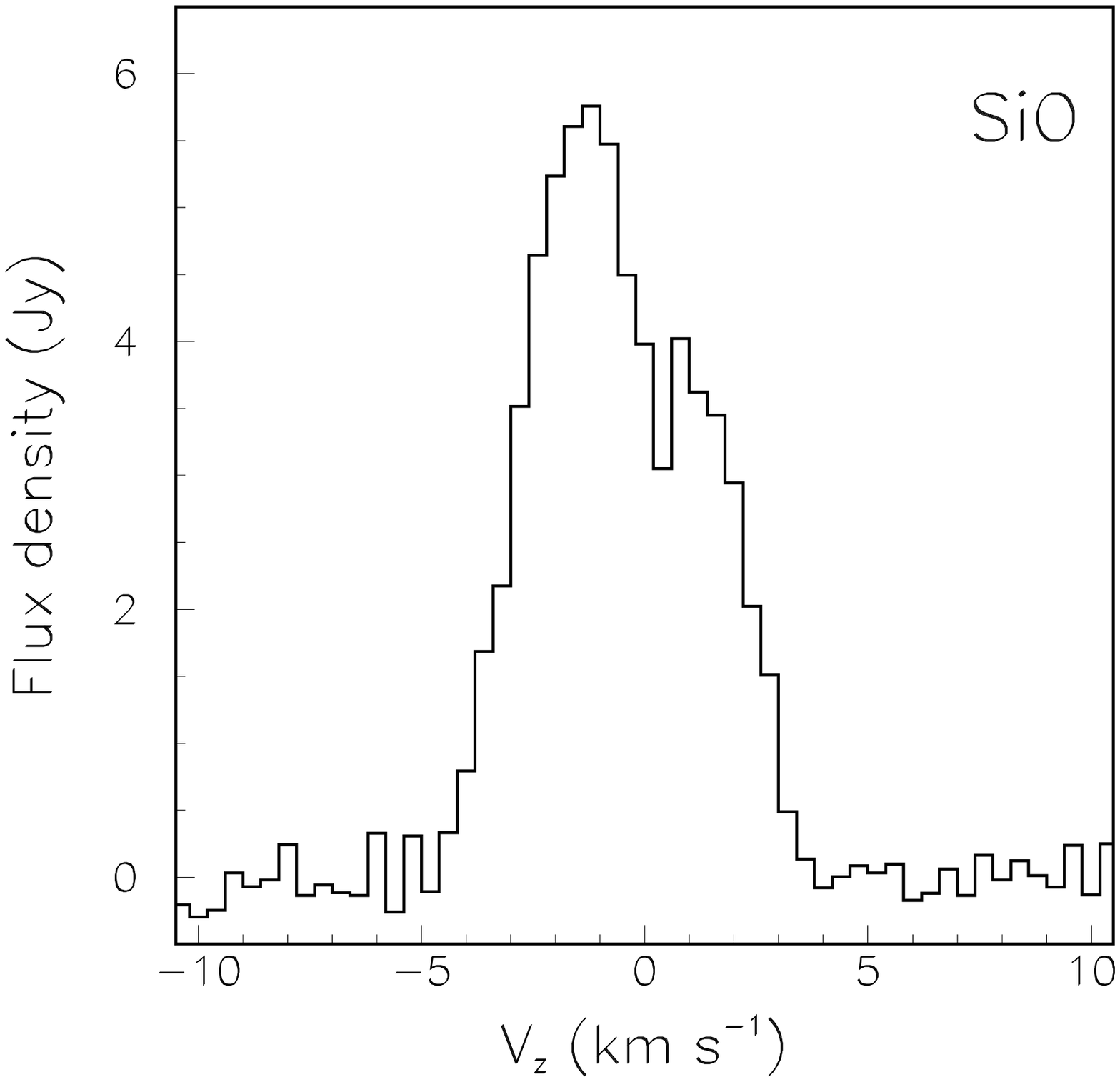}
  \caption{Left panels: distributions of the brightness measured in each data cube element for $|V_z|<10$ \kms\ integrated over squares covering $\pm3.7$ arcsec in ($x,y$) for CO and SiO. The curve shown for SiO is a Gaussian having a mean of 0.14 mJy\,beam$^{-1}$ and a $\sigma$ of 0.66 mJy\,beam$^{-1}$. Right panels: Doppler velocity spectra for $R>1$ arcsec and both $|x|$ and $|y|$ smaller than 3.7 arcsec measured with respect to the Mira A systemic velocity (47.7 \kms).}
  \label{fig3}
\end{figure*}

Figure \ref{fig3} (first two panels) shows the distributions of the brightness measured in each data cube element (pixel [$x,y$]$\times$velocity channel $V_z$) of CO and SiO emissions for $|x|$ and $|y|$ smaller than 3.7 arcsec and $|V_z|< 10$ \kms. A noise level (1$\sigma$) of 0.66 mJy\,beam$^{-1}$ is obtained for SiO emission. In the case of CO emission, as discussed in the previous section, we can only define a brightness cut of 0.08 Jy\,beam$^{-1}$, below which the measured brightness distribution averages to $-$0.5 mJy\,beam$^{-1}$ with an rms of 32 mJy\,beam$^{-1}$ equivalent to 1$\sigma$ of effective noise.  

\begin{table*}
  \caption{Some results of the study of the performance of the cleaning algorithm after 20,000 iterations (only 6000 iterations for the case of a 10 mJy beam$^{-1}$ threshold). Thresh is the threshold value in mJy\,beam$^{-1}$, Weight is the weighting scheme, Beam is the beam area of the synthesized beam and $f_i$ (in mJy) is measured inside 0.4$\times$0.4 arcsec$^2$ squares centred on $(x,y)=(0.5,0.5)$, ($-$2.4,$-$2.3), ($-$1.7,$-$0.7) and ($-$1.8, 1.1) arcsec for $i=1$ to 4 respectively (Figure \ref{fig2} left).}
  \label{tab3}
  \begin{tabular}{|c|c|c|c|c|c|c|c|c|}
    \hline
    Code&Thresh&Mask&Weight&Beam (arcsec$^2$)&$f_1$&$f_2$&$f_3$&$f_4$\\
      \hline
      GILDAS&\multirow{5}{*}{0}&\multirow{2}{*}{Circle, 6$''$ radius}&\multirow{4}{*}{Natural}&\multirow{4}{*}{0.39$\times$0.36}&$-$43&$-$42&740&$-$27\\
        \cline{1-1}\cline{6-9}
        CASA&&&&&$-$89&$-$35&673&$-$29\\
        \cline{1-1}\cline{3-3}\cline{6-9}
        GILDAS&&\multirow{4}{*}{3$\sigma$ masks}&&&$-$16&$-$10&838&20\\
        \cline{1-1}\cline{6-9}
        CASA&&&&&$-$70&$-$56&859&$-$33\\
        \cline{1-1}\cline{4-9}
        GILDAS&&& Robust 1.0&0.31$\times$0.29&$-$36&$-$19&832&$-$14\\
        \cline{1-2}\cline{4-9}
        GILDAS&10&&Natural&0.39$\times$0.36&$-$34&$-$15&824&$-$24\\
        \hline
        \multicolumn{5}{|c|}{Dirty map}&$-$192&$-$116&619&$-$145\\
        \hline
  \end{tabular}
\end{table*}

\begin{figure*}
  \includegraphics[height=5.cm,trim=0.5cm 1cm 0.5cm 2.cm,clip]{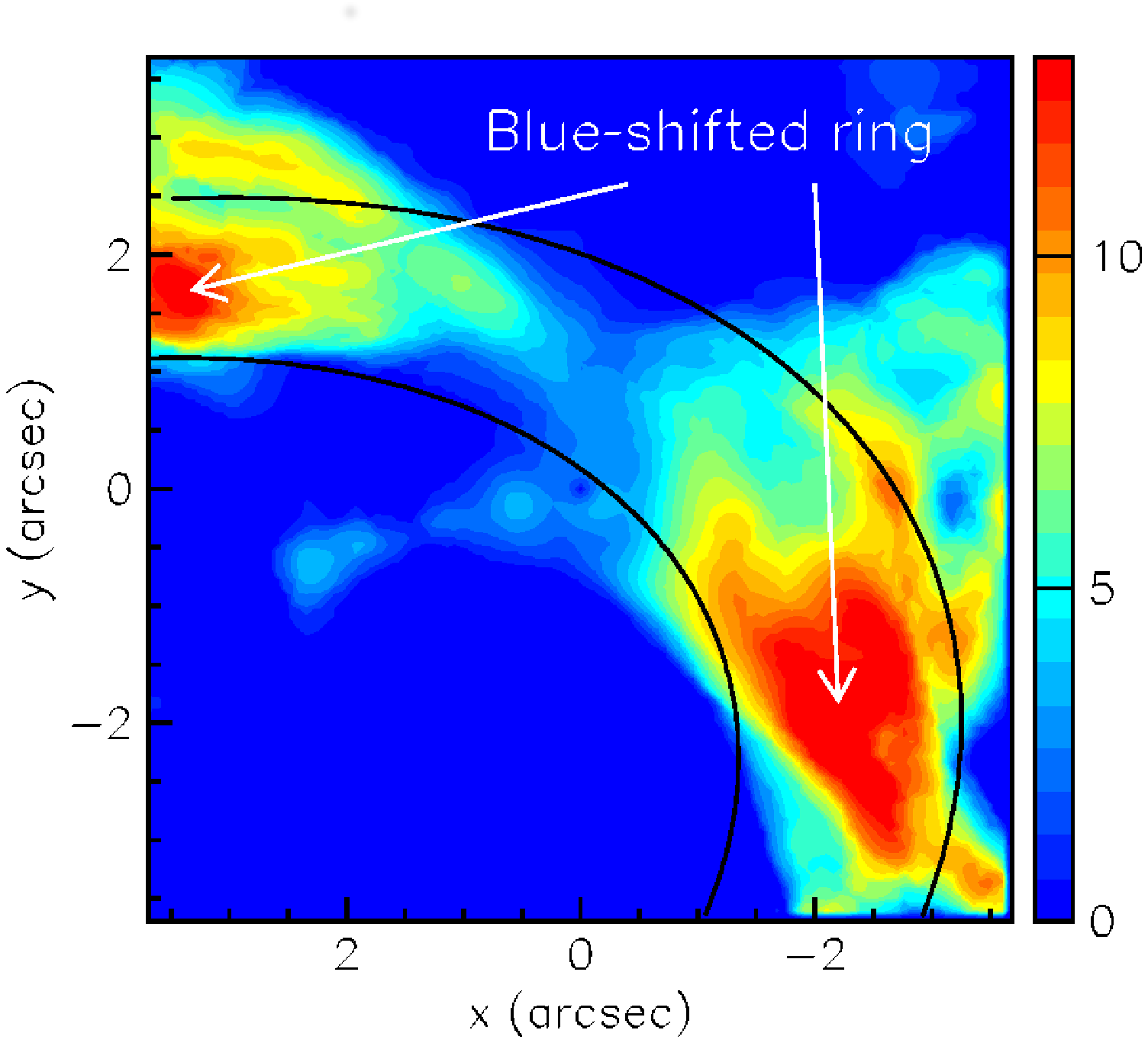}
  \includegraphics[height=5.cm,trim=0.5cm 1cm 0.5cm 2.cm,clip]{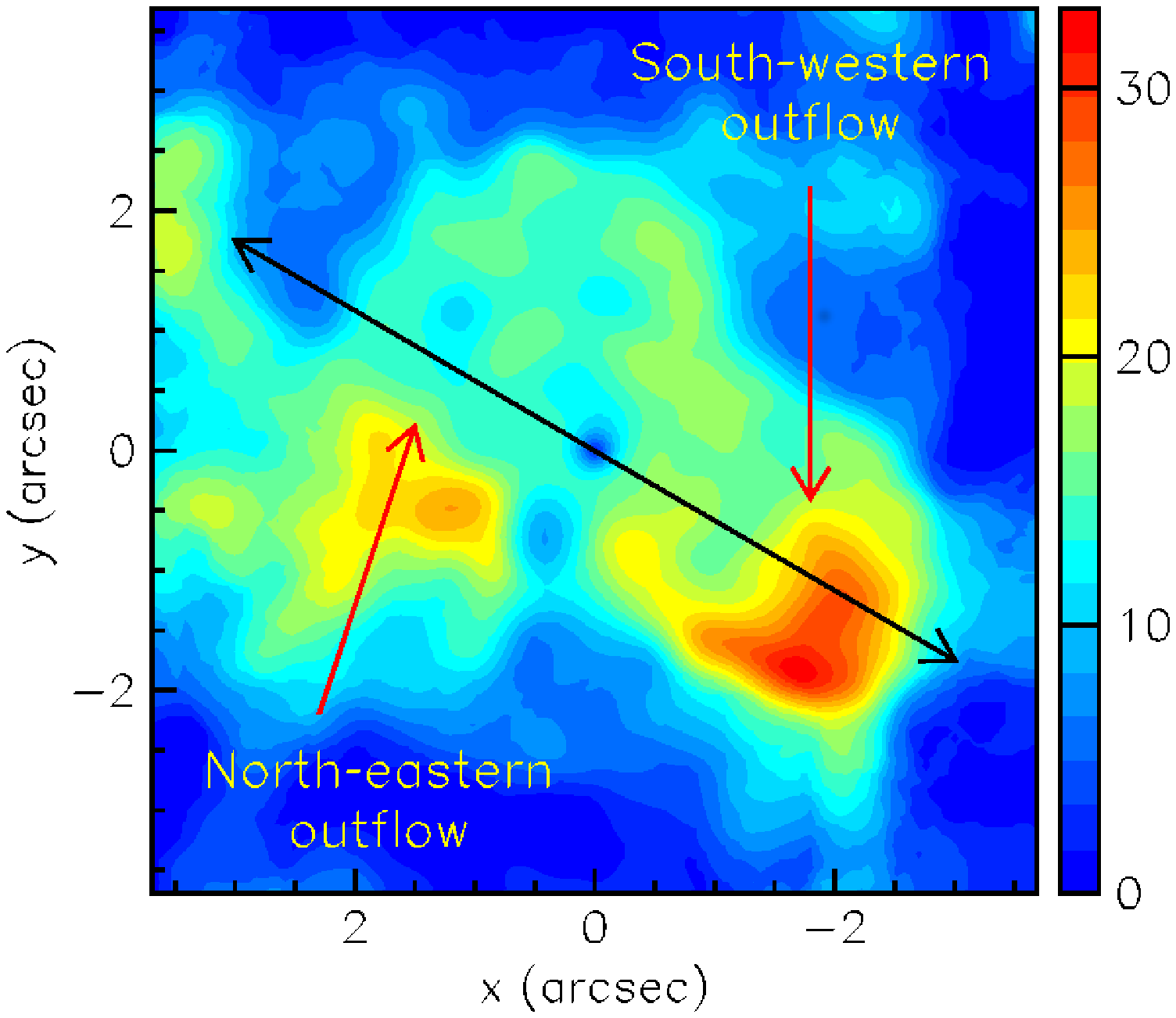}
  \includegraphics[height=5.cm,trim=0.5cm 1cm 0.5cm 2.cm,clip]{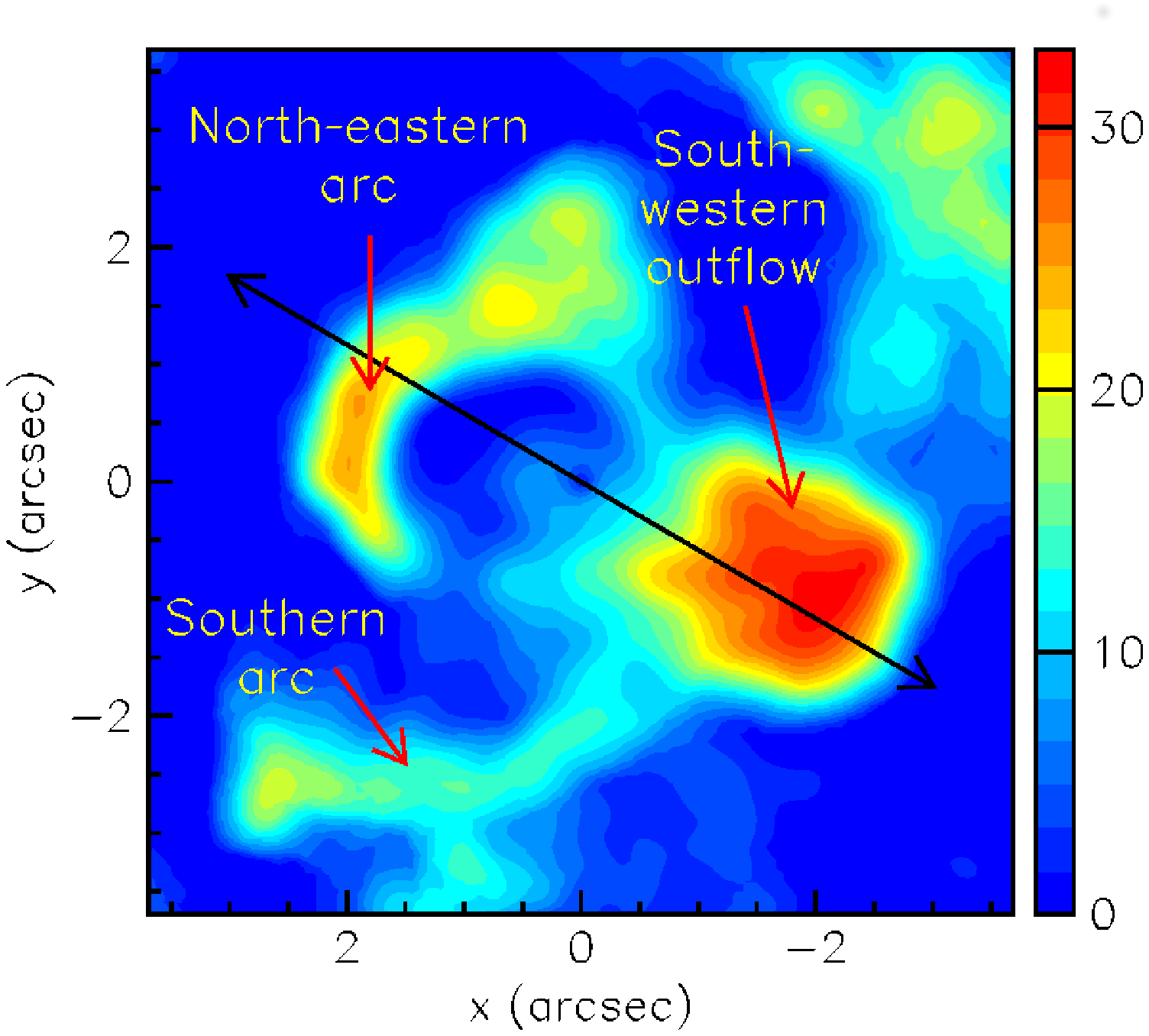}
  \caption{Left to right: intensity maps of CO emission multiplied by $R$ and integrated over $-9<V_z<-5$ \kms\ (left), $-4<V_z<0$ \kms\ (centre) and $0<V_z<4$ \kms\ (right). The arcs of ellipses shown in the left panel are from Figure \ref{fig7} of \citet{Nhung2016} and are meant to locate the emission of the Blue-shifted ring. The black arrows point to 60\dego\ and 240\dego\ (Figure \ref{fig5}c). The colour scales are in units of Jy\,arcsec$^{-1}$\,\kms. }
  \label{fig4}
\end{figure*}

\begin{figure*}
  \includegraphics[height=3.8cm,trim=0.5cm 1cm 0.5cm 2.cm,clip]{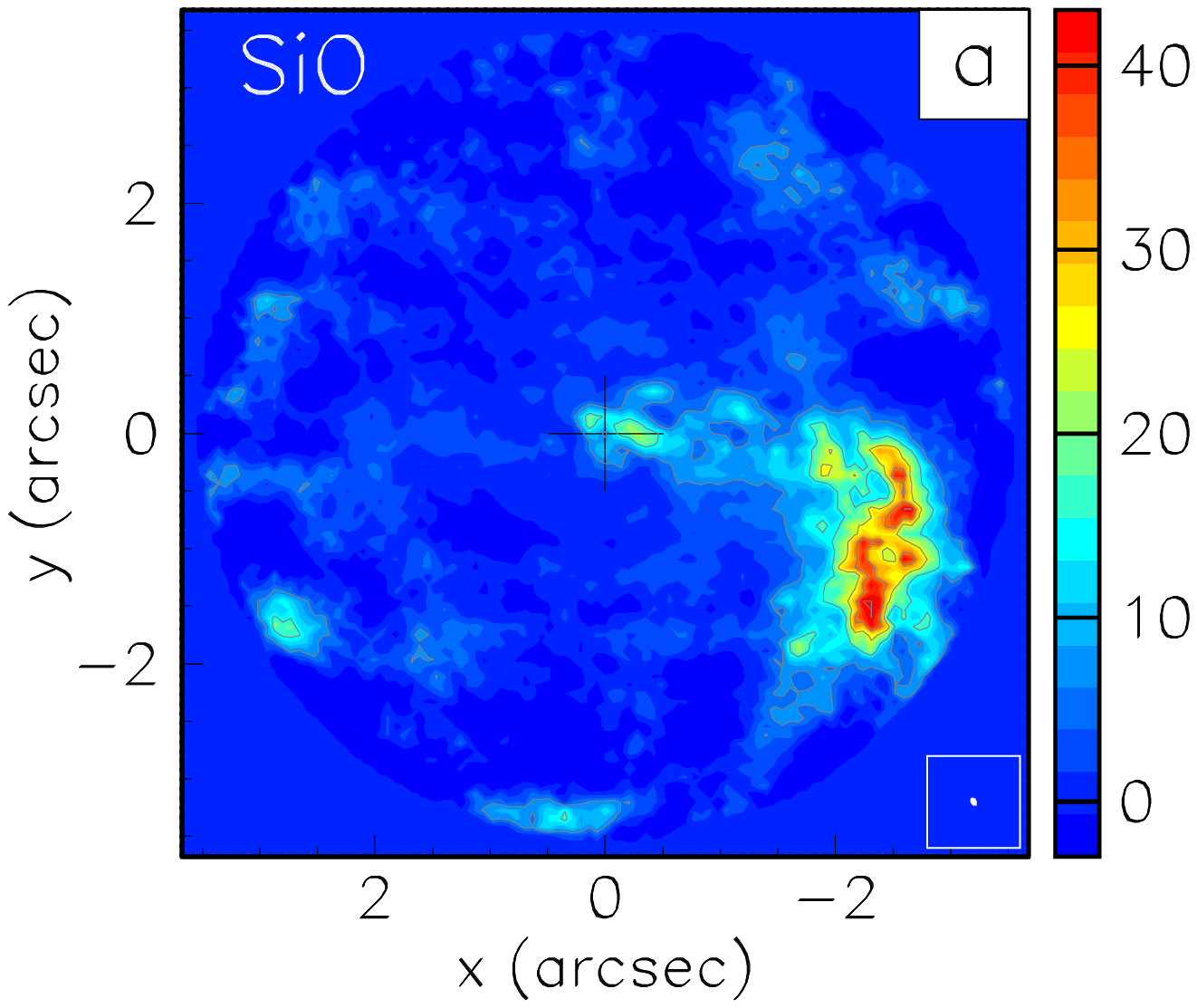}
  \includegraphics[height=3.8cm,trim=0.5cm 1cm 0.5cm 2.cm,clip]{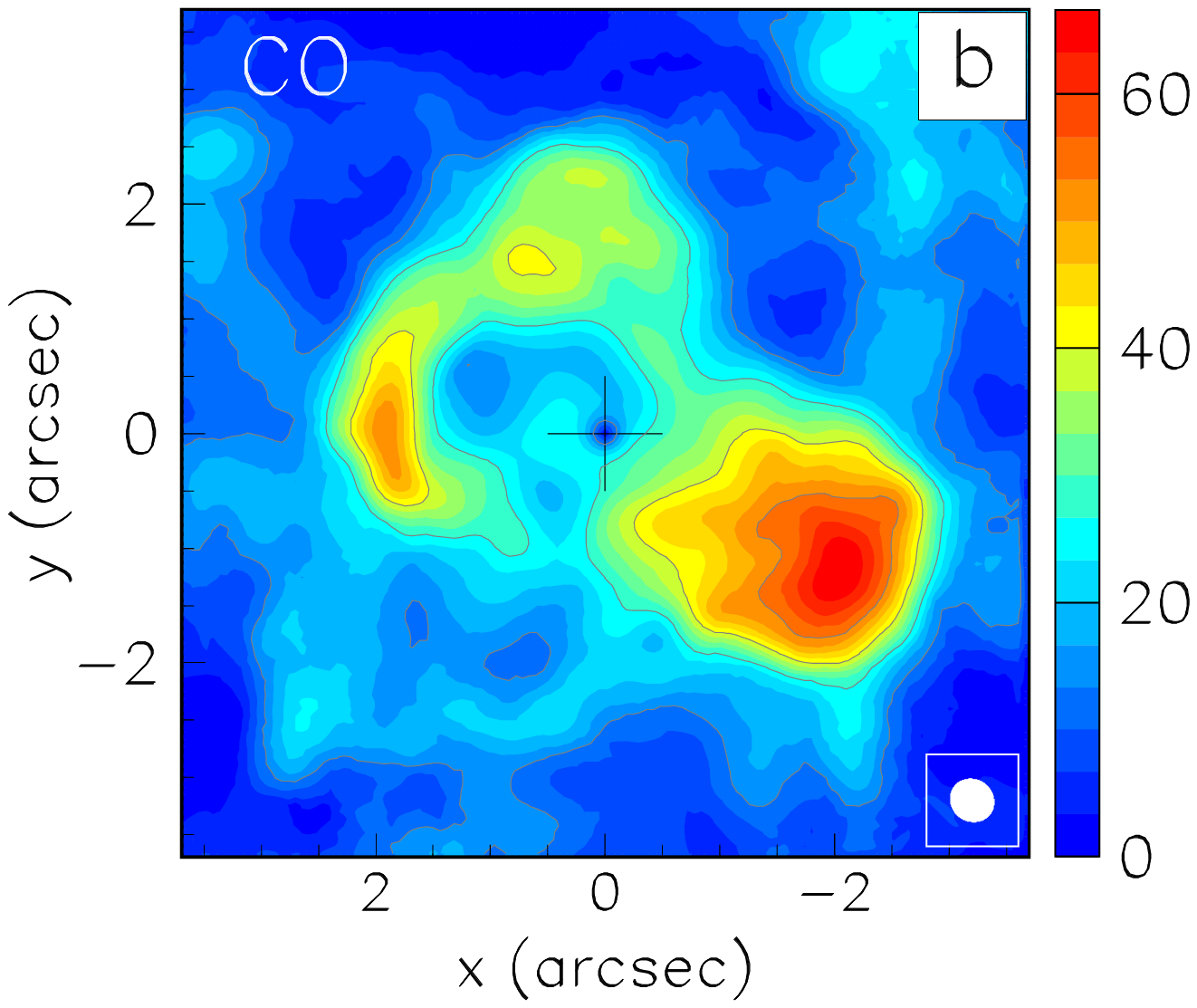}
  \includegraphics[height=3.8cm,trim=0.5cm 1cm 2.1cm 2.cm,clip]{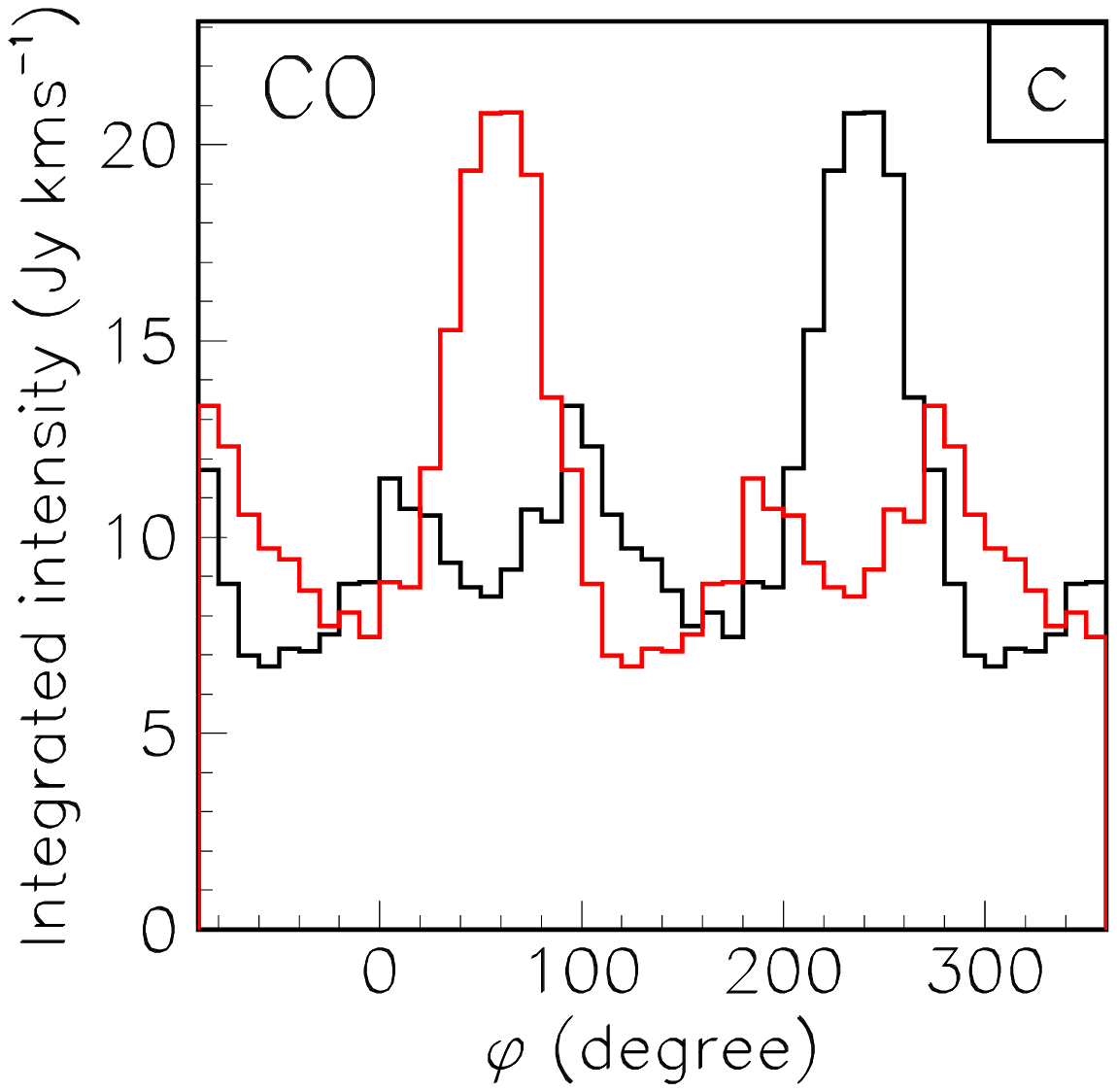}
  \includegraphics[height=3.8cm,trim=0.4cm 1cm 2.1cm 2.cm,clip]{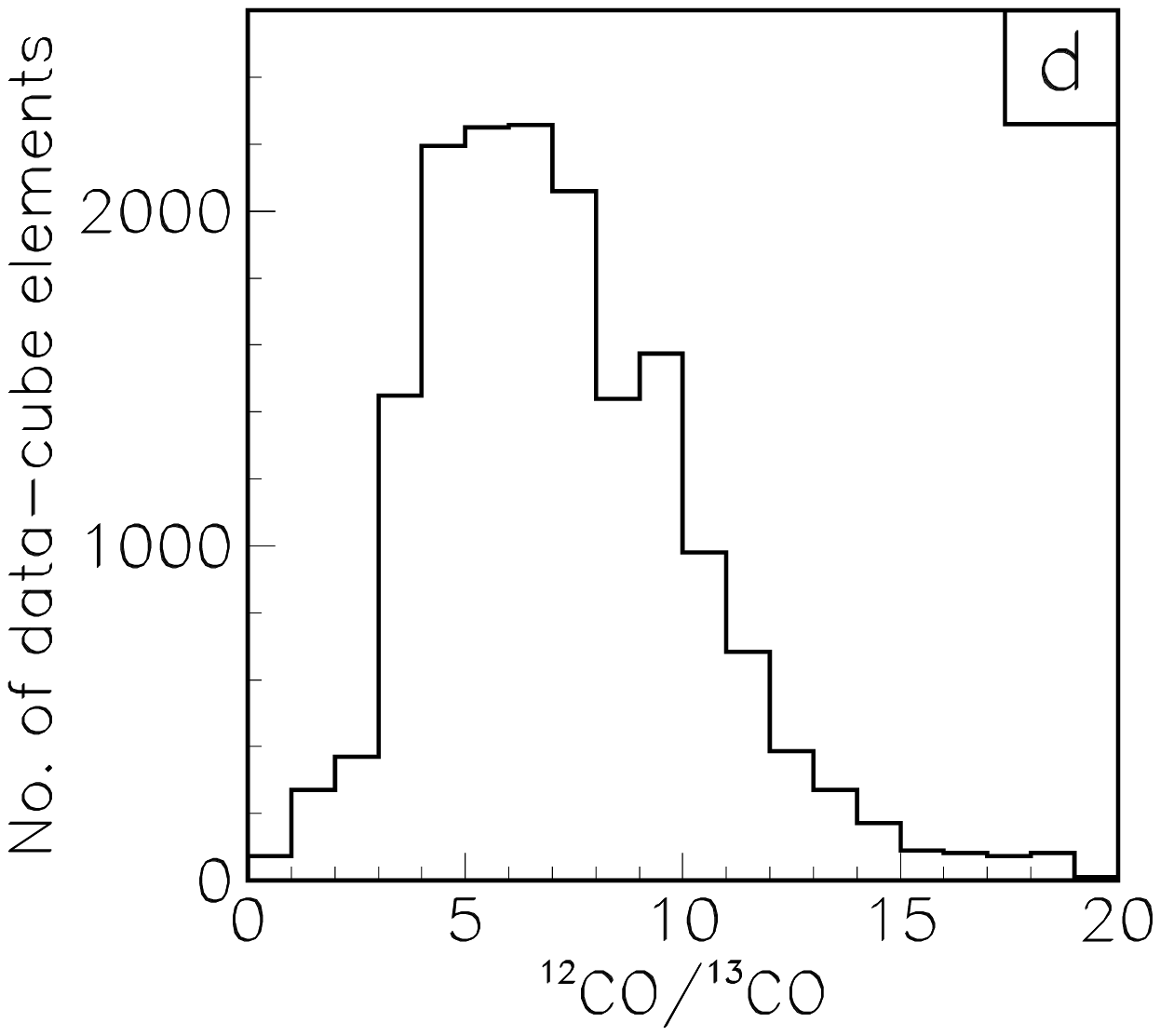}
 \caption{From left to right: maps of the intensity multiplied by $R$ (Jy\,arcsec$^{-1}$\,\kms) for SiO (a) and CO (b) emissions integrated between $-$5 and 5 \kms; (c) dependence on $\varphi$ (black) of the CO intensity integrated between $-$5 and 5 \kms\ in $V_z$ and between 1 and 3.7 arcsec in $R$. The red histogram is obtained from the black histogram by rotation of 180\dego\ about Mira A; (d) distribution of the $^{12}$CO/$^{13}$CO brightness ratio for regions of large column density; the mean value is 7.3 and its spread between the four fragments listed in Table \ref{fig5} is 0.8.}
  \label{fig5}
\end{figure*}

Figure \ref{fig3} (last two panels) displays Doppler velocity spectra of the CO and SiO emissions for $R>1$ arcsec and both $|x|$ and $|y|$ smaller than 3.7 arcsec. Apart from CO emission from the Blue-shifted ring (see below) between $-$10 and $-$5 \kms, both SiO and CO emissions are essentially confined to the interval $|V_z|<5$ \kms. Figure \ref{fig4} displays separate intensity maps of the CO emission for each of the three intervals of Doppler velocity, $-9<V_z<-5$ \kms, $-4<V_z<0$ \kms\ and $0<V_z<4$ \kms. Global maps of both SiO and CO emissions, integrated between $V_z=-5$ \kms\ and $V_z=+5$ \kms\ are displayed in Figure \ref{fig5}a,b. In both Figures \ref{fig4} and \ref{fig5}a,b, and in most intensity maps in the remaining of this article (except Figure \ref{fig9}), the intensity has been multiplied by $R$, the projected distance from Mira A, which qualitatively accounts for the radial dilution of a steady flow. The blue-shifted map of CO emission is dominantly associated with the feature described as a bubble by \citet{Ramstedt2014} to which we refer as Blue-shifted ring \citep{Nhung2016}. In the present article we shall not add anything to the descriptions and interpretations of this feature that were given in these earlier publications and we concentrate instead on the region defined as $|V_z|<5$ \kms\ and $1<R<3.7$ arcsec.

The SiO map shows significant emission confined to a south-western wedge with a strong enhancement below $R\sim3$ arcsec. It had been shown earlier by \citet{Wong2016} but had not been studied nor even commented upon.

The CO maps display clear anisotropy with an enhancement of emission in the south-western quadrant. This is further illustrated in Figure \ref{fig5}c that displays the dependence on $\varphi$ of the intensity: south-western and north-eastern emissions are back to back, $\varphi \sim240$\dego\ and $\sim$60\dego but the south-western emission is confined to a single peak while the north-eastern emission is split into two similar halves. This immediately raises two major questions that need to be addressed in the remaining of the article, none of which will be given a definitive answer: is there a relation between the south-western and north-eastern emissions or are they back to back (when projected on the sky plane) by pure coincidence? And, if there is a relation, is the split of the north-eastern emission genuine, namely has it been originally produced in such a split form, or has a hole of emission somehow been carved in the middle of an originally uniform and broad emission?      

\begin{figure*}
  \includegraphics[height=7.cm,trim=0.5cm 1cm 0.cm 2.cm,clip]{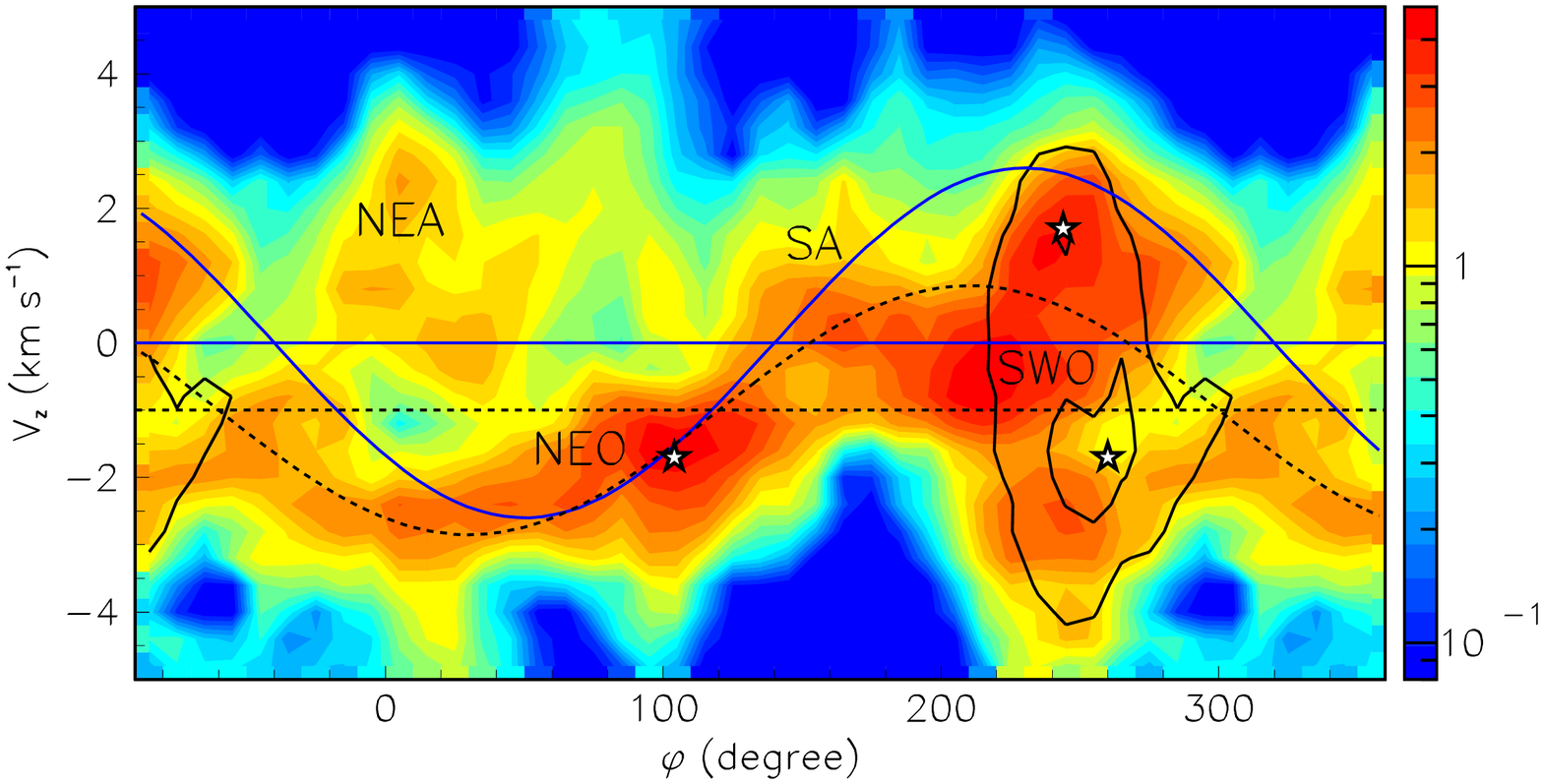}
   \caption{PV map of CO (colour) and SiO (contours at 0.25 and 0.85 Jy) in the $V_z$ vs $\varphi$ plane integrated over $1<R<3.7$ arcsec. The colour scale is in Jy. NEA, SWO, SA and NEO stand for North-eastern arc, South-western outflow, Southern arc and North-eastern outflow, respectively. The stars show the location of the centres of the radial outflows listed in Table \ref{tab4}. The full (dashed) sinusoidal lines are the traces of an isotropic radial wind of 2.8 (2.0) \kms\ velocity blowing in the plane of Mira B orbit using systemic velocities of 47.7 (46.7) \kms\ (see text).}
  \label{fig5pai}
\end{figure*}

Figure \ref{fig5pai} displays a position-velocity map (PV map) integrated between 1 and 3.7 arcsec in $R$. In the remaining of the article we call PV maps distributions of the intensity in a plane having $V_z$ as axis of ordinate and a sky coordinate, $x$, $y$, $R$ or $\varphi$ as axis of abscissa. They are projections of the data cube on a plane perpendicular to the plane of the sky as usual intensity maps are projections of the data cube on the plane of the sky. They are extensions of the usual PV diagrams that cover only a slit on the sky plane. In Figure \ref{fig5pai} the axis of abscissa is the position angle $\varphi$. In such maps expansion at constant radial velocity $V$ in a given direction is imaged as a point at fixed values of $V_z$ and $\varphi$ independently from the value of $R$; moreover the surface element $\mbox{d}S=\mbox{d} V_z\mbox{d}\varphi$ is proportional to the associated volume element in space $(R^2/V)\mbox{d}R\mbox{d}S$  independently from the values of $V_z$ and $\varphi$. Displaying the PV map of Figure \ref{fig5pai} in separate smaller intervals of $R$ reveals this way three significant radial outflows, two for CO emission and one for SiO emission (see Figure \ref{figa3} of the appendix). The data cube coordinates of their centres are listed in Table \ref{tab4} and indicated by stars on Figure \ref{fig5pai}. We show in Figure \ref{fig5pai} the trace of a 2.8 \kms\ wind blowing isotropically in the plane of the Mira B orbit \citep{Prieur2002}, suggesting its correlation with an enhancement of emissivity; a better match is obtained using a systemic velocity of 46.7 \kms\ (instead of 47.7 \kms), an orbit intersecting the plane of the sky at a position angle of $\sim$120\dego\ (instead of 140\dego), both of which are within uncertainties, and a wind velocity of $\sim$2.0 \kms\ (instead of 2.8 \kms). The possible relevance of this feature to the dynamics at stake is discussed in Sections \ref{sec5} and \ref{sec7}.

In the following four sections we define volumes of the data cubes as intervals in $R$, $\varphi$ and $V_z$ aimed at selecting different components in order to study separately their morpho-kinematics. The parameters defining these volumes are listed in Table \ref{tab5}. While the strong inhomogeneity of the measured brightness across the data cube invites to select such fragments, there exists some arbitrariness in defining different components to be studied separately: the choice of four components retained in Table \ref{tab5} is not unique. It is based on the PV map of Figure \ref{fig5pai} and includes a South-western outflow, a North-eastern outflow, a Southern arc and a North-eastern arc.

We might have considered separately the blue-shifted and red-shifted parts of the South-western outflow when looking at CO emission alone. We refrained from doing so because the projections of the SiO emission on both the ($V_z,\varphi$) and ($V_z,R$) planes are not split and its anti-correlation with CO emission suggests that they should be studied together (see Figure \ref{fig5pai}, region marked as SWO, and Figure \ref{fig6}b below).

Similarly, we might have split the North-eastern outflow into two halves and have considered separately the northern part ($\varphi<60$\dego) and the southern part ($\varphi>60$\dego) as suggested by the $\varphi$ distribution displayed in Figure \ref{fig5}c. We refrained from doing so because it was shown earlier by \citet{Nhung2016} and illustrated in their Figure 12 that part of the Mira A wind is focused by Mira B in a way reminiscent of the predictions of Roche Lobe Overflow models \citep{Mohamed2007,Mohamed2012}: it seemed therefore wiser to study the blue-shifted part of eastern emission (marked as NEA and NEO in Figure \ref{fig5pai}) as a single component, including both what is focused by Mira B and what is not. Moreover, the approximate coincidence of the North-eastern outflow with the trace of the Mira B orbit displayed in the PV map of Figure \ref{fig5pai}, would rather suggest extending the outflow to the whole 360\dego\ $\varphi$ range.

Alternatively, we might have considered together the blue-shifted and red-shifted parts of eastern emission, in analogy with what is done for the south-western outflow. We refrained from doing so because SiO emission is absent in the eastern hemisphere and the North-eastern arc has such a strong identity that it seemed more reasonable to study it separately. We recall in this context that the strong north-eastern depletion observed within this arc (right panel of Figure \ref{fig4}) was interpreted by \citet{Ramstedt2014} as the wake of Mira B's accretion in the Mira A wind.

In any case, the arbitrariness of the adopted separation in four components must be kept in mind: only after having studied them separately in some detail, may we hope to have clearer ideas about their identity.

Finally, we estimate the importance of absorption by modelling the impact of optical thickness in locations of large column density. In the centre of the south-western outflow, assuming a uniform radial expansion velocity of 7 \kms\ and a temperature of 100 K, a simple radiative transfer calculation gives an absorption reaching $\sim$37\%. The result is quite insensitive to the precise values of the expansion velocity and of the temperature. As far as the results obtained in the present article are concerned, the effect is small enough to be ignored. As a further check we reduced and analysed data of the $^{13}$CO($\nu$=0, $J$=3-2) line emission from the same project as for the $^{12}$CO data. The ratio of the $^{12}$CO to $^{13}$CO brightness measured over regions of large column density is illustrated in Figure \ref{fig5}d. Its average value is 7.3 with an rms deviation with respect to the mean of 3.1. The spread between the values obtained for the South-western outflow (8.6), North-eastern outflow (6.9), North-eastern arc (7.5) and Southern arc (6.4) is only $\pm$0.8, confirming the small impact of absorption. Correcting for the different absorption in $^{12}$CO and $^{13}$CO data boosts the brightness ratio to $\sim$10.5. Accounting for the different excitation energies and frequencies this result translates into a $^{12}$CO/$^{13}$CO abundance ratio of 12$\pm$2, consistent with values observed for M stars \citep{Ramstedt2014b}. The uncertainty is a crude estimate based on the robustness of the result over various changes of parameters and hypotheses that we have explored.

Channel maps of CO emission are provided in Figure \ref{figa1} of the appendix.

\section{The South-western outflow}\label{sec4}

\begin{figure*}
  \includegraphics[height=5.cm,trim=0.5cm 1cm 0.5cm 2.cm,clip]{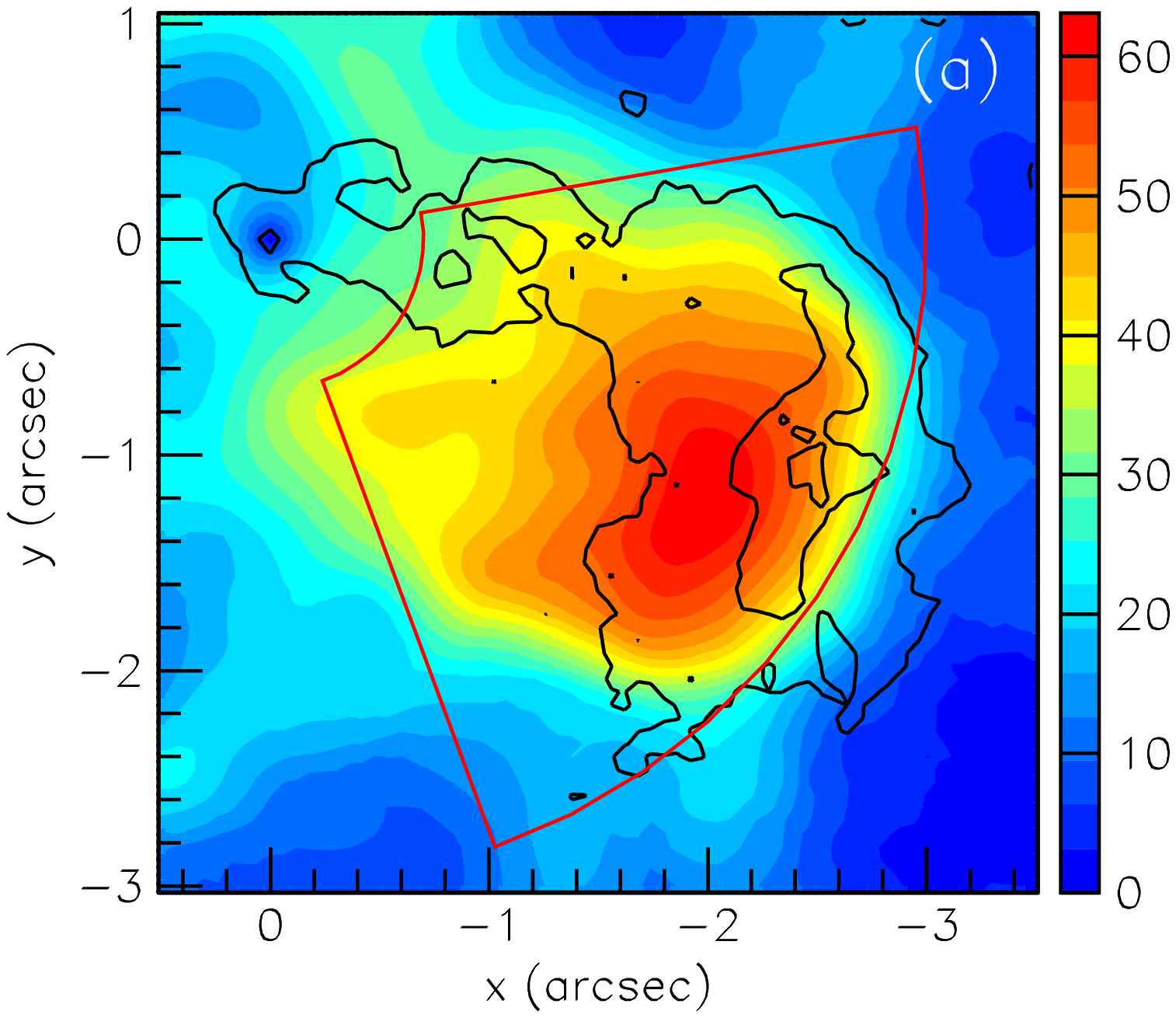}
  \includegraphics[height=5.cm,trim=0.5cm 1cm 0.5cm 2.cm,clip]{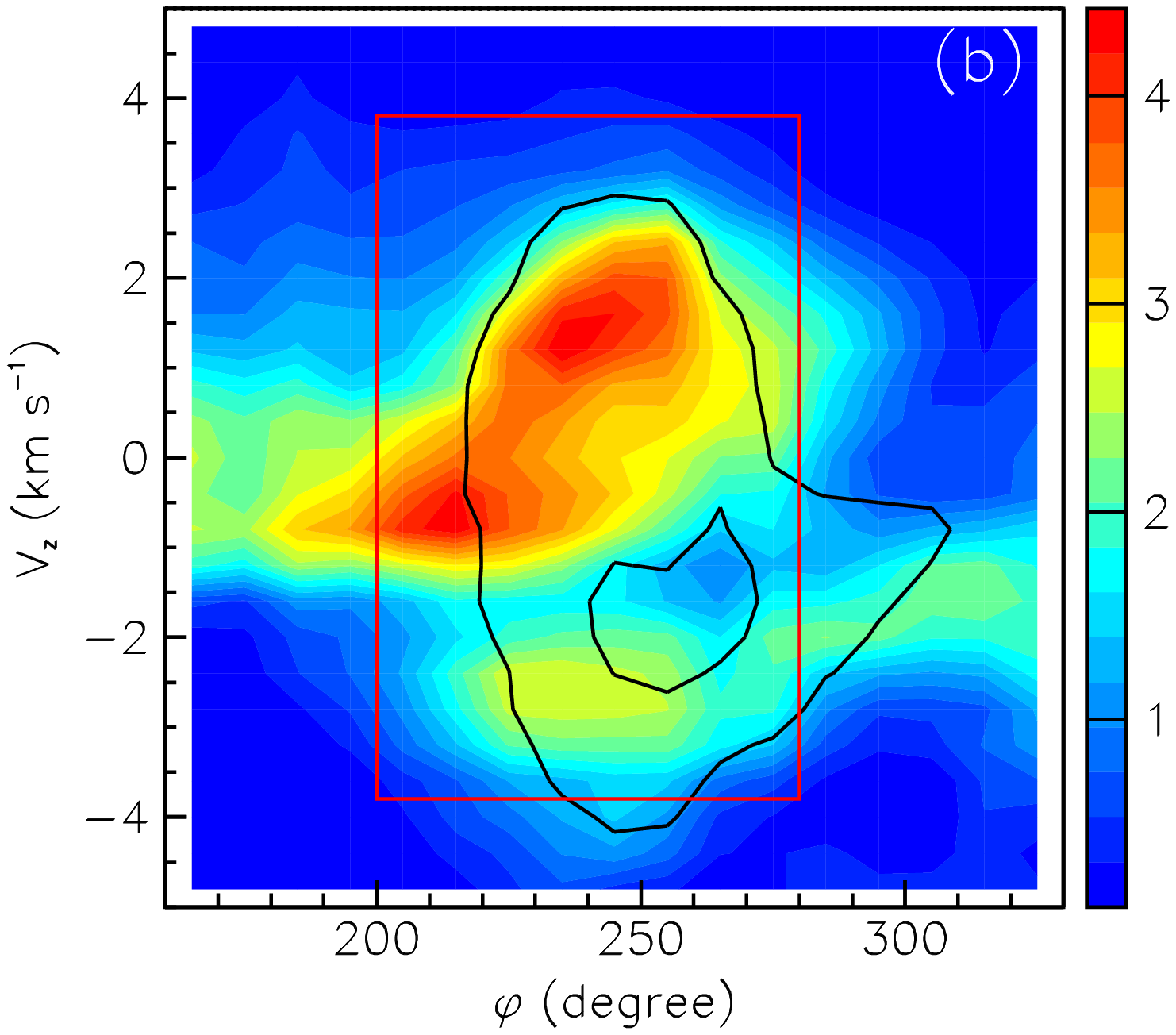}
  \includegraphics[height=5.cm,trim=0.5cm 1cm 0.5cm 2.cm,clip]{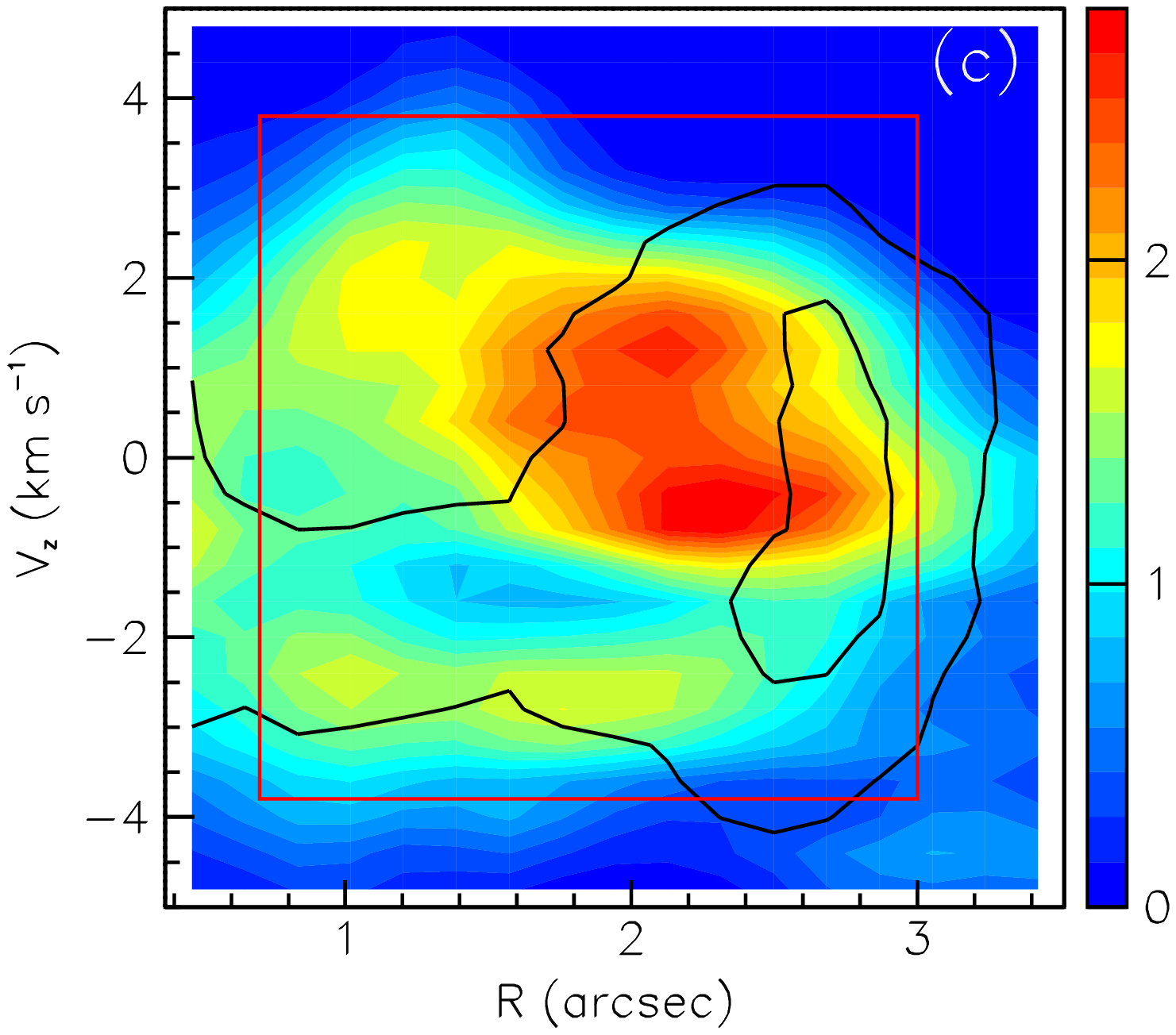}
  \includegraphics[height=5.cm,trim=0.5cm 1cm 0.5cm 2.cm,clip]{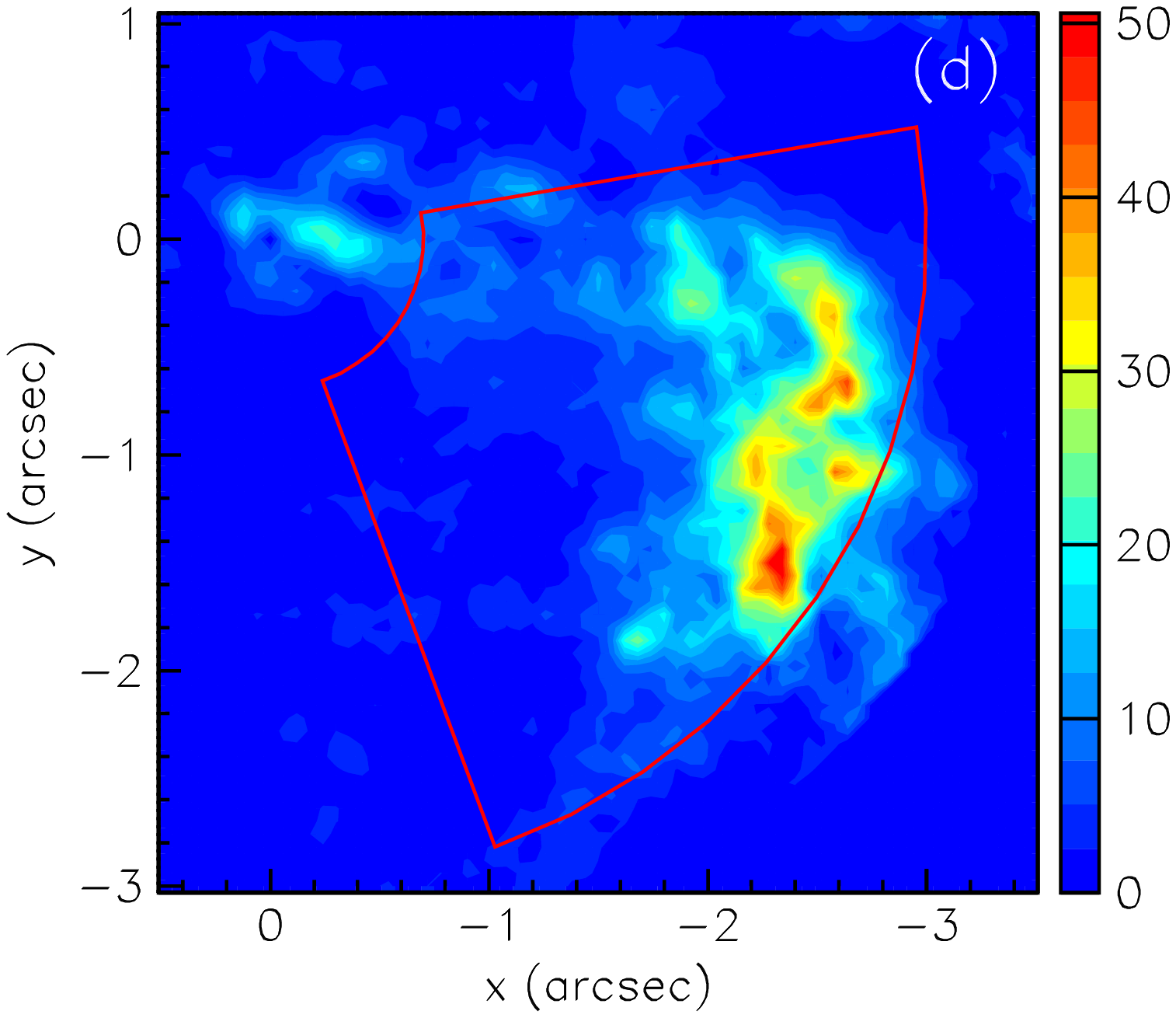}
  \includegraphics[height=5.cm,trim=0.5cm 1cm 0.5cm 2.cm,clip]{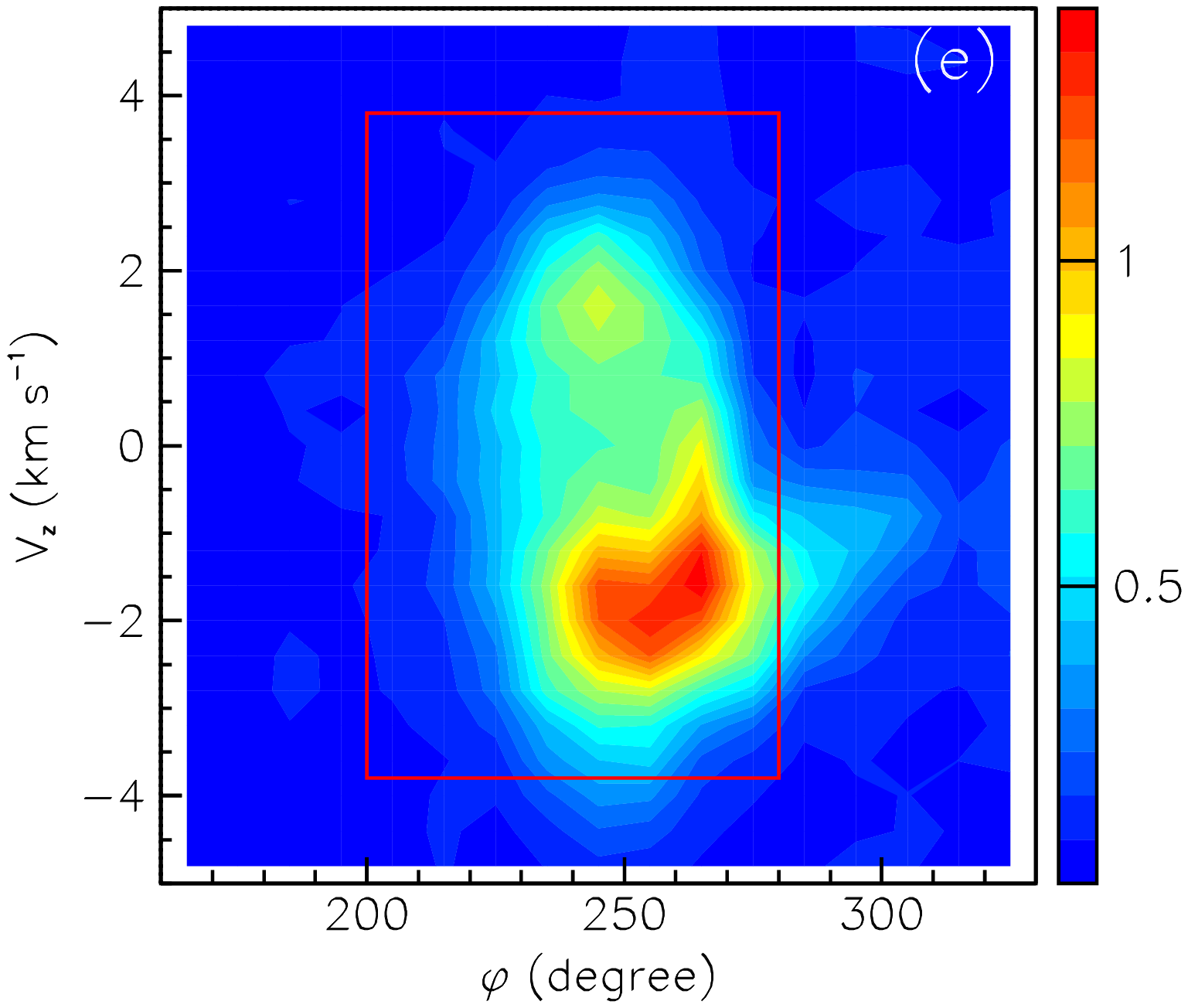}
  \includegraphics[height=5.cm,trim=0.5cm 1cm 0.5cm 2.cm,clip]{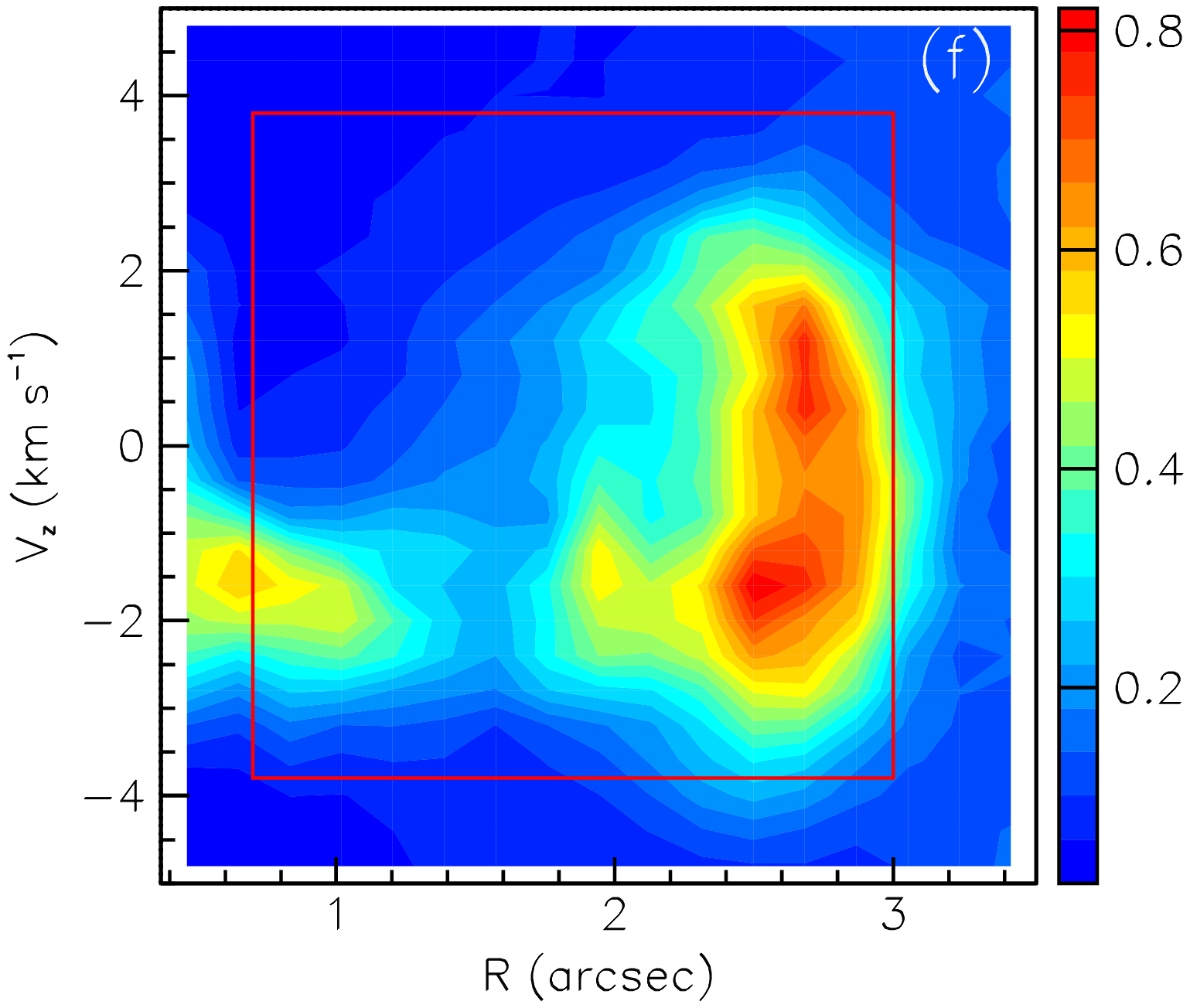}
\caption{Projections of the South-western outflow region of the data cube (see Table \ref{tab5}) on the ($x,y$) plane (left, multiplied by $R$), the ($V_z,\varphi$) plane (centre) and the ($V_z,R$) plane (right). Units for the left panels are Jy\,arcsec$^{-1}$\,\kms\ and for the other panels Jy. Red lines delimitate the selected volumes. The upper panels are for CO, with SiO contours at 6 and 25, 0.3 and 1.0, and 0.2 and 0.6 units respectively; the lower panels are for SiO.}
 \label{fig6}
\end{figure*}

The definition of the South-western outflow adopted in Table \ref{tab5} and illustrated in Figure \ref{fig6} is meant to cover both SiO and CO emissions, motivated by a possible relation between the two, but in spite of the split of CO emission between the blue-shifted and red-shifted hemispheres.

\begin{figure*}
  \includegraphics[height=5.5cm,trim=0.cm 1.5cm 1.cm 1.5cm,clip]{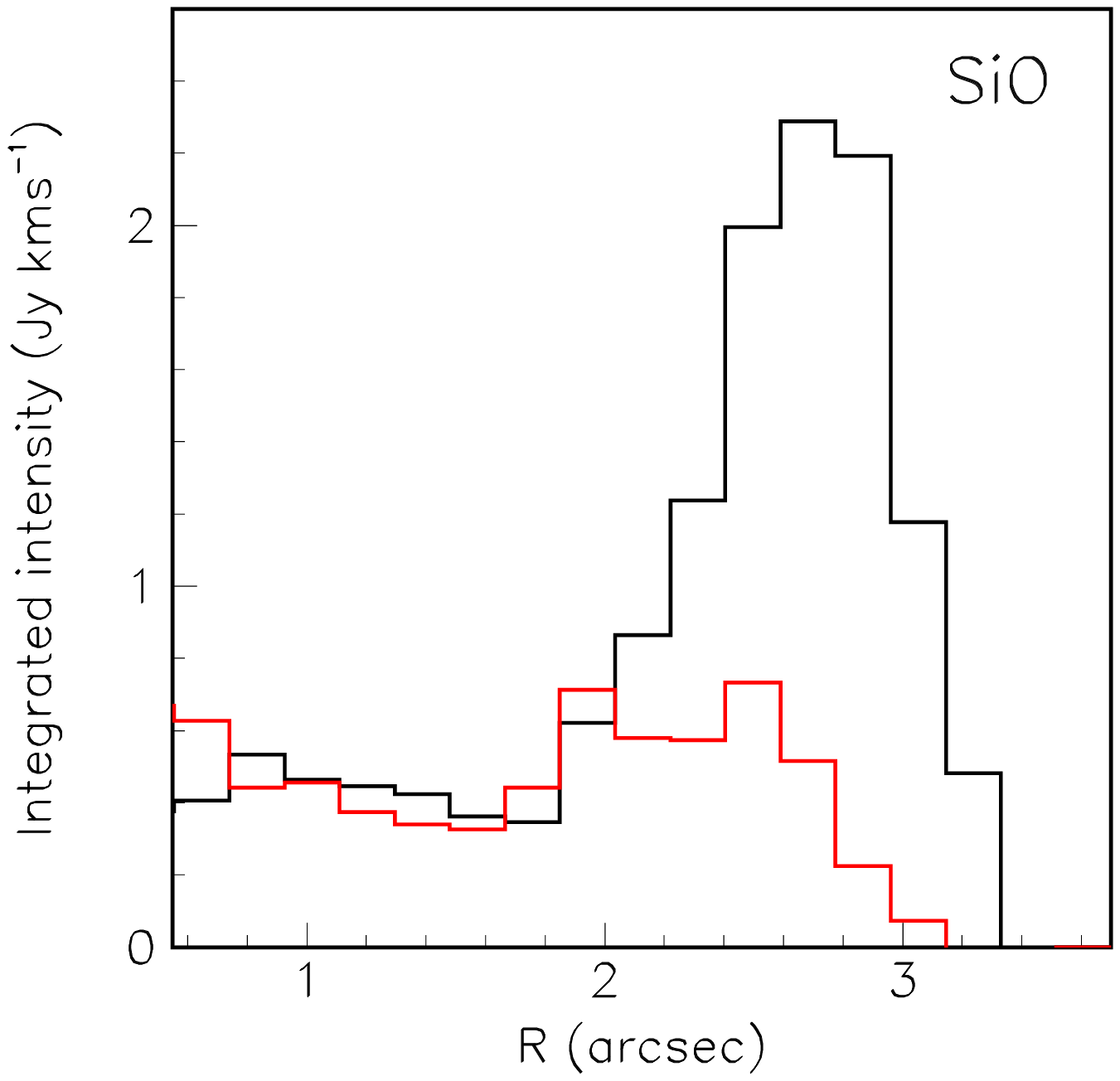}
  \includegraphics[height=5.5cm,trim=0.cm 1.5cm 1.cm 1.5cm,clip]{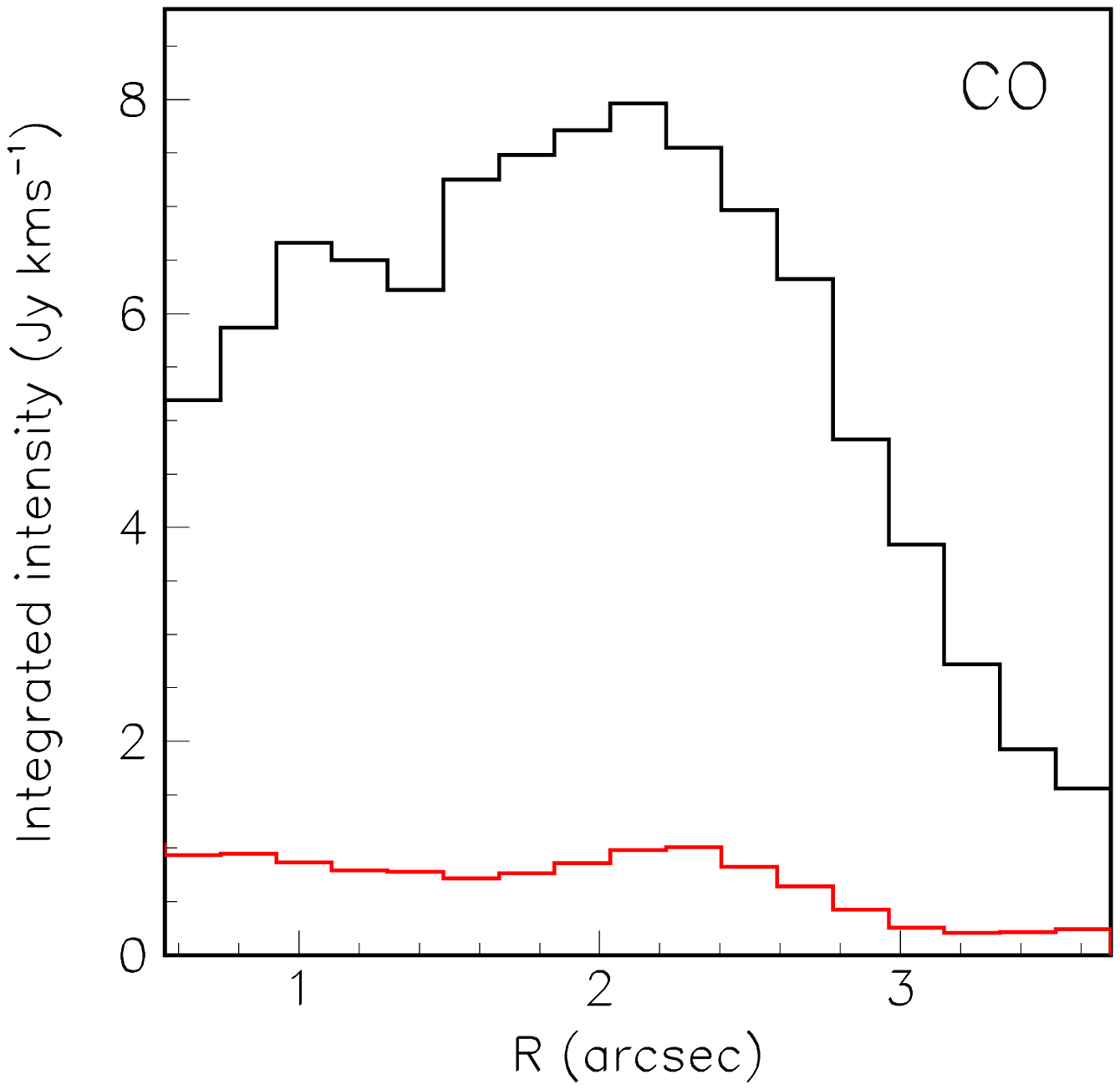}
  \caption{Left: radial profile of SiO and CO emissions integrated in the regions corresponding to the first ($-2.7<V_z<-0.7$ \kms, 250\dego$<\varphi<$280\dego, red) and second ($-4<V_z<4$ \kms, 200\dego$<\varphi<$250\dego, black) components of SiO emission (see text).}
  \label{fig7}
\end{figure*}

The morphology of SiO emission is simple and consists of two components, one confined to a narrow region of the ($\varphi,V_z$) plane and the other confined to a narrow range of $R$ (Figure \ref{fig6}e,f). The first component covers approximately 2 \kms\ and 40\dego\ centred on $V_z\sim-1.7$ \kms\ and $\varphi\sim265$\dego. Its radial profile is shown in the left panel of Figure \ref{fig7}. The second component covers a broad range of $\varphi$ and $V_z$, at steradian scale if one assumes that its extensions in the sky plane and perpendicularly to it are similar (lower panels of Figures \ref{fig6}). 

At distances from Mira A exceeding some 100 au, one expects the vast majority of SiO molecules to have condensed into grains \citep{Wong2016, Khouri2018}. The presence of SiO molecules in the gas phase revealed by millimetre emission should therefore trace their extraction from dust grains as the combined effect of temperature and turbulence, such as associated with shocks \citep{Schilke1997, Flower2003, Burkhardt2019,Gusdorf2008}. Indeed the radial profile of SiO emission, shown in the left panel of Figure \ref{fig7}, displays a sharp decrease at larger radii as expected from a shock wave. A possible interpretation is therefore of a mass ejection having produced the first component some time ago, accompanied by a shock wave covering most of the South-western outflow, and producing the second component; in this interpretation, the outflow pre-exists to the mass ejection and covers a broad solid angle on either side of the sky plane as the emission covers both negative and positive $V_z$ values. If mass ejection took place on the occasion of the December 2003 X-ray flare \citep{Karovska2005}, it has travelled some 300 au, namely 4.5\,10$^{10}$ km, in 11 years, namely $3.4\,10^8$ s, meaning a velocity of $\sim$130 \kms, typical for coronal mass ejections of MS stars. The shock front has an angular extent of $\sim$50\dego, between 220\dego\ and 270\dego\ in $\varphi$, and a velocity range of $\sim$7 \kms, between $-$4 and 3 \kms\ in $V_z$ (Figure \ref{fig6}e). Assuming that it is approximately circular and is roughly symmetric about the plane of the sky, the velocity range would correspond to an angular extent of 50\dego\ normal to the plane of the sky, namely $\pm$25\dego\ approximately and the expansion velocity in the sky plane would then be 3.5/tan25\dego$\sim$7 \kms. The trailing behind the shock wave would then suggest that dust grains condense again within typically one arcsec, meaning 40 years or so.

The shock front shows a hot spot at its intersection with the first component associated with the trajectory of the mass ejection and with a clear suppression of CO emission at $(V_z,\varphi)=(-1.7\,\mbox{\kms}, 260\mbox{\dego})$ (Figures \ref{fig5pai} and \ref{fig6}b) suggesting that CO molecules have been expelled and/or dissociated along the trajectory of the mass ejection; this implies in turn that it would take more than 100 years for CO molecules to recombine in this region of the circumbinary envelope where the temperature is expected to take values between 100 and 200 K \citep{Wong2016}, probably because the density has been lowered by the mass ejection. Similarly, along the trajectory of the mass ejection, an important fraction of the SiO molecules has remained in the gas phase, revealing a slower condensation on dust grains than behind the shock wave.

CO emission covers approximately between $-$4 and 4 \kms\ but is enhanced above $-$1 \kms\ (Figure \ref{fig6}b,c). In the scenario described in the preceding paragraphs, an outflow was therefore present before the 2003 event and extended over some 240\dego$\pm$40\dego\ in $\varphi$ and between $\sim -4$ and 4 \kms\ in $V_z$. The proposed scenario offers no explanation for the formation of such an outflow, which implies a significant breaking of spherical symmetry; it is unrelated to the subsequent mass ejection, which happened to occur near the edge of the outflow ($\sim$15\dego\ south of the northern edge and $\sim$2.3 \kms\ red-ward from the blue-shifted edge). The validation of such a scenario would require that the observed morphology of CO emission, in particular its radial dependence, be explained by dissociation and subsequent recombination of CO molecules.

\begin{table*}
  \caption{Radial outflows expanding at approximately constant velocity. The coordinates in the ($V_z,\varphi$) plane of the centres of the outflows illustrated in Figure \ref{figa3} are listed for 4 bins of equal width covering between 1 and 3 arcsec in $R$.}
  \label{tab4}
  \begin{tabular}{ccc}
    \hline
    $R$ (arcsec)&$V_z$ (\kms)&$\varphi$ (degrees)\\
    \hline
    \multicolumn{3}{c}{CO: North-eastern Outflow}\\
    \hline
    1-1.5&$-$1.7&109\\
    \hline
    1.5-2&$-$1.8&99\\
    \hline
    2-2.5&$-$1.8&105\\
    \hline
    2.5-3&$-$1.6&103\\
    \hline
    Mean (no weight)&$-$1.7&104\\
    \hline
    \multicolumn{3}{c}{CO: South-western Outflow}\\
    \hline
    1-1.5&2.4&253\\
    \hline
    1.5-2&1.5&239\\
    \hline
    2-2.5&1.5&241\\
    \hline
    2.5-3&1.2&242\\
    \hline
    Mean (no weight)&1.7&244\\
    \hline
    \multicolumn{3}{c}{SiO: South-western Outflow}\\
    \hline
    1-1.5&$-$1.5&275\\
    \hline
    1.5-2&$-$1.3&270\\
    \hline
    2-2.5&$-$2.0&255\\
    \hline
    2.5-3&$-$1.5&242\\
    \hline
    Mean (no weight)&$-$1.7&260\\
    \hline
\end{tabular}
\end{table*}

\begin{table*}
  \caption{Parameters defining the CO emission fragments studied in Sections \ref{sec4} to \ref{sec7}.}
  \label{tab5}
  \begin{tabular}{cccccc}
   \hline
    &\makecell{$R$ interval\\(arcsec)}&\makecell{$\varphi$ interval\\(degree)}&\makecell{$V_z$ interval\\(\kms)}&\makecell{Integrated flux density\\(Jy\,\kms)}&\makecell{Volume\\(arcsec$^2$\,\kms)}\\
    \hline
    South-western outflow&[0.7, 3]&[200, 280]&[$-$3.8, 3.8]&117&34\\
    \hline
    North-eastern outflow&[1, 3.7]&[$-$30, 150]&[$-$3.8, $-$0.6]&98&46\\
    \hline
    North-eastern arc&[1.3, 2.3]&[$-$30, 110]&[$-$1, 3.4]&44&18\\
    \hline
    Southern arc&\makecell{(see section \ref{sec7})}&[120, 180]&[$-$0.6, 3.4]&13&10\\
    \hline
    Total&$-$&$-$&$-$&272&108\\
    \hline
    Reference&[1., 3.5]&[0, 360]&[$-$5, 5]&375&179\\
    \hline
  \end{tabular}\\
  
\end{table*}

The radial profiles of CO emission in the same ranges of $V_z$ and $\varphi$ as for SiO are displayed in the right panel of Figure \ref{fig7}. Emission is nearly suppressed along the first SiO component and the intensity in the region of the second SiO component starts to decrease beyond $R\sim2$ arcsec at the same place where the SiO intensity starts to rise. We note that in the region studied here the temperature dependent factors that scale the CO emissivity, of the form $(2.8 \mbox{K}/T)\exp(-33.2 \mbox{K}/T)$, do not vary rapidly: they reach a maximum at 33.2 K and decrease slowly with increasing $T$, reaching half-maximum at $T\sim150$ K and a tenth of it at $T\sim1000$K. It is therefore unlikely that temperature alone be responsible for shaping CO emission; it is more natural to interpret it as a combined result of turbulence and lower density, along the lines of the mass ejection/shock wave scenario described above for SiO emission.

Early observations of CO emission \citep{Fong2006, Josselin2000,Planesas1990} were made before the occurrence of the X ray burst. It is difficult to tell if they do or do not detect the outflow because their angular resolution is not good enough but they do show an east-west enhancement. We remark that at larger distances from the stars \citep{Ramstedt2014} the South-western outflow is seen to extend up to some 10 arcsec in the form of a nearly detached shell, interpreted by these authors as suggesting the occurrence of an earlier event of enhanced mass loss, possibly associated with the ejection of the Blue-shifted ring or bubble.

\section{The North-eastern outflow}\label{sec5}

\begin{figure*}
  \includegraphics[height=4.2cm,trim=.9cm 1.cm .5cm 1.5cm,clip]{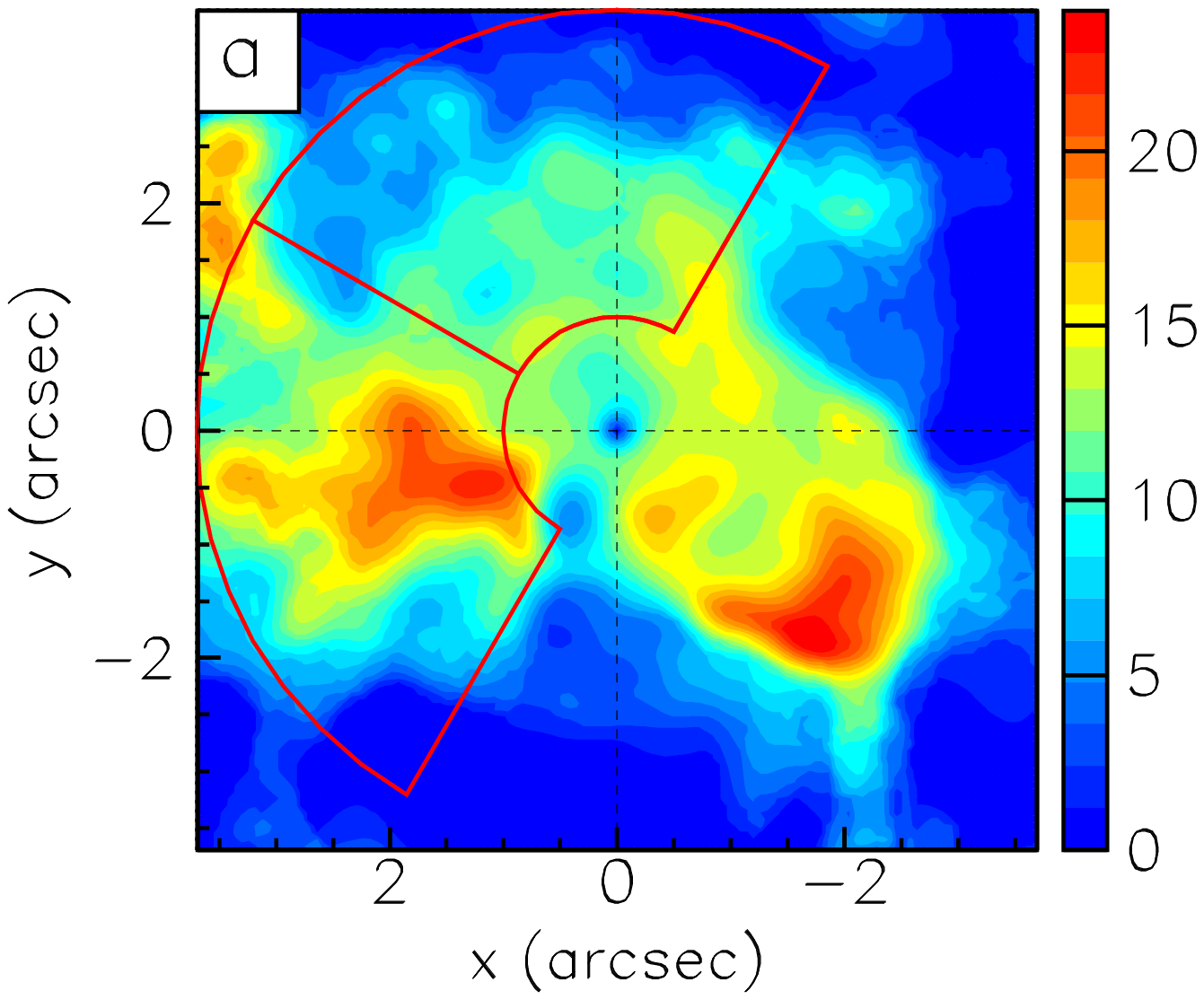}
  \includegraphics[height=4.2cm,trim=0.7cm 1.cm .5cm 1.5cm,clip]{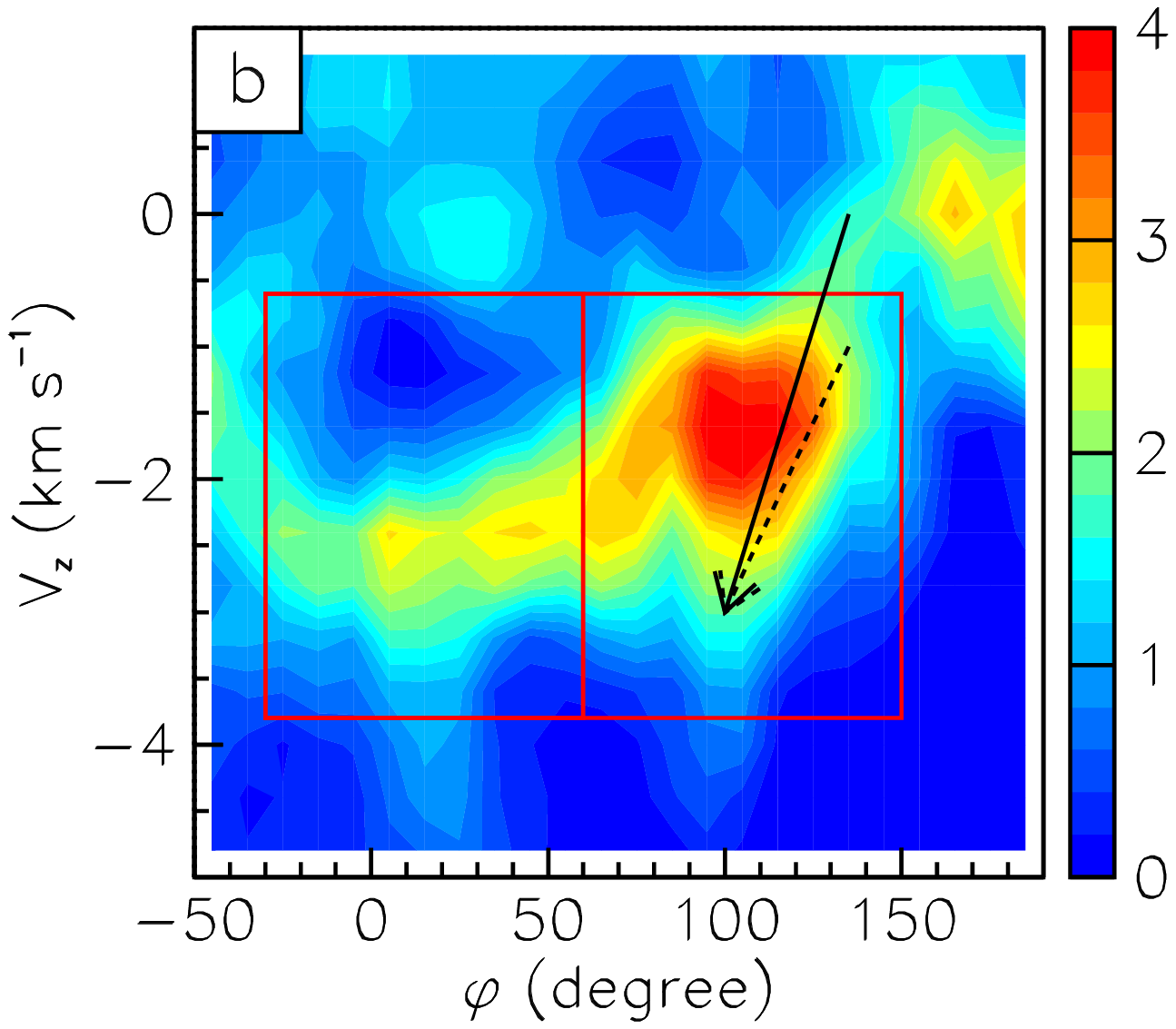}
  \includegraphics[height=4.2cm,trim=0.7cm 1.cm 2.3cm 1.5cm,clip]{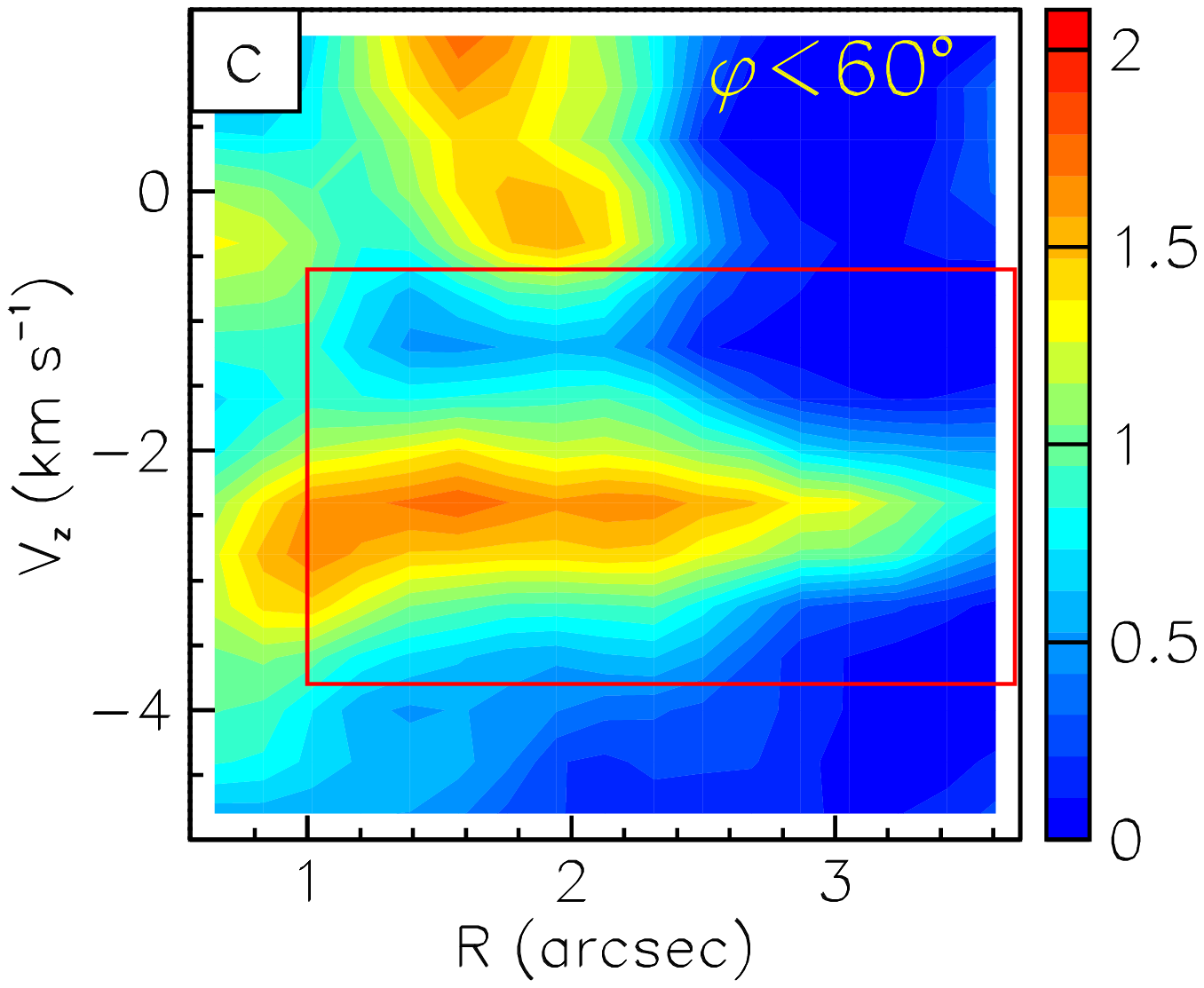}
  \includegraphics[height=4.2cm,trim=2.8cm 1.cm .4cm 1.5cm,clip]{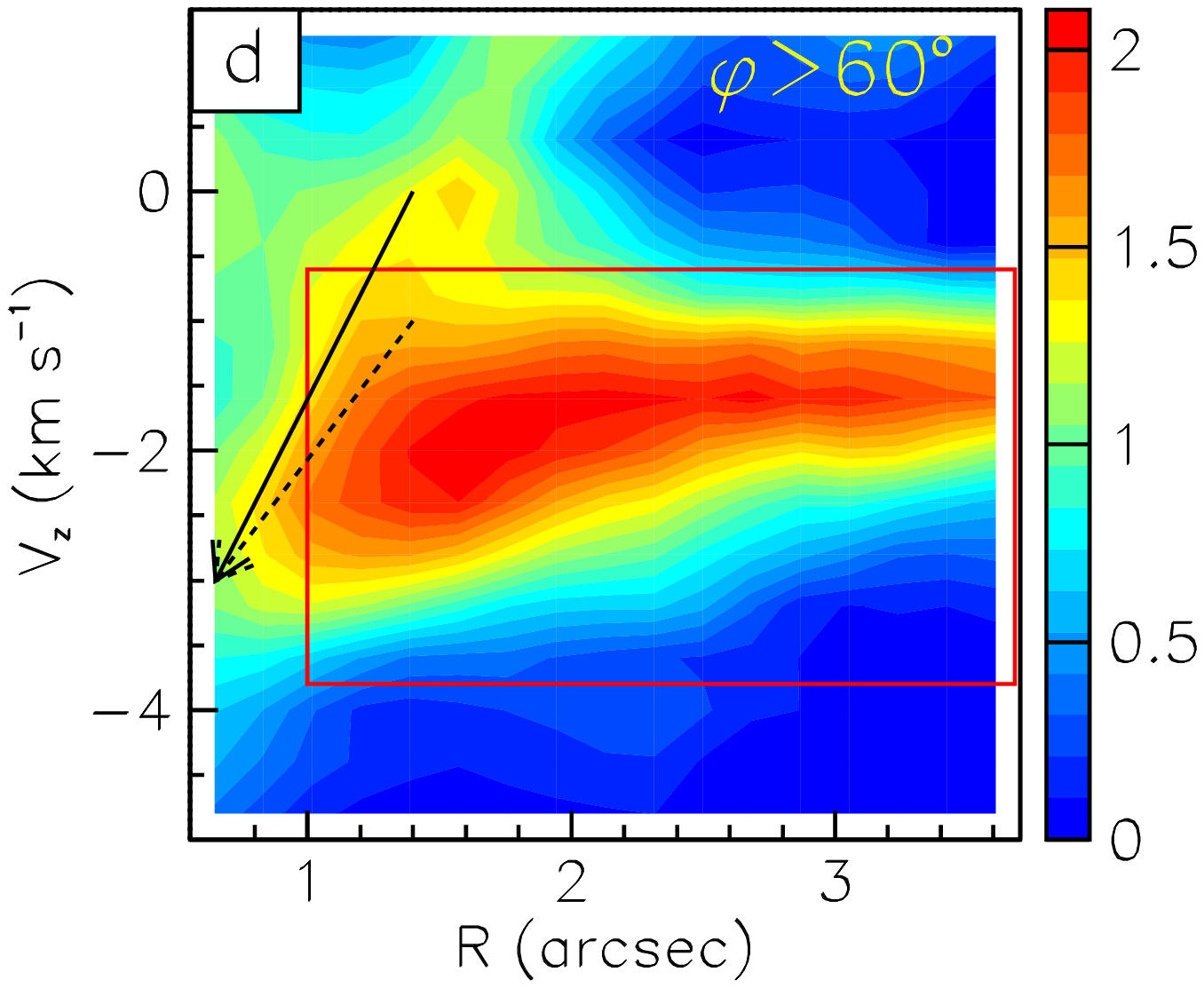}
  \caption{Projections of the North-eastern outflow region of the data cube (see Table \ref{tab5}) on the ($x,y$) plane (panel a, multiplied by $R$), the ($V_z,\varphi$) plane (panel b) and the $(V_z,R)$ plane for two $\varphi$ ranges (panel c and d). Red lines delimitate the selected volumes. The arrows in panels b and d illustrate the motion of the Mira B focused wind between 1923 and the time of observation (see text). They are shown for systemic velocities of both 47.7 \kms\ (full) and 46.7 \kms\ (dashed). They start at the position of the wind that was flowing from Mira A to Mira B in 1923 and end at the position of the wind that was flowing from Mira A to Mira B in 2014, namely pointing to smaller values of $R$. The colour scales are in units of Jy\,arcsec$^{-1}$\,\kms\ for panel a and Jy for the other panels.}
    \label{fig8}
\end{figure*}

\begin{figure*}
  \includegraphics[height=4.8cm,trim=0.cm 1.cm 0.cm 1.5cm,clip]{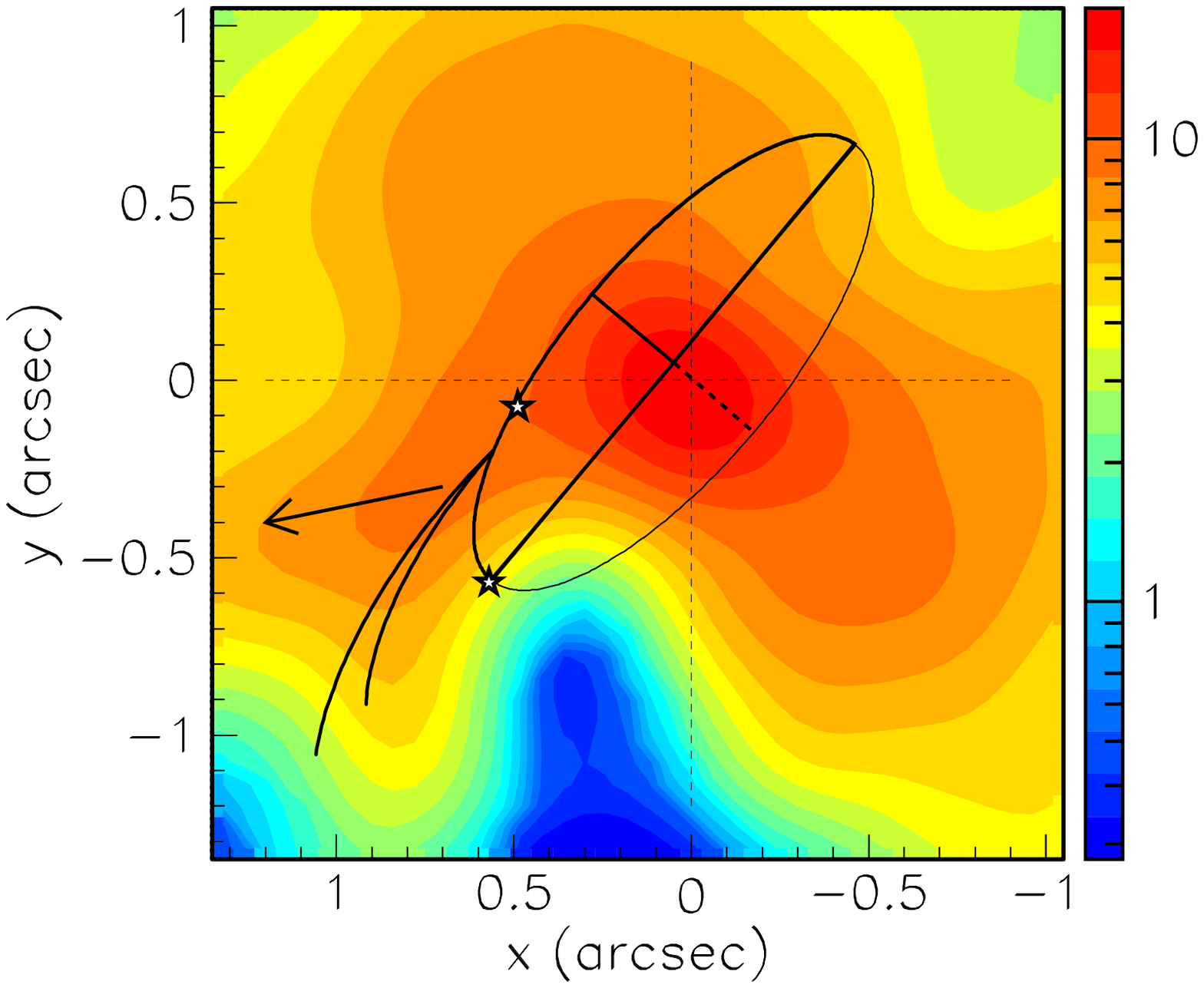}
  \includegraphics[height=4.8cm,trim=0.cm 1.cm 0.cm 1.5cm,clip]{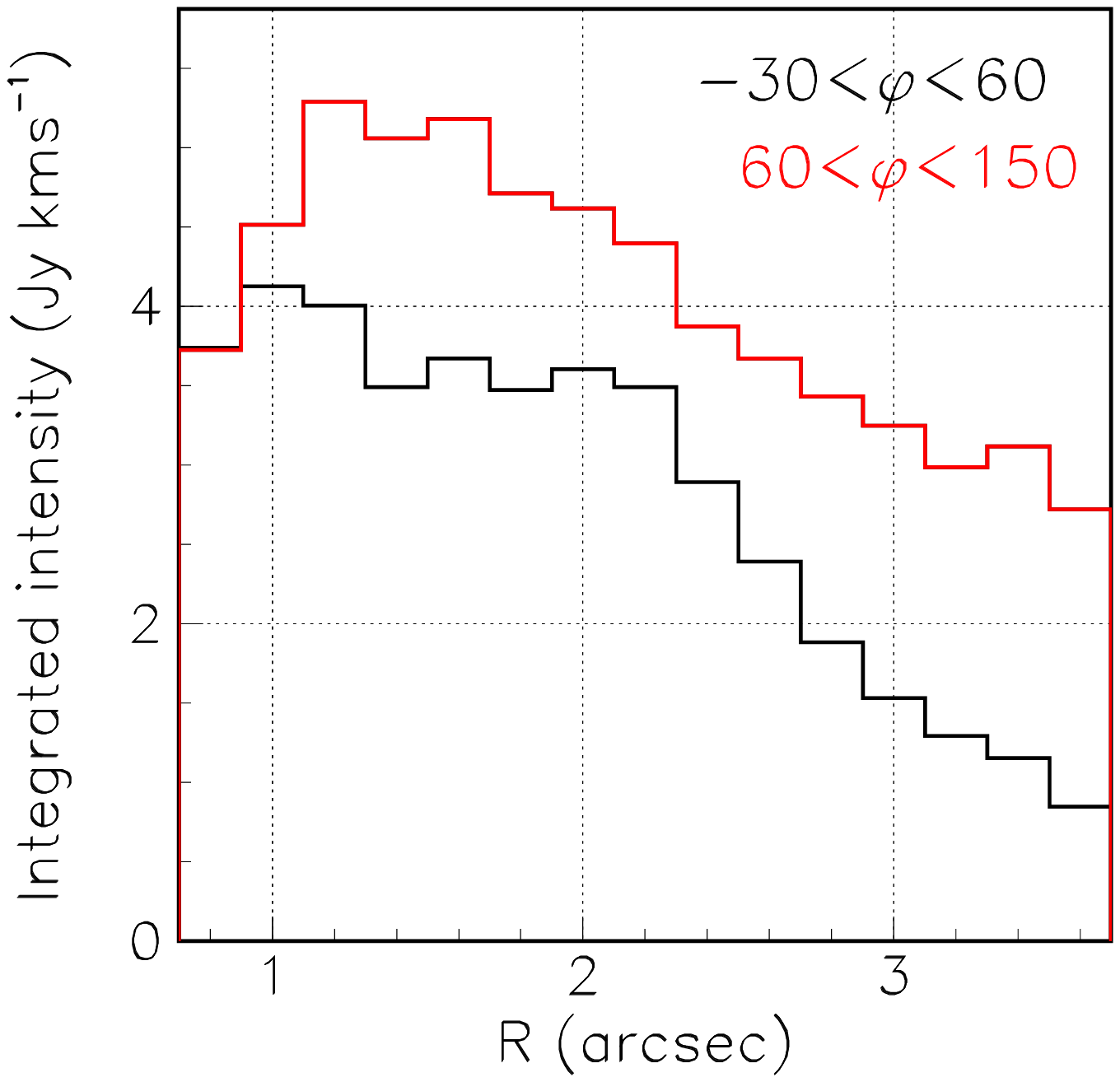}
  \includegraphics[height=4.8cm,trim=0.cm 1.cm 0.cm 1.5cm,clip]{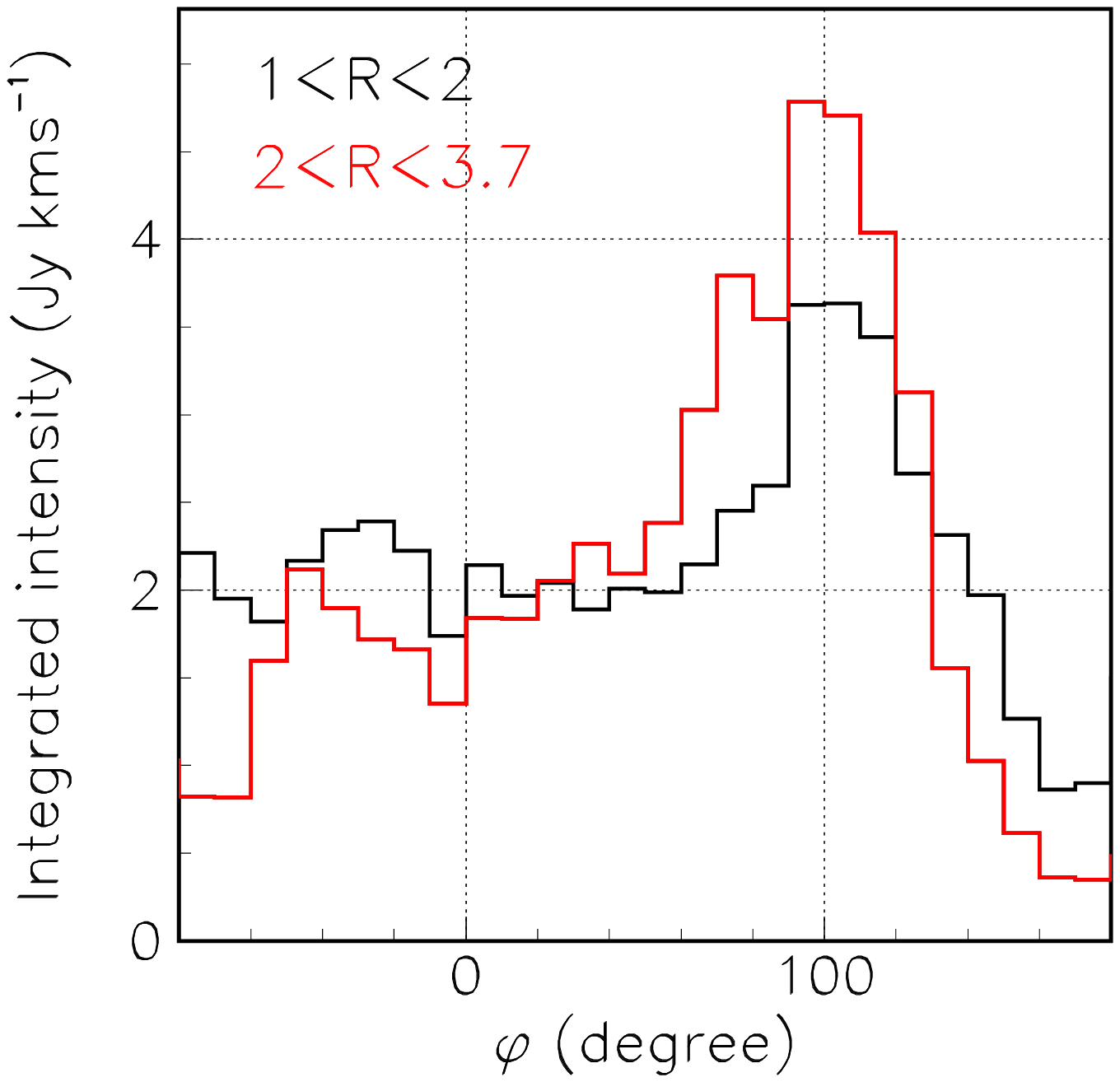}
  \caption{Left: intensity map of CO emission for $-3.4<V_z<-2.2$ \kms\ (Jy arcsec$^{-2}$\kms). The ellipse is the projection of the Mira B orbit \citep{Prieur2002}. The curves show the expected positions of a gas volume flowing in the Mira A to Mira B direction with a velocity of 3.9 (less curved) and respectively 2.6 (more curved) \kms. The stars show the positions of Mira B in 1923 (more south) and at the time of observation (more north). The arrow shows an unexplained eastern arm. Centre: dependence on $R$ of the intensity of the North-eastern outflow integrated over the region $-3.8<V_z<-0.6$ \kms\ for $-$30\dego$<\varphi<60$\dego\ (black) and 60\dego$<\varphi<150$\dego (red). Right: dependence on $\varphi$ of the intensity of the North-eastern outflow integrated over the region $-3.8<V_z<-0.6$ \kms\ and $1<R<2$ arcsec (black) or $2<R<3.7$ arcsec (red).}
 \label{fig9}
\end{figure*}

The North-eastern outflow, as defined in Table \ref{tab5}, is illustrated in Figure \ref{fig8}. It includes gas flowing in the direction of Mira B that is gravitationally  focused by it, a fraction of which being probably accreted, as discussed earlier in \citet{Nhung2016} and illustrated in the left panel of Figure \ref{fig9} for $-3.4<V_z<-2.2$ \kms. For this reason, part of the discussion of the present section addresses properties of the morpho-kinematics at shorter distances from Mira A than implied by the values quoted in Table \ref{tab5}, in the direction of Mira B, between  approximately 0.5 and 1 arcsec from Mira A.

The orbit of Mira B has been calculated by \citet{Prieur2002}, however with large uncertainties; \citet{Planesas2016} have commented on the need for improving their result in the light of more recent measurements. Measurements of the Mira B position projected on the plane of the sky cover the past century starting in 1923. The calculation produces a nearly circular orbit with a mass of 4.4 M$_{\odot}$ for the AB pair, while the improved analysis of \citet{Planesas2016} prefers a pair mass of 2.7 M$_{\odot}$. The result is displayed in the left panel of Figure \ref{fig9}. We lack sufficient information to model reliably the morpho-kinematics of the part of the Mira A wind that is focused by Mira B, but we can perform a very crude evaluation of its main features. We describe the focused wind as a beam of constant velocity $V$ flowing along the line linking Mira A to Mira B. It spans the whole sky plane in the manner of a light-house beam and is alternatively blue- and red-shifted with a period of some 500 years \citep{Prieur2002} depending on the relative positions of Mira A and Mira B. In particular it started being blue shifted near 1923, corresponding to the major axis of the projection of the orbit on the sky plane, and will remain so for some 150 years. Such a model makes only sense to the extent that the projected radial velocity is larger than the projected orbital velocity. Being radial, the flow projects as a point on the ($V_z$ vs $\varphi$) PV map: the history of the focused part of the Mira A wind can therefore be read from the sinusoidal trajectory of this point on this PV map. Such trajectories were displayed in Figure \ref{fig5pai}. Within the plane of the orbit, emission should be enhanced on a spiral following the motion of Mira B about Mira A. Such a spiral has a radius that increases counter-clockwise when projected on the sky plane, namely in the opposite direction of the clockwise rotation of Mira B about Mira A. This result should remain valid when using more realistic models. Indeed the model described by \citet{Mohamed2012} obeys this general law (note that in their Figures 1 and 2 Mira B is on the right, meaning that east points right and therefore Mira B rotates counter-clockwise; in their Figure 1 where Mira B is fixed the sky rotates clockwise as it should and in their Figure 3 the spiral has a radius increasing clockwise, as it should). Figures \ref{fig8}b and \ref{fig9} show indeed enhanced emission in the expected region, but Figure \ref{fig9} suggests the presence of another arm bending east.

Within the plane of the orbit, the velocity $V$ of the flow governs the relation between $V_z$ and $z$, $V_z/V=z/\sqrt{R^2+z^2}$. In particular, in 1923, date of the first accurate measurement, Mira B was approximately in the plane of the sky containing Mira A and both $z$ and $V_z$ cancelled. At the time of the present observations, some 90 years later, we estimate from the CO channel maps \citep[][and left panel of Figure \ref{fig9}]{Nhung2016} that $V_z\sim-3.0\pm1.0$ \kms. At that time, in 2014, the line joining Mira A to Mira B was inclined by $\sim$50\dego\ with respect to the plane of the sky (pointing toward us) and projected on it at a position angle of $\sim$100\dego. We infer from this that $V=V_z/\sin50$\dego$\sim3.9\pm1.3$ \kms. In 90 years, the wind that was focused by Mira B in 1923 and was blowing in the plane of the sky has therefore spanned $\sim0.7\pm0.2$ arcsec. The slope of the arrow displayed in Figure \ref{fig8}d, spanning 3 \kms\ in $V_z$ and 0.7 arcsec in $R$ is again consistent with observation. The implied trajectory displayed in the left panel of Figure \ref{fig9} bends in the right direction but prefers the lower value of the wind velocity, 2.6 rather than 3.9 \kms, meaning 0.5 rather than 0.7 arcsec for the distance covered since 1923. We also show in Figure \ref{fig8} arrows obtained for a systemic velocity of 46.7 \kms. We underline the crudeness of these arguments: they show the consistency between the model and the observations but do not guarantee that the proposed interpretation is unique.

The central and right panels of Figure \ref{fig9} display the dependence on $R$ and $\varphi$ of the brightness of the North-eastern outflow integrated between $-$3.8 and $-$0.6 \kms. They show continuity over the whole $\varphi$ and $R$ ranges of the North-eastern outflow.

An important result of this analysis is that, at the scale of a century or a few centuries, the effect of Mira B focusing the wind of Mira A can account for only a small perturbation of the North-eastern outflow. Most of it extends over a broad range of $R$, well beyond 3 arcsec and a broad range of position angles, from $-$30\dego\ to 150\dego, and covers nearly 2 \kms\ in the blue-shifted hemisphere. For it to be related to Mira B focusing, it would have to be formed from the accumulation of a slow focused wind over several orbital periods and would require enough dilution of the density to produce a uniform distribution. In such a scenario, the North-eastern outflow would only be the blue-shifted part of the enhancement that has been identified in the PV map of Figure \ref{fig5pai}. Its red-shifted part would overlap the South-western outflow and it would be difficult to identify it. At a velocity of 1 to 2 \kms, the wind would travel between 1 and 2 arcsec per orbital period: to produce the uniform brightness distribution illustrated in the $V_z$ vs $R$ maps of Figure \ref{fig8} would require important dilution and velocity spread.  Yet, without more detailed modelling, such a scenario cannot be excluded and would have the advantage of providing an explanation for the observed morphology.

In summary, the sinusoidal lines of enhanced emission displayed in the PV map of Figure \ref{fig5pai} suggest the presence of an isotropic slow wind confined to the vicinity of the orbital plane but, within this plane, only a small fraction of it can reliably be assigned to Mira B focusing and has twice as high a velocity. However, the possibility that the bulk of it be the result of accumulation over several orbital periods of a slow focused wind cannot be excluded. Another plausible scenario is that the bulk of the North-eastern outflow has no relation with Mira B, in which case the question of what caused its formation remains open.  

\section{The North-eastern arc} \label{sec6}

\begin{figure*}
  \includegraphics[height=4.9cm,trim=0.cm 1.cm 0.5cm 1.5cm,clip]{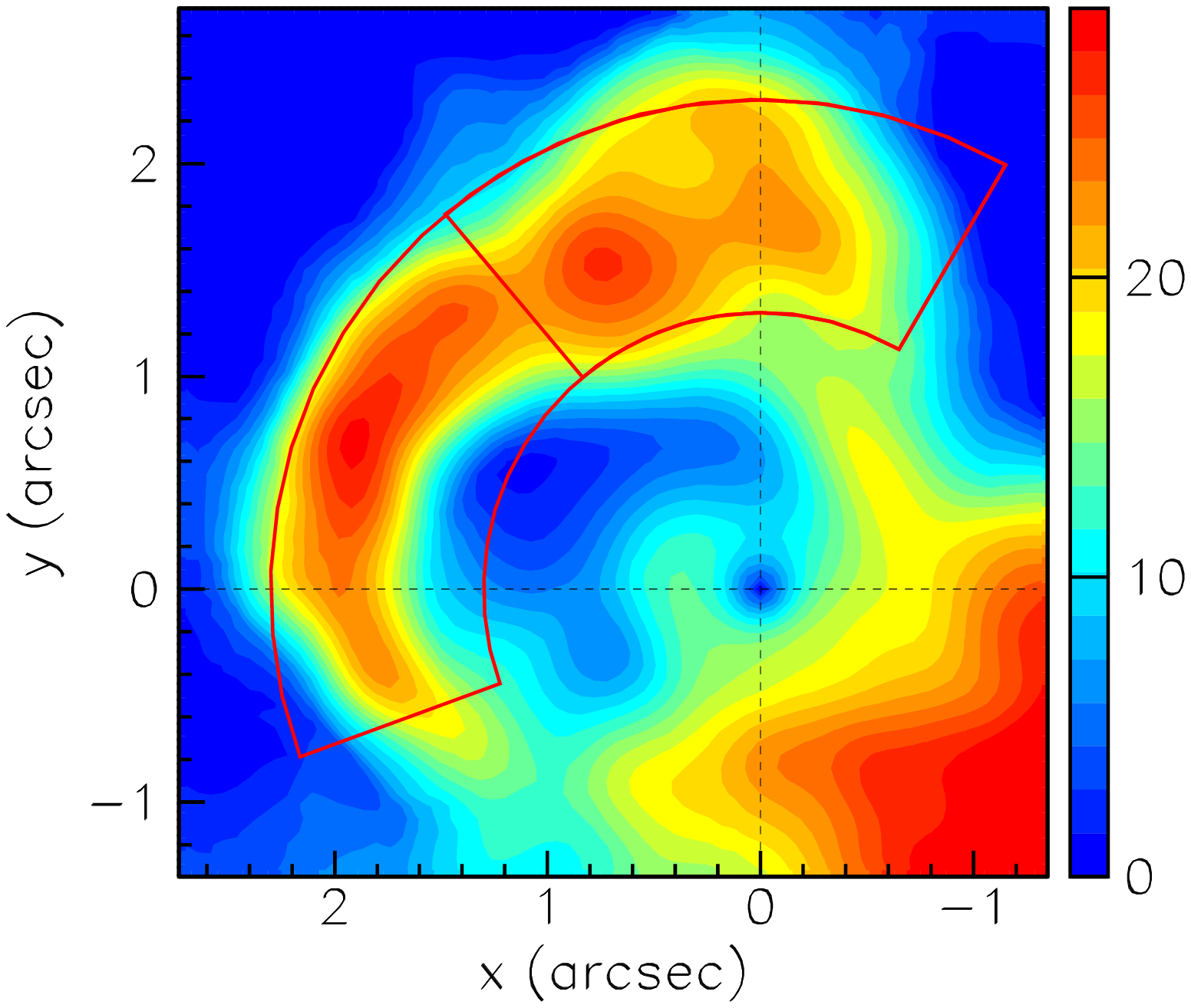}
  \includegraphics[height=4.9cm,trim=0.cm 1.cm 0.5cm 1.5cm,clip]{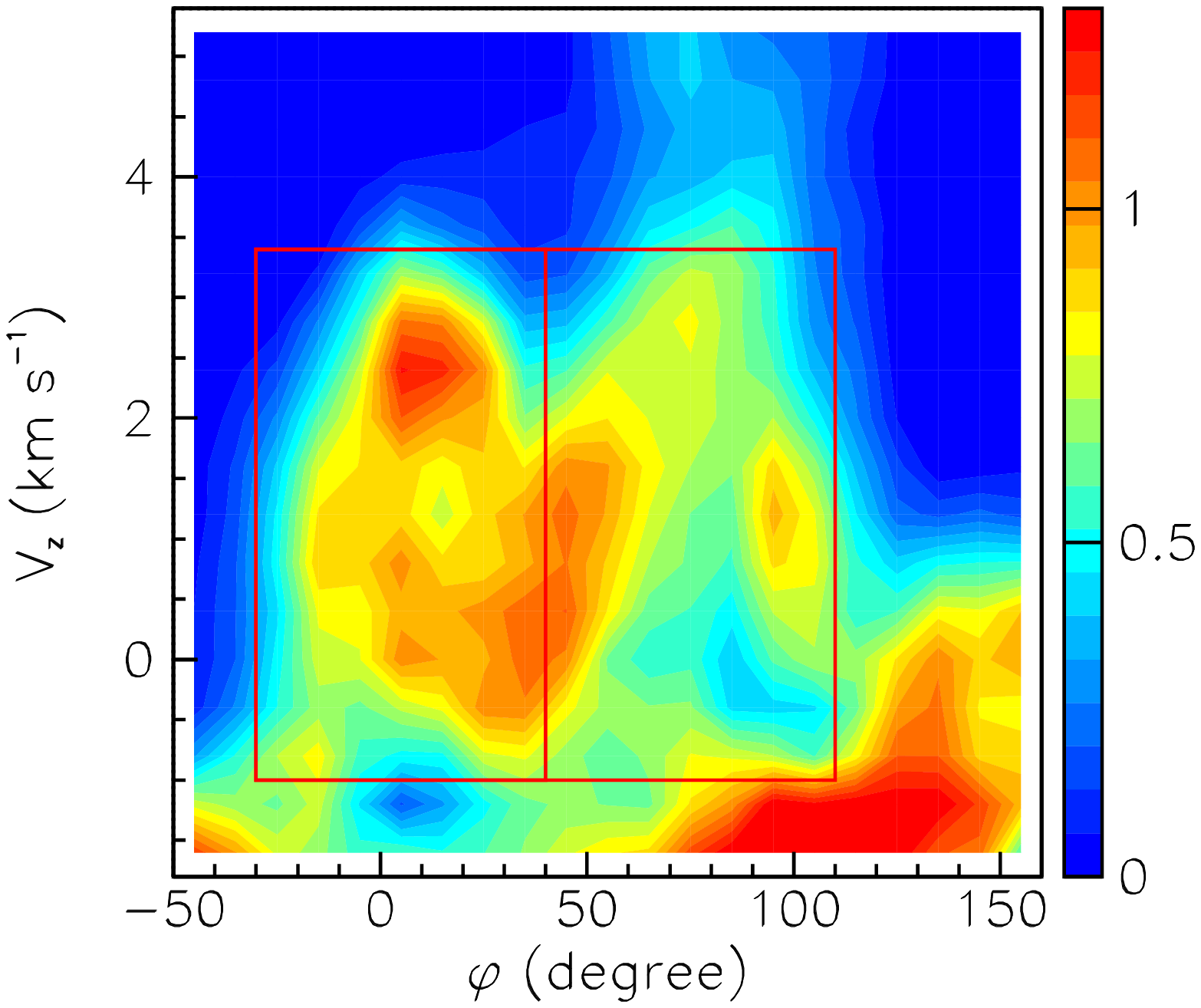}
  \includegraphics[height=4.9cm,trim=0.cm 1.cm .5cm 1.5cm,clip]{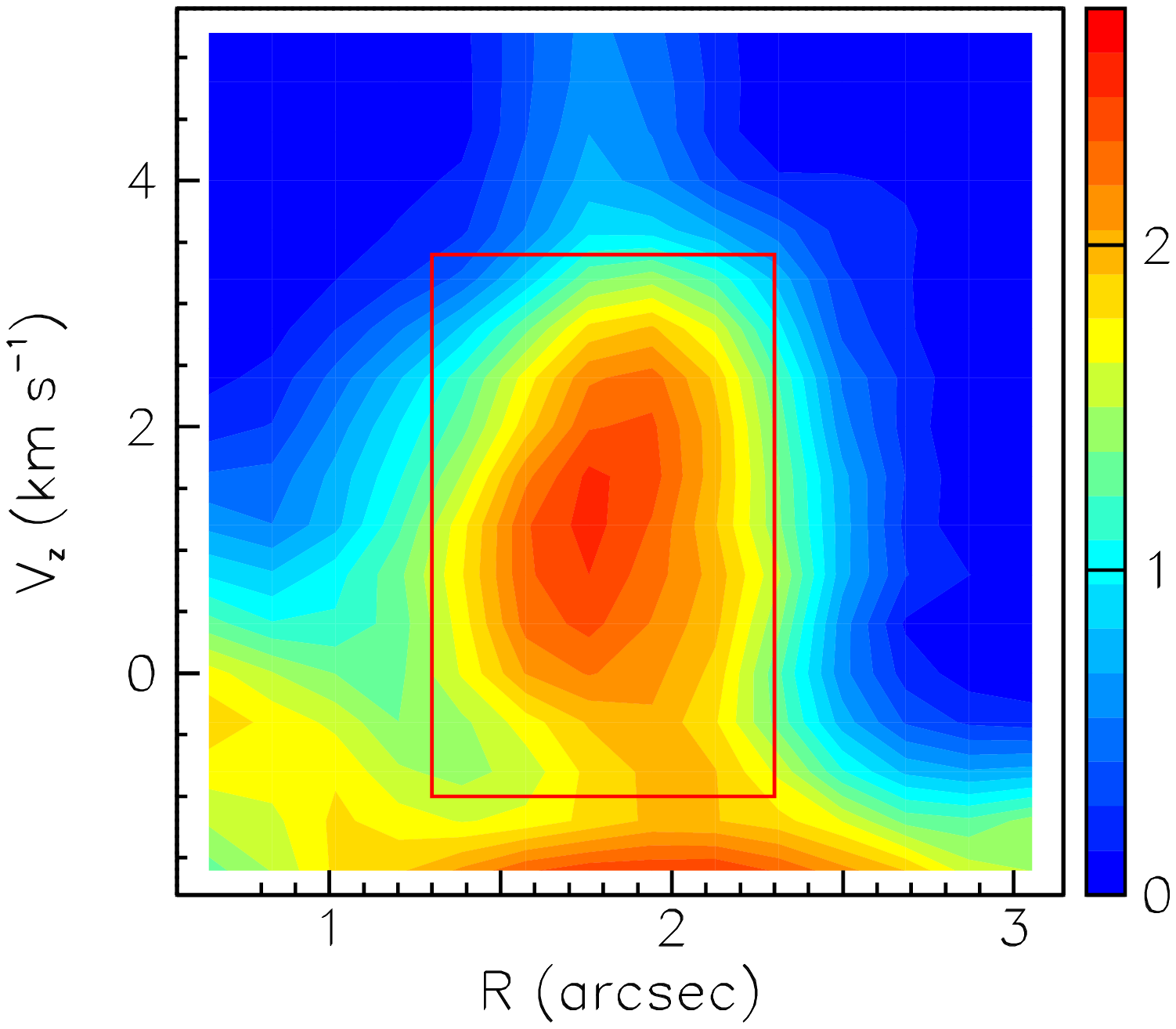}
  \caption{Projections of the North-eastern arc region of the data cube (see Table \ref{tab5}) on the $(x,y)$ plane (left, multiplied by $R$), the ($V_z,\varphi$) plane (centre) and the ($V_z,R$) plane (right). Red lines delimitate the selected volumes. The colour scales are in units of Jy\,arcsec$^{-1}$\,\kms\ for the left panel and Jy for the other panels.}
  \label{fig10}
\end{figure*}

\begin{figure*}
  \includegraphics[height=4.3cm,trim=0.7cm 1.cm 2.cm 1.5cm,clip]{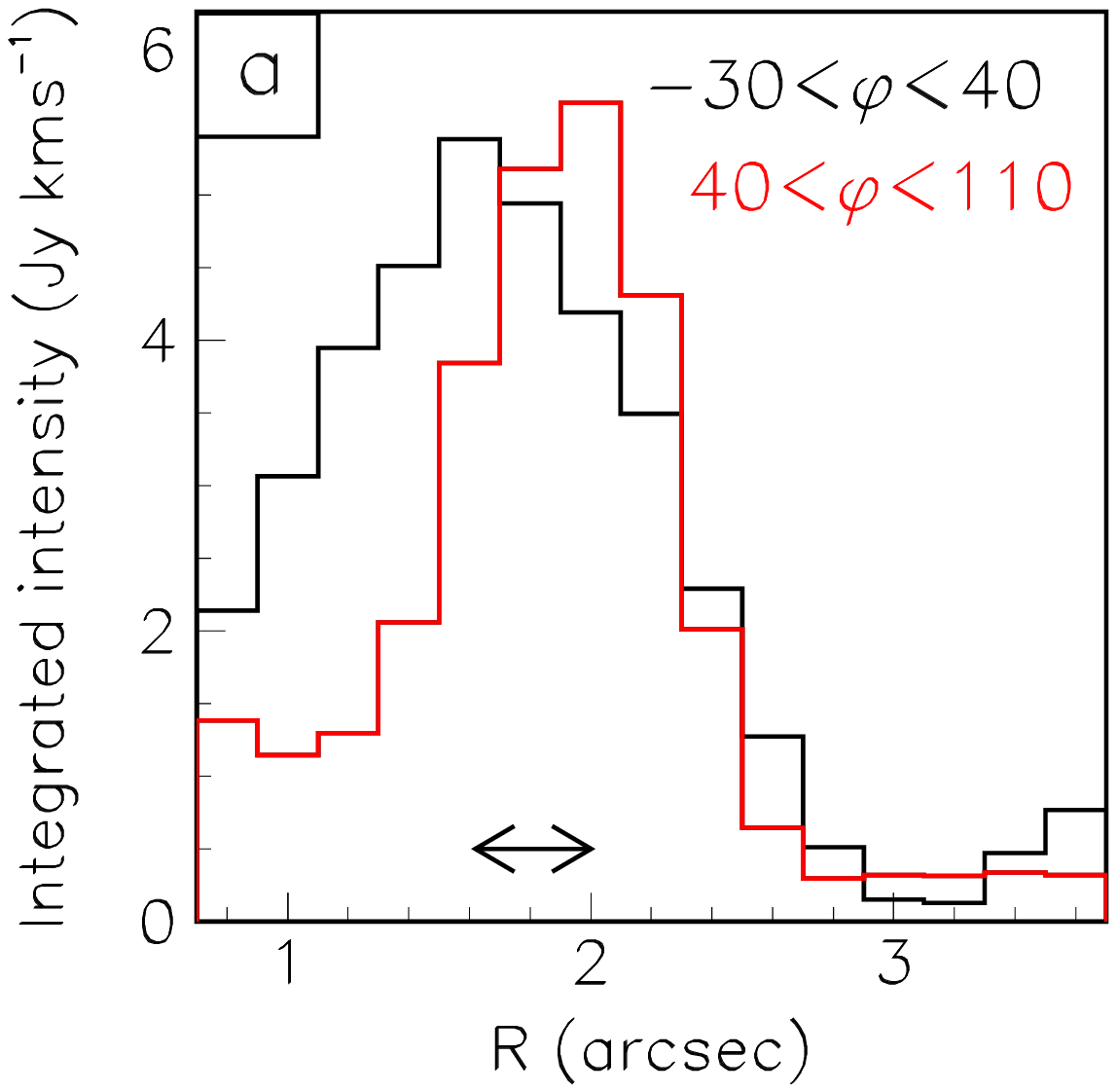}
  \includegraphics[height=4.3cm,trim=0.7cm 1.cm 2.cm 1.5cm,clip]{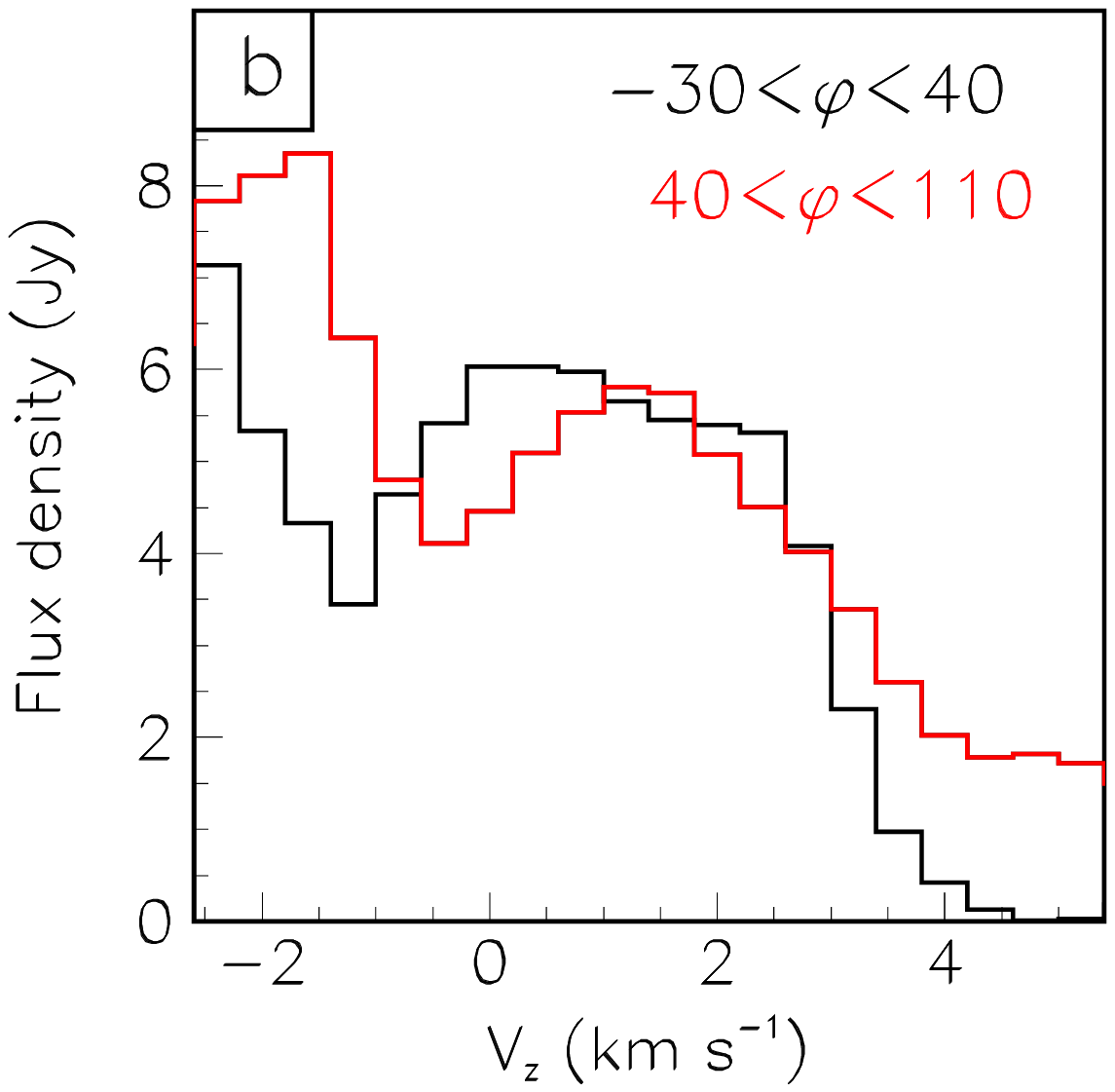}
  \includegraphics[height=4.3cm,trim=0.7cm 1.cm 2.cm 1.5cm,clip]{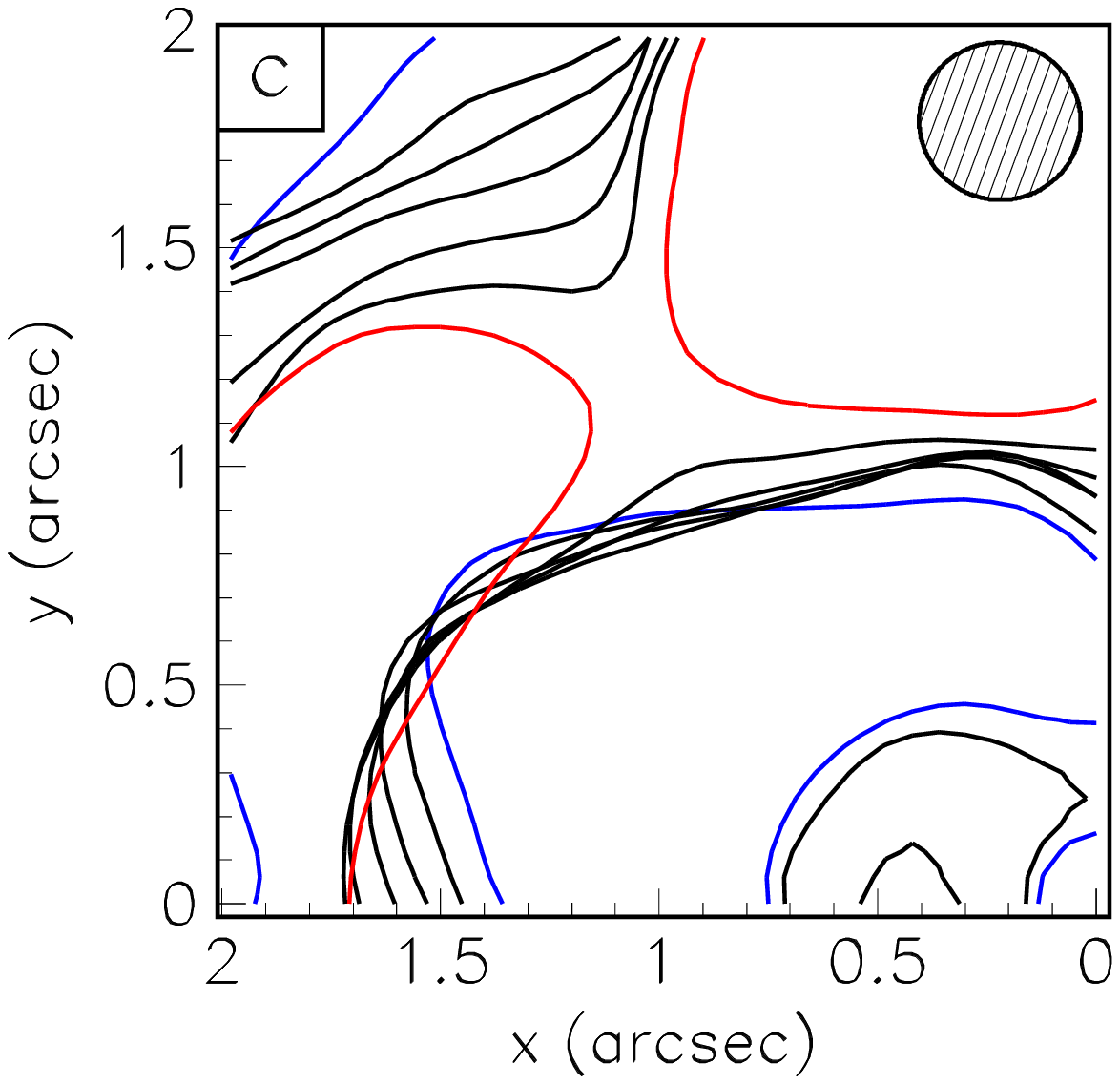}
  \includegraphics[height=4.3cm,trim=0.7cm 1.cm 2.cm 1.5cm,clip]{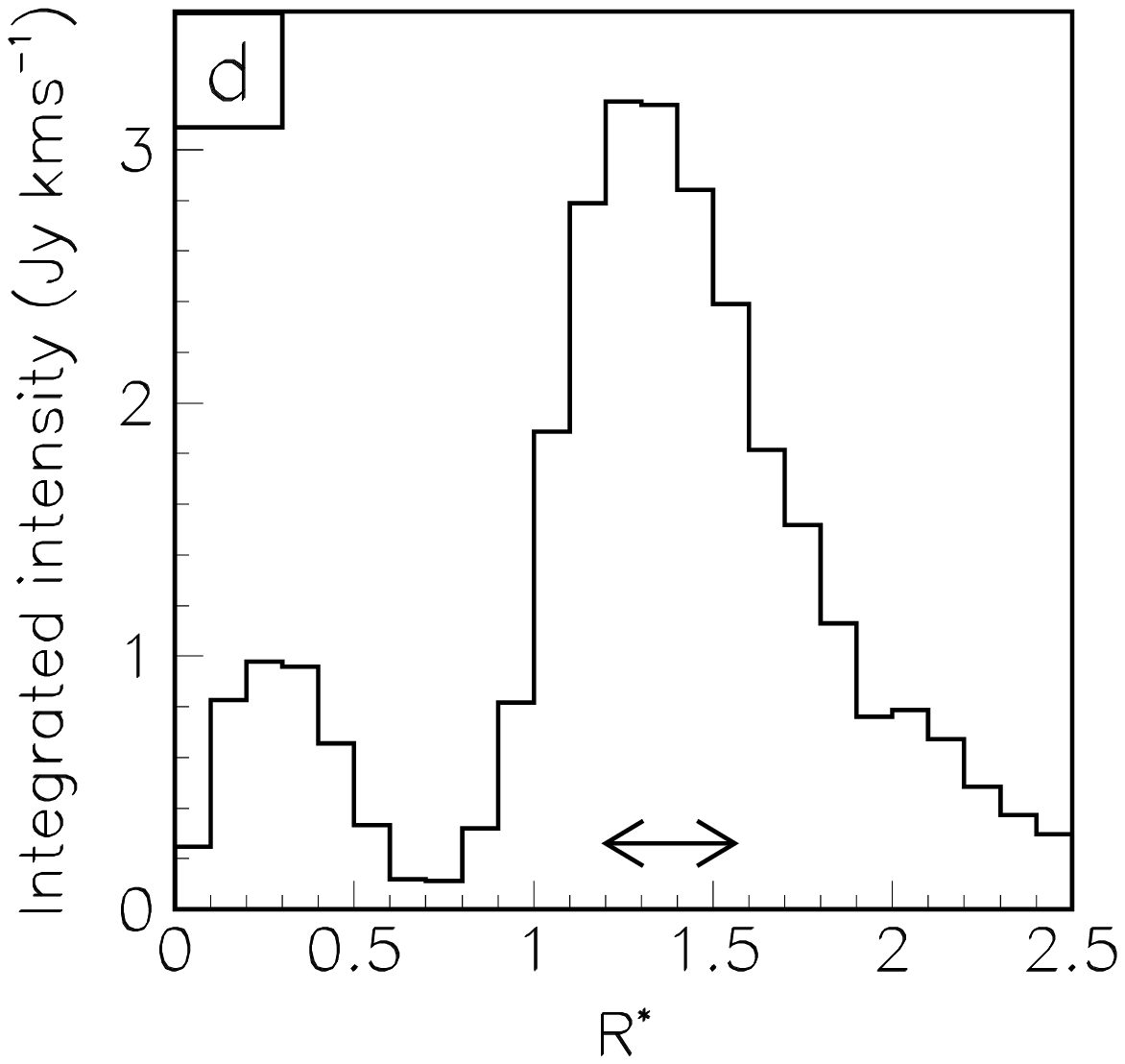}
  \caption{North-eastern arc. Radial (a) and  $V_z$ (b) profiles for either $-30$\dego$<\varphi<$40\dego (black) or 40\dego$<\varphi<$110\dego (red). (c) the CO intensity contours at $\sim$1/3 of maximum in the arc are plotted in bins of 0.4 \kms\ of $V_z$ between 0.8 and 2.4 \kms\ in black, at 0.4 \kms\ in blue and at 2.8 \kms\ in red. (d) distribution of $f$ on $R^*=\sqrt{(x/1.72)^2+(y/1.02)^2}$ for $0.2<V_z<3.0$ \kms\ and $0<\varphi<$90\dego. The beam size is shown in the upper right corner of panel c, its scales in $R$ and $R^*$ are shown as arrows in panels a and d respectively.}
    \label{fig11}
\end{figure*}

In contrast with radial outflows that are confined to narrow regions of the $(V_z,\varphi)$ plane and extend over a large $R$ range, the North-eastern arc covers a narrow region of the $(V_z,R)$ plane and extends over a large $\varphi$ range (Figure \ref{fig10}). In analogy with the South-western outflow, one may think of a wind expanding at a velocity of a few \kms\ and covering a solid angle of a few steradians, the blue part of which is focused by Mira B. However, the limited extension of the arc in the $(V_z,R)$ plane and the small contribution of Mira B focusing found in the study of the North-eastern outflow (Section \ref{sec5}) make it difficult to support such an interpretation. As the intensity map of the arc displays two maxima, we show separately in Figure \ref{fig11}a,b the $R$ and $V_z$ profiles of each half of the arc, defined as $-$30\dego$<\varphi<$40\dego\ and 40\dego$<\varphi<$110\dego, respectively. The suppression of CO emission within the arc, for $R<\sim1$ arcsec, is unlikely to be due to a pure temperature effect and is more likely the result of low density and/or of dissociation of the CO molecules. However, the absence of significant SiO emission, in contrast with the South-western outflow, makes it difficult to conceive a sensible model of the dynamics at stake. In order to describe more precisely the nature of the suppression of CO emission within the arc, we explore the structure of its sharp inner edge by inspecting successive channel maps in bins of 0.4 \kms. Figure \ref{fig11}c displays CO intensity contours set at approximately one third of the maximal intensity measured on the arc. We find that they are essentially invariant when $V_z$ spans between $\sim$0.4 and $\sim$2.8 \kms\ and $\varphi$ between $\sim$0\dego and $\sim$90\dego (black contours). They have an approximately elliptical shape centred on Mira A and having semi-major and semi-minor axes along the $x$ and $y$ axes of 1.72 and 1.02 arcsec respectively. The radial profile of the arc, measured as a function of $R^*=\sqrt{(x/1.72)^2+(y/1.02)^2}$, which is equal to unity on the ellipse, is displayed in Figure \ref{fig11}d.  It gives clear evidence for the sharpness of the inner edge. This is at variance with what is observed in the South-western outflow (Figure \ref{fig7}): together with the absence of SiO emission, this seems to exclude an interpretation in terms of a shock wave, such as was suggested in this former case (Section \ref{sec4}).

\citet{Ramstedt2014} suggest that the inner arc depletion has been produced by what they call an accretion wake: they say that ``the two companions seem surrounded by a common dense molecular envelope, and at red-shifted velocities, there appears to be an accretion wake behind Mira B''. It is difficult to understand what they mean precisely. The North-eastern arc is red-shifted and Mira B is blue-shifted at the time of observation. An interpretation in terms of a spiral related with a focusing of the Mira A wind in the direction of Mira B, as expected for example in WRLOF models, is considered in detail in Section \ref{sec7} and found difficult to reconcile with observations. \citet{Wood2002} have made a detailed study of possible mechanisms for the creation of a cavity in the vicinity of Mira A and/or Mira B. These include a wind interaction shock produced by an interaction between Mira A and Mira B winds and a possible photo-dissociation front. However, none of these interpretations predicts a morphology compatible with the observed morphology of the North-eastern arc. These remarks illustrate the difficulty to conceive a sensible model of the North-eastern arc, however without proposing a better interpretation of its formation. 

Finally, we remark the similarity between the North-eastern arc and the Blue-shifted ring \citep{Nhung2016} described as a bubble by \citet{Ramstedt2014}. Both cover a limited range of Doppler velocity ($\sim -1$ to 3.5 \kms\ for the former and $\sim-10$ to  $-5$ \kms\ for the latter) and a narrow radial range ($< 1$ arcsec), both display a sharp inner edge. Both are elliptic with similar aspect ratios (1.7 for the former and 1.5 for the latter), which may however be accidental if they are projections of circular rings; the radius of the Blue-shifted ring is four times as large as that of the North-eastern arc.

\section{The Southern arc} \label{sec7}

\begin{figure*}
  \includegraphics[height=5.5cm,trim=0.7cm 1.cm .5cm 1.5cm,clip]{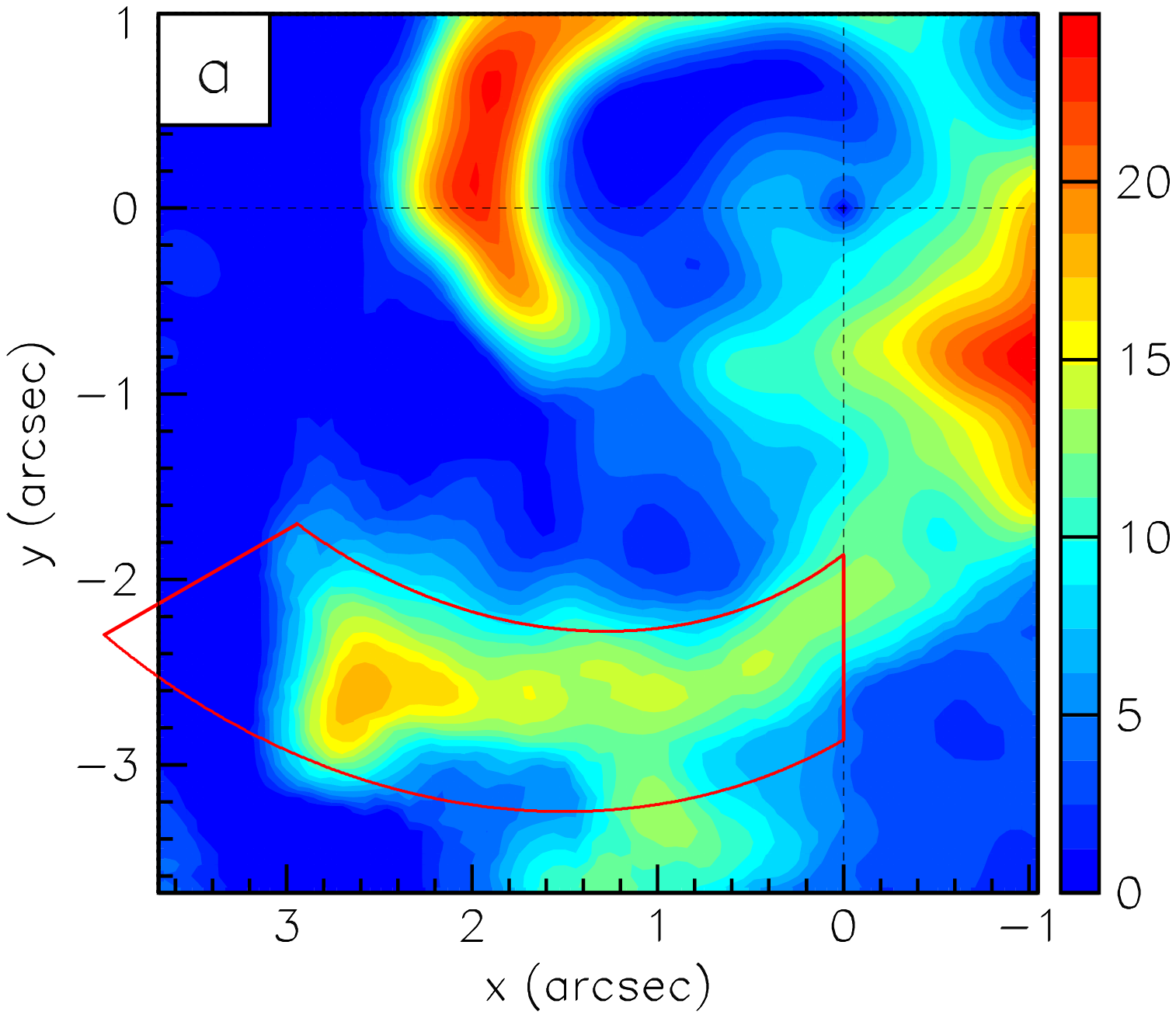}
  \includegraphics[height=5.5cm,trim=0.7cm 1.cm .5cm 1.5cm,clip]{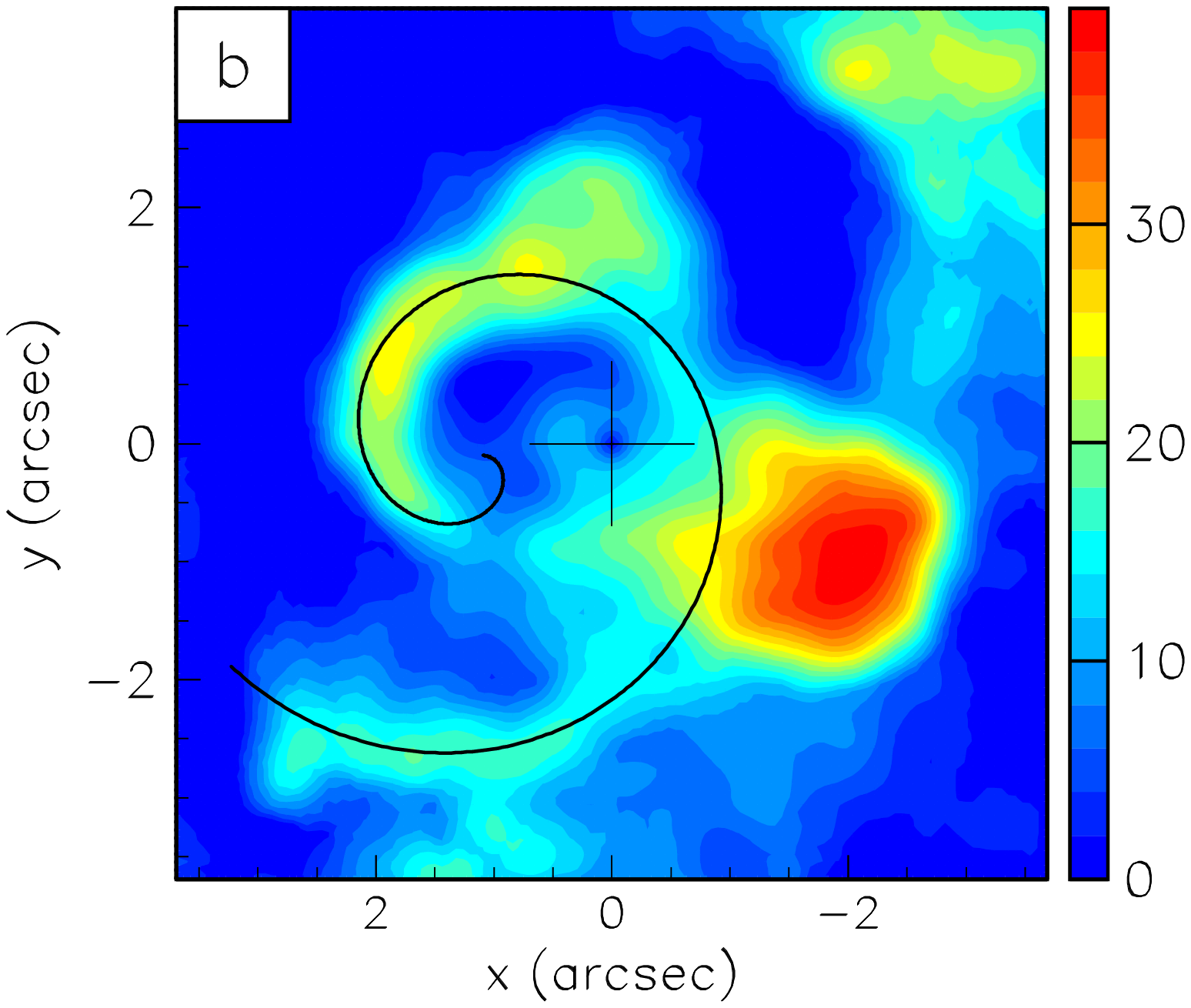}
  \includegraphics[height=5.5cm,trim=0.7cm 1.cm 2.1cm 1.5cm,clip]{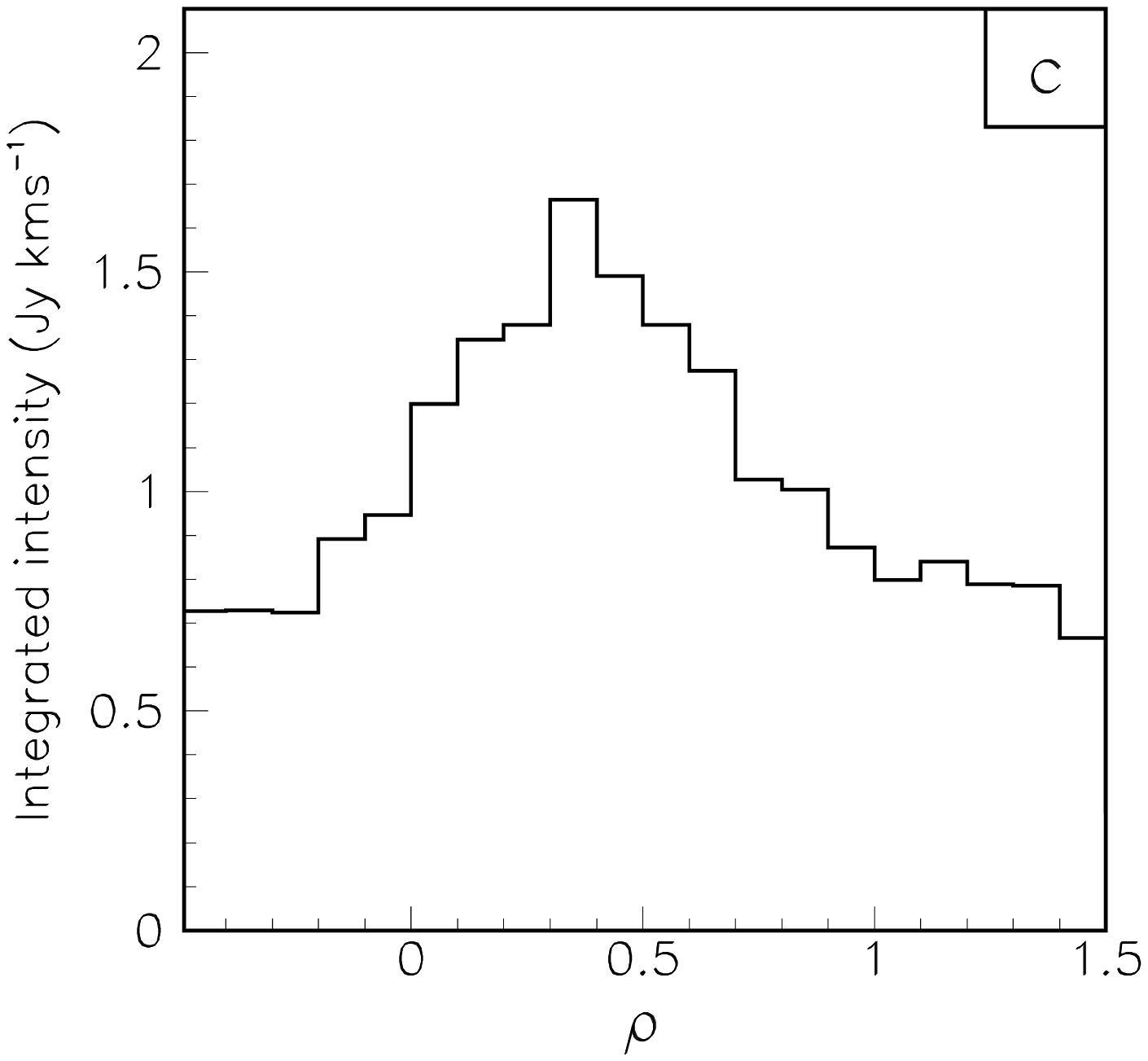}
  \includegraphics[height=6cm,trim=0.7cm 1.cm 0.cm 1.5cm,clip]{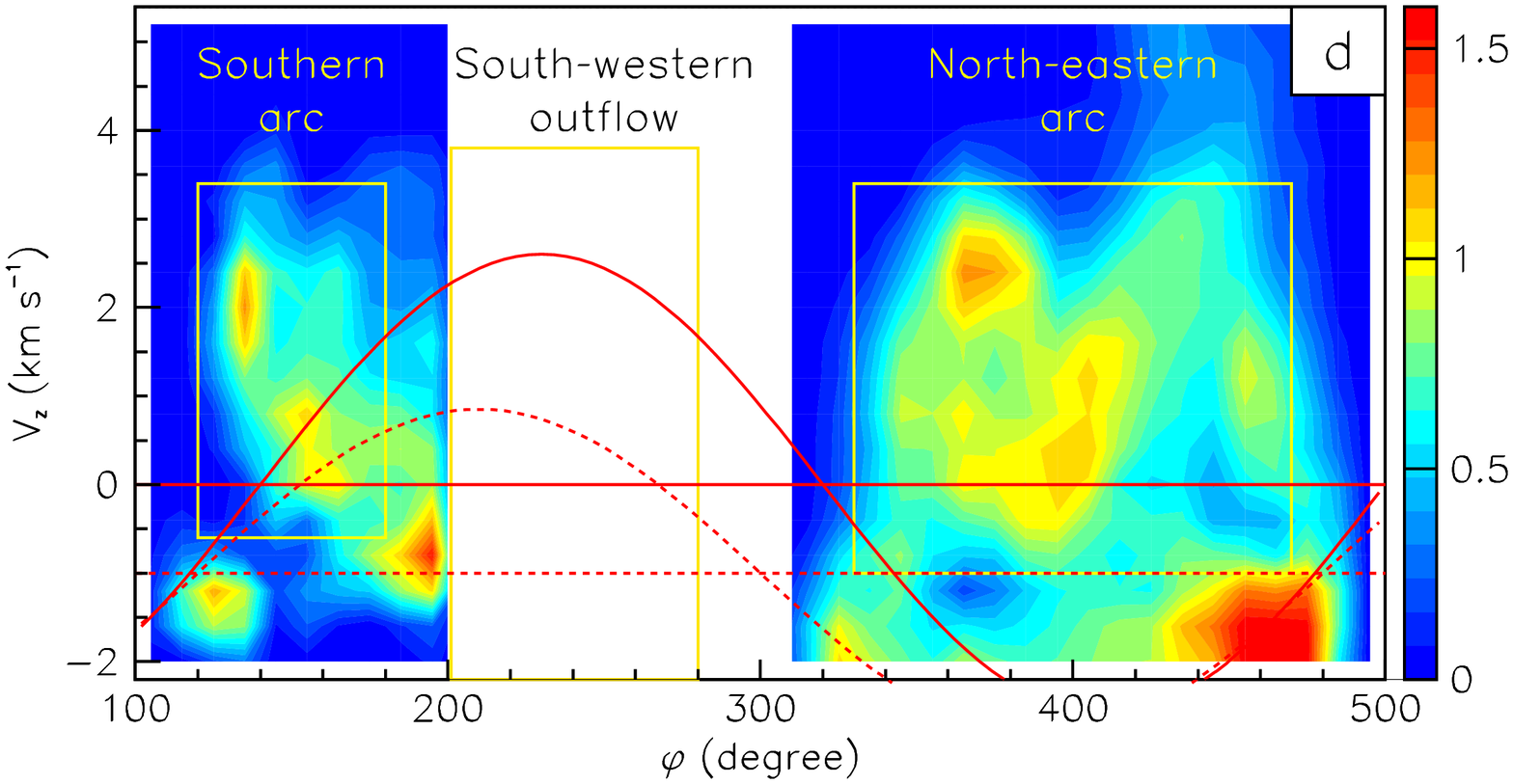}
  \caption{(a) CO intensity, multiplied by $R$, and integrated between $V_z=-0.6$ and 3.4 \kms. The red lines define the region of the sky plane covered by the Southern arc (Table \ref{tab5}).  (b) Same as panel a but covering a larger solid angle and displaying the spiral linking the Southern and North-eastern arcs. (c) dependence on $\rho$ (see text) of the intensity of the Southern arc. (d) PV map of CO emission in the $V_z$ vs $\varphi$ plane. The $\varphi$ interval [110\dego, 200\dego] is integrated between the spirals $R_1$ and $R_2$ used to define the Southern arc (panel a); the $\varphi$ interval [200\dego, 310\dego], hosting the South-western outflow, is not shown; the $\varphi$ interval [310\dego, 500\dego] is integrated between $R=1.3$ and 2.3 arcsec (Figure \ref{fig10} left). The yellow rectangles delimit fragments defined in Table \ref{tab5}. The red lines show the sinusoidal lines and associated systemic velocities displayed in Figure \ref{fig5pai}. }
  \label{fig12}
\end{figure*}

In addition to the Blue-shifted ring and the three major lumps of CO emission mentioned above, South-western outflow, North-eastern outflow and North-eastern arc, one more can easily be identified in the region of the data cube under study in this work: an arc of emission interpreted by \citet{Ramstedt2014} as a spiral; in their Figure A2 they connect together a number of CO emission lumps oriented in the north-east/south-west direction and compare them with a spiral having an arm separation exceeding 5 arcsec, which they interpret as an effect of binarity. We included this arc in the list of emission fragments given in Table \ref{tab5} and we display the CO intensity integrated between $-0.6$ and 3.4 \kms\ in Figure \ref{fig12}a. We use two spirals to delimit its coverage in the sky plane, defined as $R_1$ (arcsec)$=3.4-2.3\Delta\varphi$ and $R_2$ (arcsec)$=4.6-2.6\Delta \varphi$ where $\Delta\varphi=(\varphi-120\mbox{\dego})/90$\dego.

As illustrated in Figure \ref{fig12}b, there seems to be continuity between the Southern arc and the North-eastern arc: they are both red-shifted and seem to join smoothly in the region covered by the South-western outflow. In order to evaluate the radial range covered by the Southern arc we define a variable $\rho=(R-R_1)/(R_2-R_1)$ which is equal to 0 on the inner limiting spiral ($R=R_1$) and to 1 on the outer limiting spiral ($R=R_2$). The dependence on $\rho$ of the intensity, integrated between $V_z=-0.6$ and $V_z=3.4$ \kms\ and between 130\dego\ and 180\dego\ in $\varphi$ is illustrated in Figure \ref{fig12}c. As in the case of the North-eastern arc, the FWHM is $\sim$0.8 arcsec.

Figure \ref{fig12}b shows that a spiral of equation $x=1.1+R_{sp}\sin\varphi$; $y=-0.1-R_{sp}\cos\varphi$; $R_{sp}=0.5+\varphi/180$\dego, where $x$, $y$ and $R_{sp}$ are measured in arcsec, matches reasonably well the observed emission integrated between Doppler velocities of $-$0.6 and 3.4 \kms. It also approximately matches the start of the spiral displayed in Figure A2 of \citet{Ramstedt2014}, at short distances from the stars. 
However, when looking  not only at the projection on the plane of the sky but also along the third $V_z$ dimension, and assuming that continuity between the Southern and North-eastern arcs implies that they are coplanar, we find it difficult to match them. This is illustrated in Figure \ref{fig12}d that shows PV maps of CO emission in the $V_z$ vs $\varphi$ plane for the radial ranges covered by the Southern and North-eastern arcs separately. One would expect the trace on the PV map of the plane containing the arcs to be approximately sinusoidal.  In the region of the Southern arc, $V_z$ is seen to decrease with $\varphi$, from $\sim$2 \kms\ at 130\dego\ to $\sim$0 \kms\ at 160\dego. In the region where the emission of the North-eastern arc is enhanced, between 0 and 60\dego, $V_z$ should therefore increase toward its maximal value. Instead, it decreases from $\sim$2.5 \kms\ to $\sim$0.5 \kms. These are crude estimates, the argument is more qualitative than quantitative, but it illustrates the difficulty to match the two arcs in space. Unfortunately, it is not possible to reveal a possible continuity in the south-western quadrant because of confusion with the emission of the South-western outflow. For this reason the region 200\dego$<\varphi<$310\dego\ has been left blank in Figure \ref{fig12}d.
 More importantly these remarks exclude an association of the Southern arc with Mira B focusing: it oscillates in phase
  opposition with respect to expectation in the $V_z$ vs $\varphi$ plane and, as discussed earlier in Section \ref{sec5},
  the spiral that fits its projection on the sky plane winds in the wrong direction. Quantitatively, it is therefore
  difficult to understand how the spiral described by \citet{Ramstedt2014} could be the effect of Mira B focusing. In addition, the abrupt interruption of the observed emission at the eastern-end of the Southern arc would pose
  a problem if the spiral is supposed to extend downstream and include several additional lumps of emission as suggested
  by these authors. We checked that this conclusion is robust with respect to two major sources of uncertainty that
  might affect it: the lack of precise knowledge of the systemic velocity and the difficulty met in cleaning the dirty map (see Section \ref{sec2.2}); the latter is particularly significant in the present case of the Southern arc for emission between 0 and 1 \kms.

Before closing this section we remark that the very complex morphology of the data cube invites numerous associations of different lumps of emission in geometric shapes such as ellipses or spirals requiring a critical evaluation of the likelihood that they have a meaningful physical identity. For example, rather than associating the Southern arc with the North-eastern arc, we might as well associate it with the arc of north-western emission visible in Figure \ref{fig12}b and associate the North-eastern arc with the eastern extension of the South-western outflow visible on the same figure. Whatever is not a circle is a spiral to first order and one needs to be very critical when claiming evidence for spiral arcs. We shall therefore refrain from mentioning some such associations that we consider not sufficiently reliable to be retained. 

\section{Large Doppler velocities}\label{sec8}

\begin{figure*}
  \includegraphics[height=4.4cm,trim=0cm 0.5cm 2.cm 1.cm,clip]{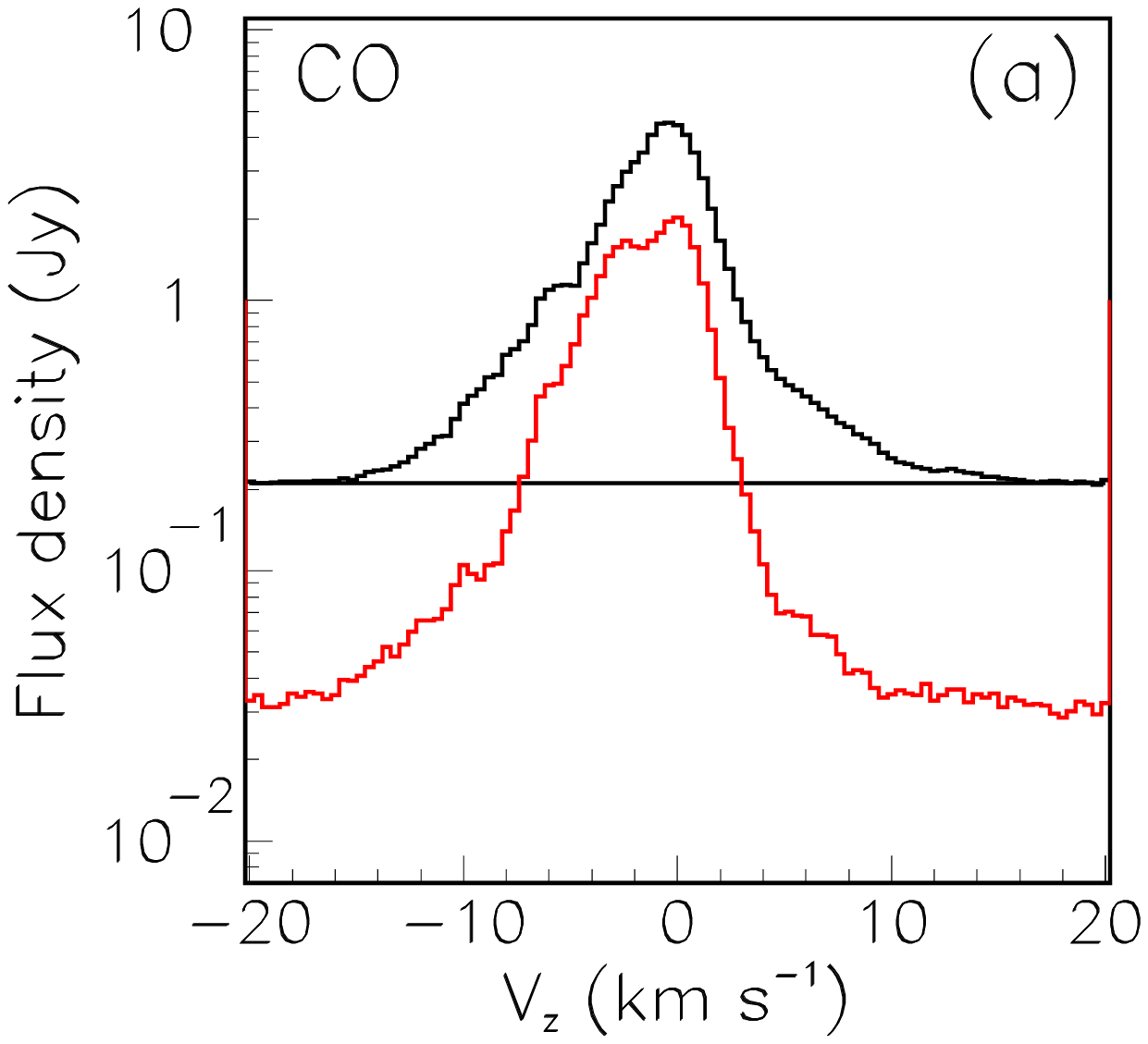}
   \includegraphics[height=4.4cm,trim=0cm 0.5cm 2.cm 1.cm,clip]{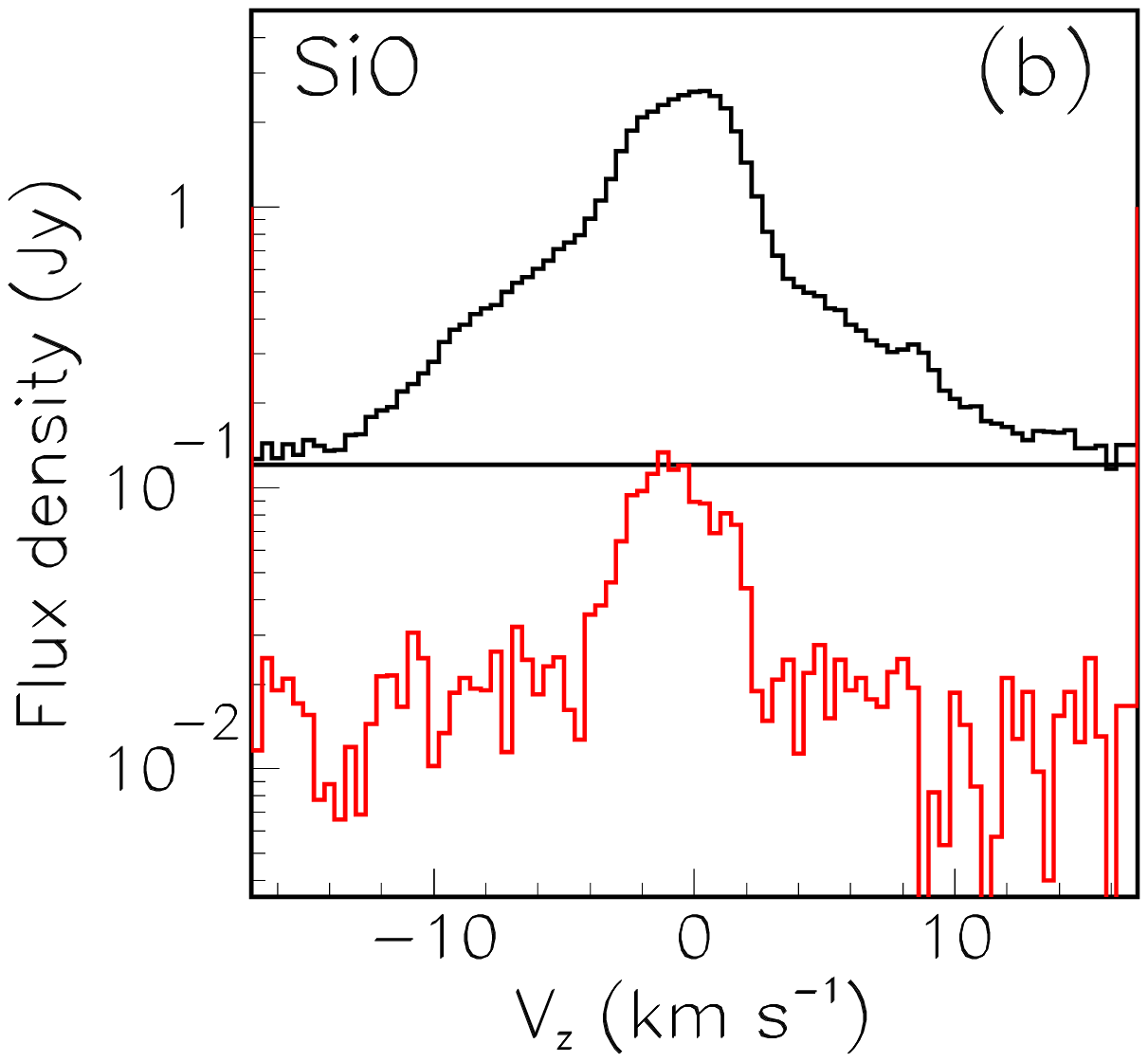}
  \includegraphics[height=4.4cm,trim=0cm 0.5cm 2.cm 1.cm,clip]{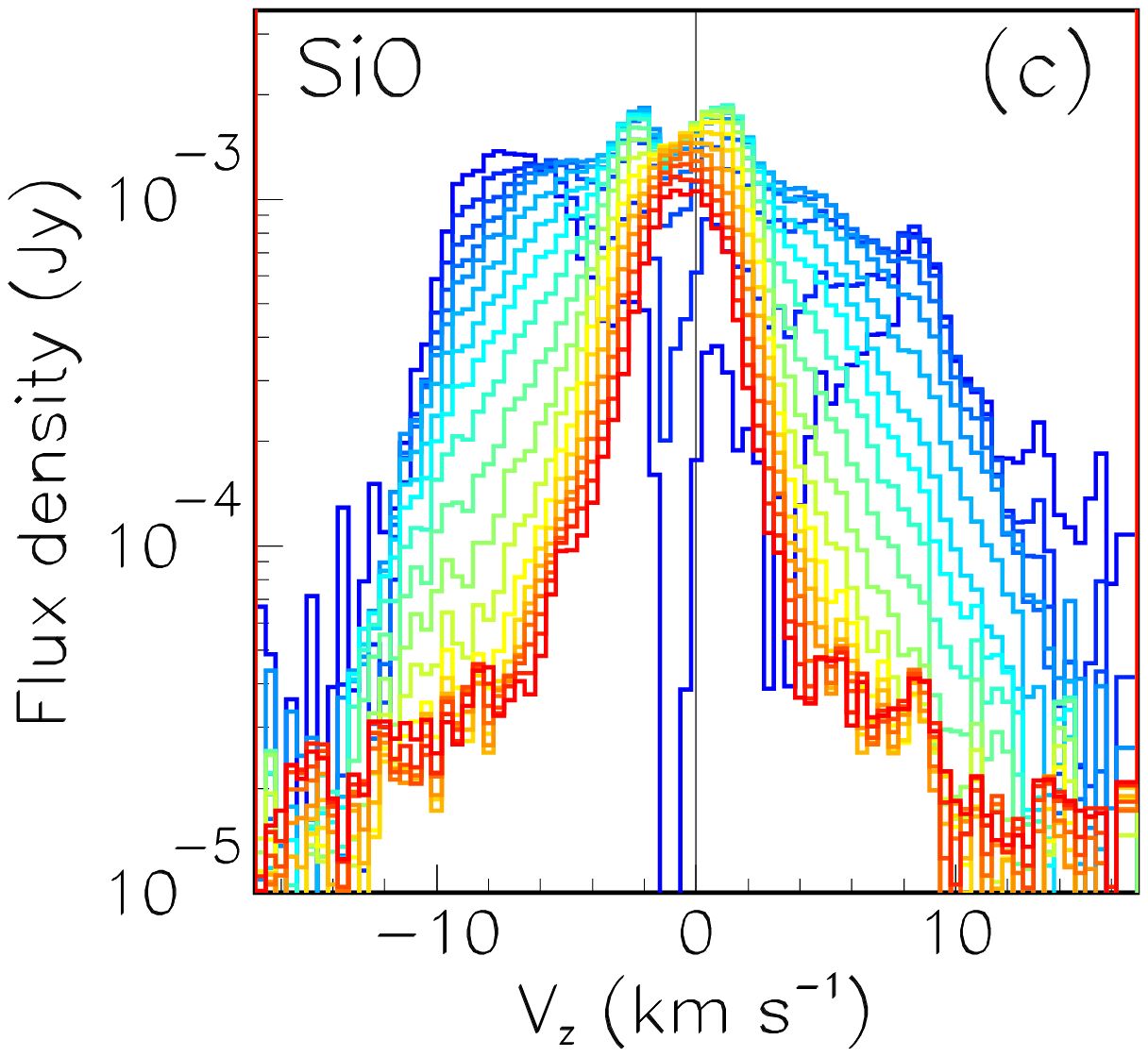}
  \includegraphics[height=4.4cm,trim=0cm 0.5cm 1.8cm 1.cm,clip]{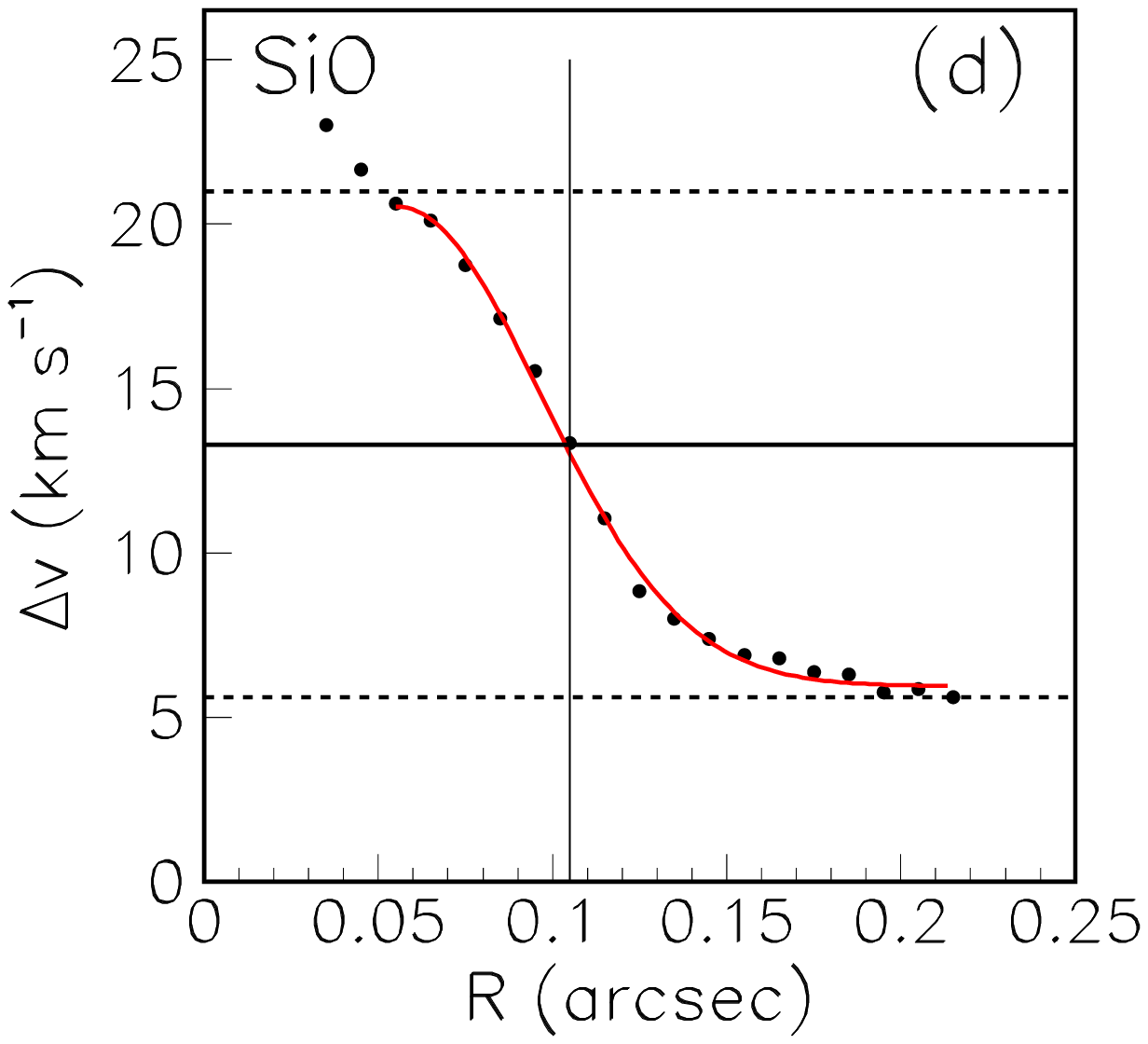}
  \caption{Doppler velocity distributions of CO (a) and SiO (b) line emission within a circle of 0.25 arcsec projected radius centred on Mira A (black) or Mira B (red). The very different beam sizes cause the apparent continuum levels to differ between the two lines. (c) Doppler velocity spectra averaged in 0.03 arcsec wide annular rings centred on Mira A with mean radii increasing from 0.025 arcsec (blue) to 0.215 arcsec (red) in steps of 0.01 arcsec. (d) dependence on $R$ of the full-width at 1/5 maximum of the spectra displayed in the panel c (see text). Note that below $R\sim0.05$ arcsec, as mentioned in the text, the effect of absorption dominates the picture. Accordingly, the spectrum of the innermost ring ($R$=0.025 arcsec) is not included in the figure.}
  \label{fig13}
\end{figure*}

Figure \ref{fig13} displays $V_z$ spectra integrated within a circle $R<0.25$ arcsec centred on Mira A or B. They reveal the presence of large values of $|V_z|$ reaching 16 \kms\ for both CO and SiO. They are more clearly marked when centred on Mira A than when centred on Mira B. Indeed, when mapping the intensity integrated over $|V_z|>6$ \kms\ one finds that
the contribution of Mira A exceeds the corresponding continuum emission significantly; that of Mira B is accounted for by continuum emission on the SiO map but receives important contribution from the Mira A emission on the CO map.

This observation raises the question of the origin of such large Doppler velocity wings. Recently, large Doppler velocities, typically reaching two to three times the terminal wind velocity, have been observed and studied in other oxygen-rich AGB stars: L$_2$ Pup \citep{Kervella2016, Homan2017, Haworth2018}, EP Aqr \citep{Hoai2019a, Homan2018a} and R Dor \citep{Hoai2019b, Decin2018, Homan2018b}. In each case, different explanations have been advanced: rapid and nearly Keplerian rotation of the gas surrounding the star, emission of gas streams along the axis of symmetry of the circumstellar envelope and possibly stellar pulsations, respectively. As none of the features described in the earlier literature suggests that the line of sight might play a particular role in the dynamics of the Mira AB pair, it would be unreasonable to retain the second scenario. Similarly, no evidence for rotation has ever been found in studies of the close environment of Mira A; the existence of an accretion disc around Mira B has been suggested by some authors \citep{Ireland2007, Reimers1985, Sokoloski2010, Vlemmings2015} implying that this disc would probably be rotating, but this would favour Mira B rather than Mira A as centre of the emission of the large Doppler velocities, contrary to observation.

This leaves pulsations as possible candidates. The visual phases of Mira A were 0.09 at the time of the CO observations \citep{Planesas2016}, namely 32\dego\ after maximum, and 0.47 at the time of the SiO observations \citep{Wong2016}, namely 11\dego\ before minimum. While there has been no systematic study of how the gas seen using ALMA in the very close vicinity of the stars is evolving within a pulsation cycle (we thank Professor Theo Khouri and the anonymous referee for clarification on this point), we know of no model that would predict the presence of large enough radial velocities to produce the observed wings of the Doppler velocity spectra. Indeed the models used by \citet{Wong2016} and \citet{Khouri2018} to describe emissions of CO($\nu$=1, $J$=3-2) and SiO($\nu$=0, $J$=5-4) display no such wings. Therefore we fail to think of a plausible interpretation of these large Doppler velocities in terms of previously considered scenarios.

In order to shed some light on this issue we take advantage of the excellent angular resolution offered by the SiO data to study the dependence of the Doppler velocity spectra on the projected distance $R$ to Mira A. Such a study would not be feasible with the CO data for which the beam size is nearly an order of magnitude larger along the Mira A to Mira B line. We do so by averaging the spectra over annular rings having a width of 0.03 arcsec and mean radii increasing between 0.025 and 0.215 arcsec in steps of 0.01 arcsec. A flat continuum contribution is subtracted from each spectrum separately. The result is illustrated in Figure \ref{fig13}c. For values of $R$ smaller than $\sim$0.05 arcsec, the effect of absorption dominates the picture, as shown by the very detailed studies performed in this region by \citet{Wong2016} and, for the CO($\nu$=1, $J$=3-2) line, by \citet{Khouri2018}. However, these authors focus on very small values of $R$ and do not significantly comment on larger values. As clearly shown in Figures \ref{fig13}c and d, the spectra evolve from a very broad profile close to the star to a nearly four times narrower profile at larger distances, a rather sharp transition occurring at $\sim$0.1 arcsec as illustrated in  Figure \ref{fig13}d. Below $R\sim0.05$ arcsec the effect of absorption becomes very important and complicates the picture, however with no significant impact on the present argument. We checked that this result remains true when considering only one quadrant (north-west, south-west, north-east or south-east, see Figure \ref{figa2} of the appendix) of the sky plane.

A natural interpretation of these results is to make the line width, rather than the wind velocity, responsible for the production of large Doppler velocity wings. As thermal broadening cannot produce such large line widths, turbulence must be invoked. The picture would then be of a shell surrounding the zone where stellar pulsation dominates and extending up to $\sim$10 au, hosting a very turbulent regime. Indeed, such a possibility was mentioned earlier by \citet{Kaminski2016a} in relation with the possible presence of very broad AlO line profiles. According to the standard picture of acceleration by transfer of momentum from dust grains to molecules of gas \citep{Hofner2018}, this is a region that hosts important shocks. At 10 au, the escape velocity from one (two) solar masses, values commonly accepted for Mira A, is $\sim$13 (19) \kms\ and the density is such that the velocity acquired by a gas molecule after its interaction with a dust grain is immediately thermalized. Such important turbulence has been observed in a different context by \citet{Falgarone1990} and discussed by \citet{Gillet1998} in the case of pulsating stars. If such a scenario were confirmed, it would require revisiting in its context the cases of the other mentioned AGB stars, L$_2$ Pup, EP Aqr and R Dor, which is, however, beyond the scope of the present work.

In the same spirit of seeking a description of high Doppler velocity wings in terms others than a simple acceleration of the wind, Homan et al. (2018a) have suggested a possible contribution of inelastic collisions in a mixed granular particle fluid. They were first, to our knowledge, to underline explicitly the need for a reliable description of such high velocity wings and for an explanation of their relative importance in SiO emission with respect to CO emission

Finally, we note that the models used by \citet{Wong2016} and \citet{Khouri2018} fail to reproduce the large Doppler velocities displayed in Figure \ref{fig13}. Indeed, these models do not account for important turbulence. If important turbulence is present between 5 and 10 au from the star, causing a broadening of the SiO line profile at the level of 20 \kms\ full width, this raises the question of the validity of models that ignore it at shorter distances and may significantly impact the conclusions reached by \citet{Wong2016} and \citet{Khouri2018} in the close neighbourhood of the star.

\section{Summary and conclusions}\label{sec9}

The above study of the morpho-kinematics of the circumbinary envelope of the Mira AB pair at distances between $\sim$100 and $\sim$350 au from the stars has drawn a picture significantly different from that which could be inferred from the recent detailed analyses of its properties at distances not exceeding $\sim$100 au. These latter studies, while recognizing the complexity and often variability of the observed emissions, usually describe the wind in terms of a slow spherical outflow, using simple radiative transfer codes and hydrodynamical models and emphasizing the effects of stellar pulsations, dust condensation and acceleration by momentum transfer from the stellar radiation. In nearly all cases, the authors underline the important role probably played by shocks associated with stellar pulsations. While these mechanisms are obviously of major importance and are likely to dominate the scene at distances below 100 au, the apparent lack of continuity with the picture obtained beyond 100 au raises questions. Any acceptable description of the Mira wind should accommodate observations over the whole radial range. The present study has no ambition to provide such a global description; instead it aims at providing a detailed and reliable picture of the main properties of the wind at distances beyond 100 au which cannot be ignored and can be used as inputs to future modelling effort.

We find that in the region of the data cube defined as $1<R<3.7$ arcsec and $-5<V_z<5$ \kms\ the observed emission is strongly fragmented. We defined a few fragments (Table \ref{tab5}) having a mean brightness of 2.6 Jy arcsec$^{-2}$ compared with only 1.3 Jy arcsec$^{-2}$ in between the fragments. We evaluate the impact of absorption as not exceeding $\sim$40\% and measure a $^{12}$CO/$^{13}$CO abundance ratio of 12$\pm$2.

Strong evidence is obtained for the presence of a South-western outflow covering a broad solid angle that extends on both sides of the sky plane. The share of the measured emissions between the blue-shifted and red-shifted hemispheres excludes that the mean direction to which it points make a large angle with respect to the sky plane. CO and SiO emissions in the South-western outflow are approximately anti-correlated, suggesting that SiO emission reveals the extraction of SiO gas from dust grains by some physical mechanism that causes the dissociation or evacuation of CO molecules. Detailed inspection of the radial profiles confirms this interpretation. A possible scenario in terms of mass ejection associated with the 2003 X-ray burst \citep{Karovska2005} and an associated shock wave has been presented but cannot be ascertained. If this scenario is real, earlier mass ejections might have occurred in the past that might have left traces on the PV map of Figure \ref{fig5pai}. In particular we note the presence of a depression of CO emission at $(V_z,\varphi)=(+0.4\,\mbox{\kms},80\mbox{\dego})$, nearly back-to-back with that of the South-western outflow (Figure \ref{figa3} of the appendix); they would be exactly back to back if the systemic velocity were 47.0 \kms\ instead of 47.7 \kms, which would be perfectly acceptable.  If the presence of this second hole were not pure coincidence it would invite an interpretation in terms of two back-to-back high velocity jets rather than in terms of mass ejection.

In the North-eastern quadrant, we find no SiO emission and CO emission is shared between two fragments. The first fragment, the North-eastern outflow, is blue-shifted and covers a broad range of angles and radial distances. It seems to be part of a global enhancement of emission associated with a slow isotropic wind blowing in the neighbourhood of the Mira B orbital plane. A possible mechanism causing such enhancement would be the accumulation of a very slow Mira B focused wind over many orbital periods, combined with significant dilution producing a uniform density distribution at the scale of a few millennia. We have given arguments showing that a small part of the outflow, probably associated with Roche lobe overflow \citep{Mohamed2007, Mohamed2012} has been focused by Mira B over the past century. But we cannot exclude the possibility that the bulk of the North-eastern outflow would have no relation to its hosting Mira B.

The second fragment of CO emission facing the South-western outflow, the North-eastern arc, takes the form of a simple red-shifted arc with limited extension in both radial distance and Doppler velocity but spanning over 90\dego\ in $\varphi$. A priori, it may be interpreted as an enhancement of emissivity, resulting for example from an increased mass loss episode, or as the remnant of a broad outflow having suffered a major event of emissivity depletion over a large solid angle, resulting for example from the dissociation of CO molecules produced by shock waves. The radial profile of the arc emission and more generally its confinement to a narrow region of the $(V_z, R)$ plane favour the first scenario over the second. The apparent continuity with the Southern arc, which was identified by \citet{Ramstedt2014} as the birth of a spiral, is not confirmed when including the dependence on Doppler velocity; moreover, we have shown that the spiral defined by \citet{Ramstedt2014} has no simple relation with Mira B and winds in the wrong direction to be interpreted as a Wind Roche Lobe Overflow spiral. The observed morpho-kinematics of three lumps of CO emission that have been identified in the present analysis, the Blue-shifted ring, the North-eastern arc and the Southern arc, share several unexplained features; in particular they appear as arcs of emissivity confined to narrow intervals of projected distance to Mira A.

Our failure to produce a sensible interpretation of the observed properties of the morphology and kinematics of the circumbinary envelope of the Mira AB system should not be used as an excuse to ignore its complexity; its impact on any model claiming to describe its properties at shorter distances from the stars should be carefully assessed. Many of these properties must have their source at short distances from the star rather than being the result of subsequent lumping: such is the case of the strong anisotropy illustrated in Figure \ref{fig5}c.  The present work has presented in as clear as possible a form what can be reliably concluded from the observation of the CO(3-2) emission at distances between $\sim$100 and $\sim$350 au from the stars. It needs to be completed by observations of other line emissions, in particular from CO molecules, in the same radial range.

Finally we have observed the presence of large Doppler velocity wings of both CO and SiO emission spectra in the vicinity of the line of sight crossing the plane of the sky at the location of Mira A. These are reminiscent of similar features observed in other low mass-loss rate, oxygen-rich AGB stars, such as EP Aqr, R Dor and L$_2$ Pup. Interpretations in terms of gas streams, stellar pulsations and near-Keplerian rotation close to the star have been proposed in these other cases, none of which provides a satisfactory scenario in the Mira AB case. We show that these large velocities are reached at typical distances of 5 to 10 au from Mira A and are caused by a large line width rather than by a large wind velocity; we suggest that they might be the effect of turbulence. If it were true, it may impact the conclusions reached in the neighbourhood of the star by \citet{Wong2016} and \citet{Khouri2018} who used models ignoring such large turbulence. This observation may encourage revisiting what we know of other AGB stars displaying similar large Doppler velocity wings in order to possibly obtain a unified picture of the physics at stake.

\section*{Acknowledgements}
We thank Dr St\'{e}phane Guilloteau for advice in dealing with specific imaging issues. We express our gratitude to the anonymous referee for many pertinent comments that helped greatly with the clarity of the presentation. This paper uses ALMA data: ADS/JAO.ALMA\#2011.0.00014.SV and ADS/JAO.ALMA\#2013.1.00047.S. ALMA is a partnership of ESO (representing its member states), NSF (USA) and NINS (Japan), together with NRC (Canada), MOST and ASIAA (Taiwan), and KASI (Republic of Korea), in cooperation with the Republic of Chile. The Joint ALMA Observatory is operated by ESO, AUI/NRAO and NAOJ. The data are retrieved from the JVO portal (http://jvo.nao.ac.jp/portal) operated by the NAOJ. We are deeply indebted to the ALMA partnership, whose open access policy means invaluable support and encouragement for Vietnamese astrophysics. Financial support from the World Laboratory, the Odon Vallet Foundation and VNSC is gratefully acknowledged. This research is funded by the Vietnam National Foundation for Science and Technology Development (NAFOSTED) under grant number 103.99-2018.325. 

%








\appendix
\section{}

\begin{figure*}
  \includegraphics[height=14cm,trim=0.5cm 1.cm .5cm 1.5cm,clip]{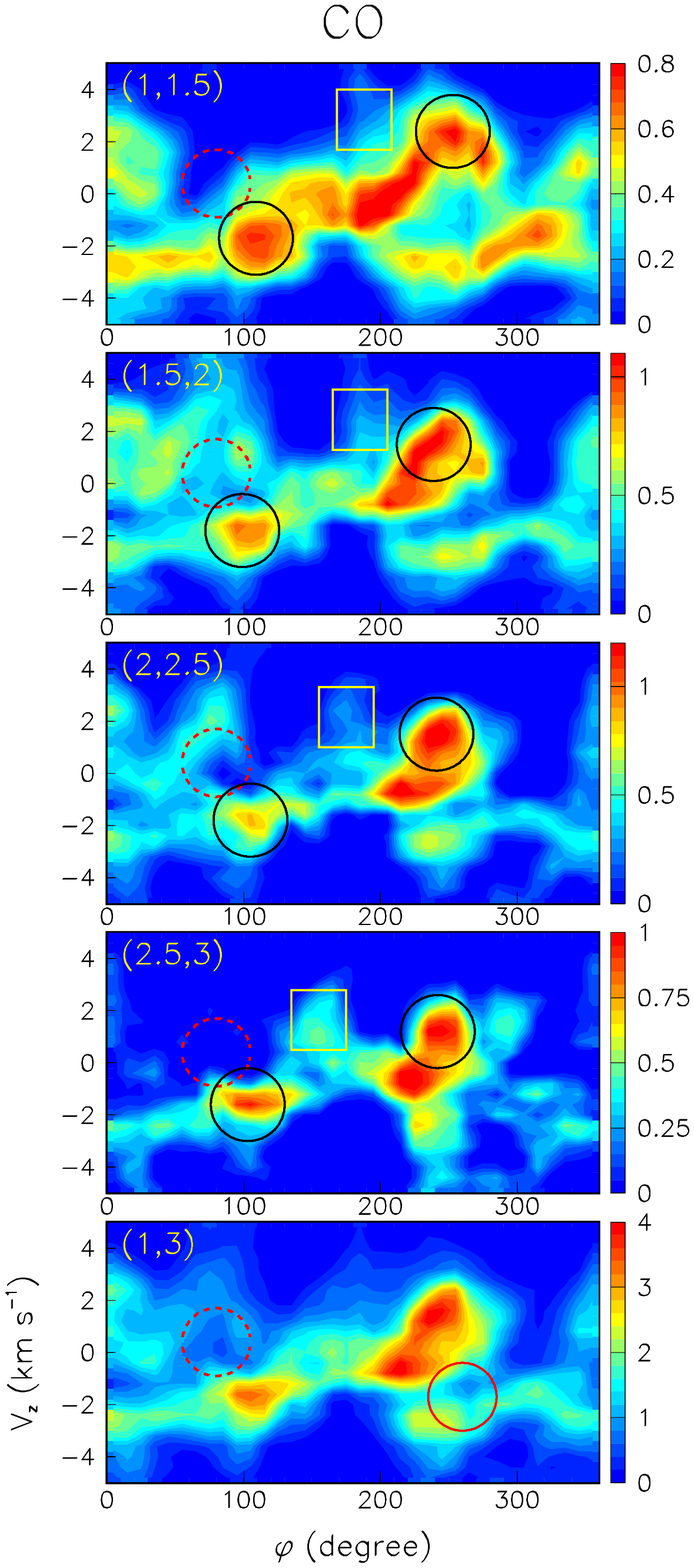}
  \includegraphics[height=14cm,trim=0.5cm 1.cm .5cm 1.5cm,clip]{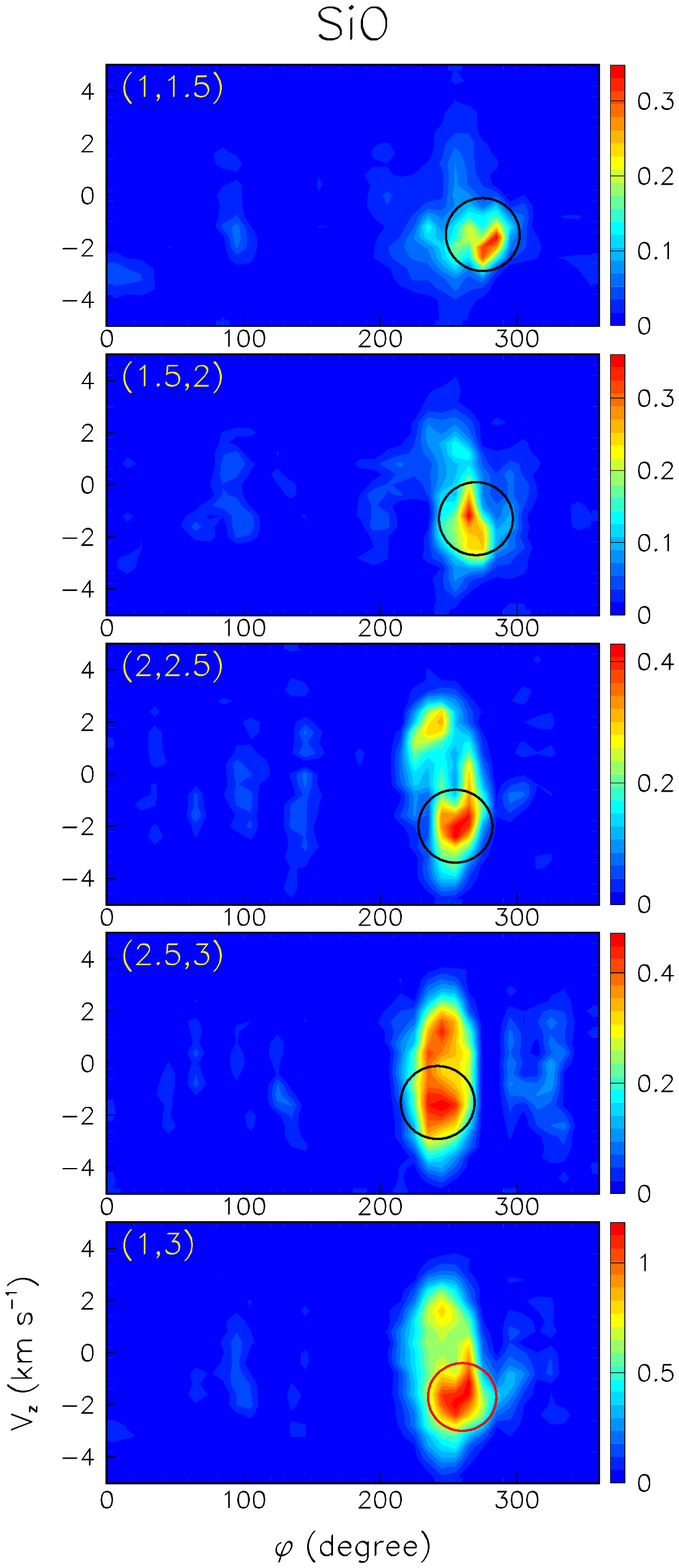}
  \caption{Position velocity maps ($V_z$ vs position angle $\varphi$) integrated
over $R$ intervals of 0.5 arcsec between 1 and 3 arcsec (from up down). The lower panels are for $1<R<3$ arcsec. CO emission is left and SiO emission is right. The black circles drawn in the upper panels indicate the radial outflows listed in Table \ref{tab4}. The yellow rectangles in the CO emission panels refers to the Southern arc (Table \ref{tab5} and Section \ref{sec7}). The red circle drawn in the lower panel of each CO and SiO emission shows the location of the suggested mass ejection (Section \ref{sec3}).  The dotted circles in the CO emission panels show the location of the depression mentioned in Section \ref{sec9} when discussing the South western outflow and possible earlier mass ejection episodes.}
  \label{figa3}
\end{figure*}

\begin{figure*}
  \includegraphics[height=5.5cm,trim=0.5cm 1.cm .5cm 1.5cm,clip]{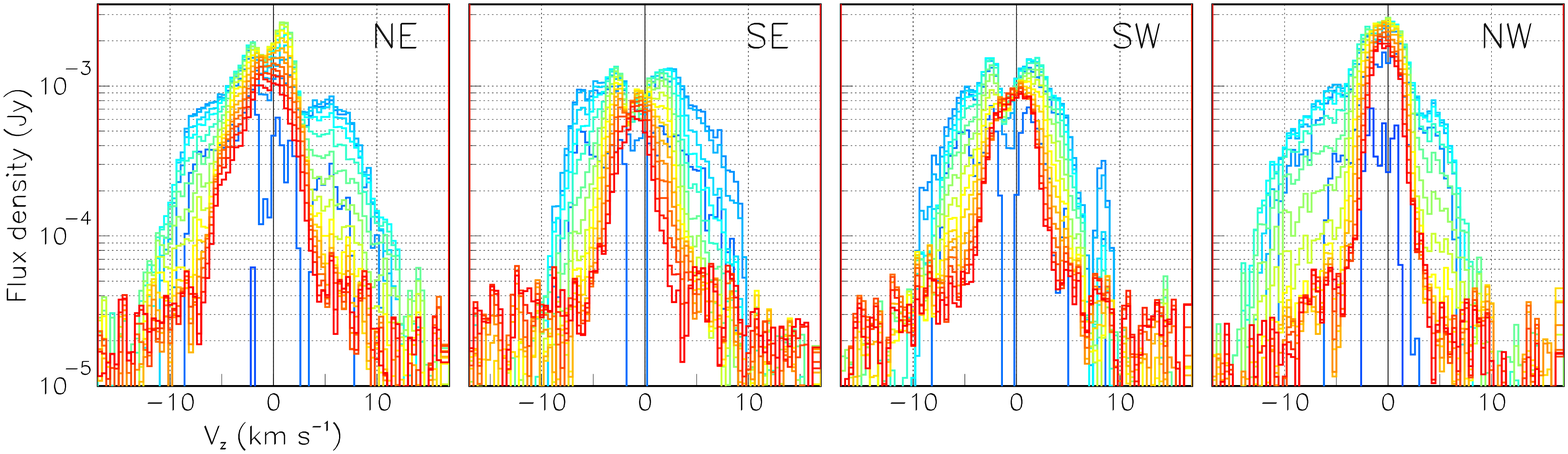}
  \caption{Same as Figure \ref{fig13}c  showing different quadrants on the sky map. Note that at the larger distances the line gets broader in the direction of the outflows (north-east and south-west).}
  \label{figa2}
\end{figure*}

\begin{figure*}
  \includegraphics[height=11.5cm,trim=-0.3cm 0cm 0cm 0cm,clip]{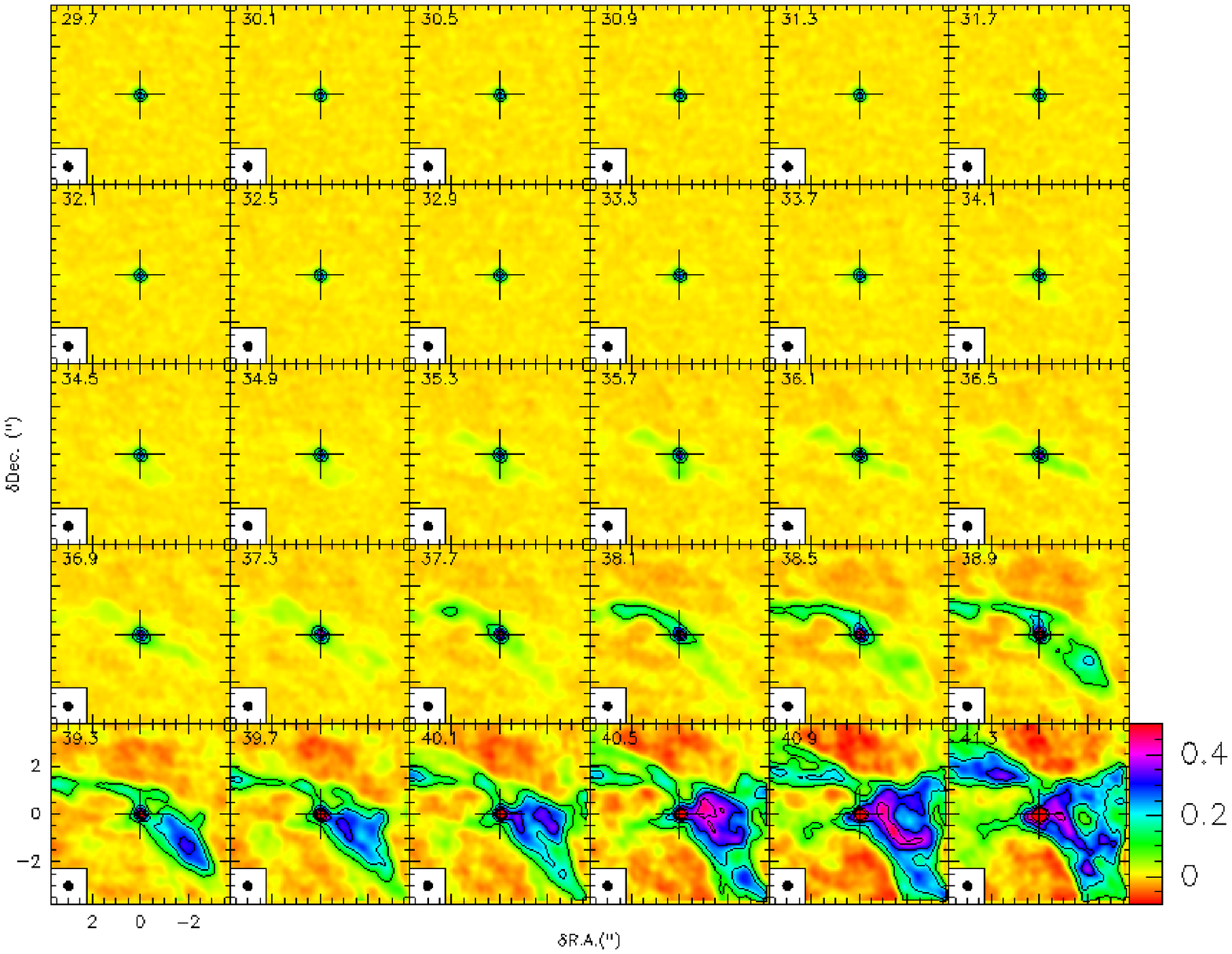}
  \includegraphics[height=11.5cm,trim=-0.1cm 0cm 0cm 0cm,clip]{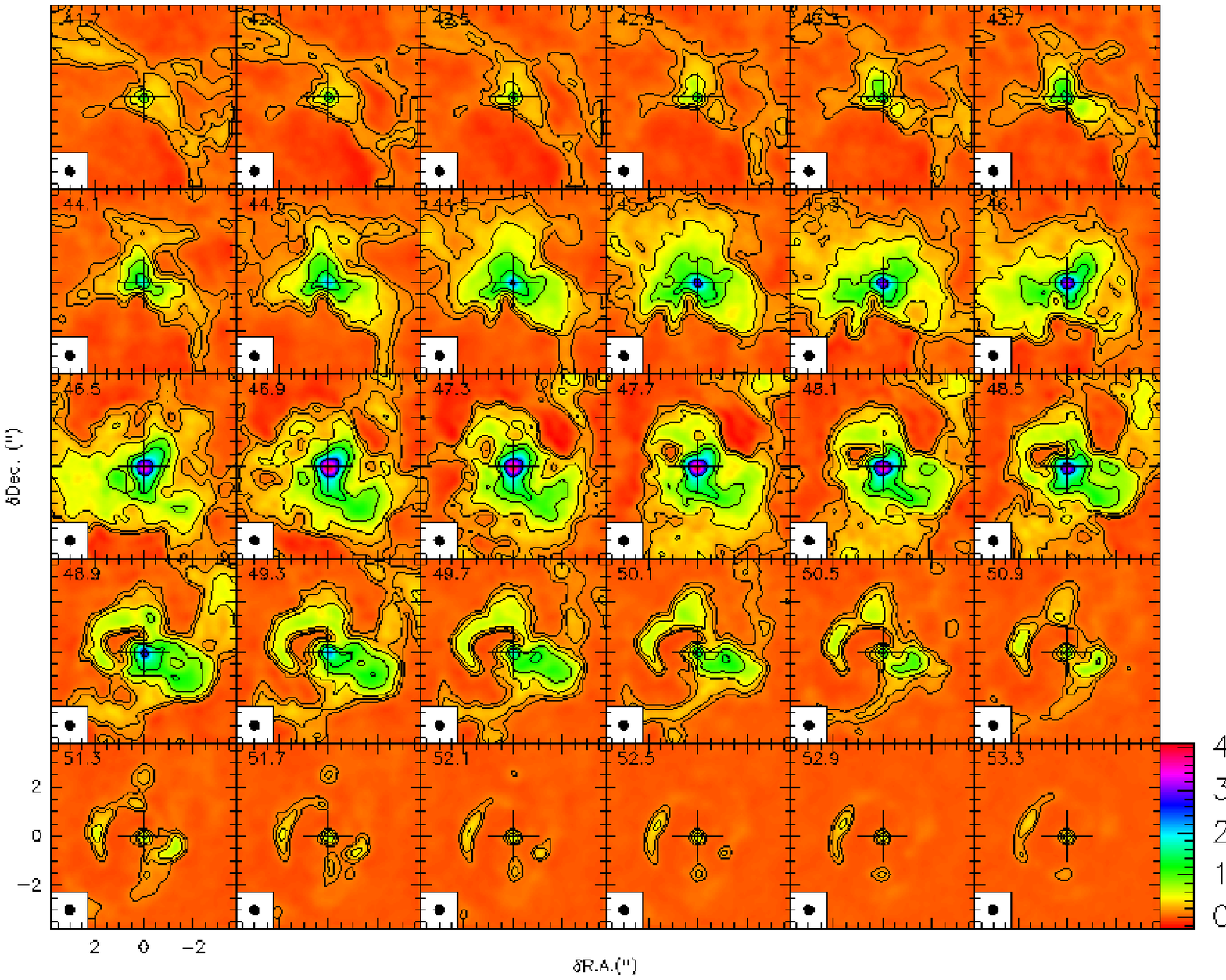}
  \caption{Channel maps of $^{12}$CO(3-2) emission. The colour scales are in units of Jy\,beam$^{-1}$.}
  \label{figa1}
\end{figure*}
\begin{figure*}
  \includegraphics[height=11.5cm,trim=-0.3cm 0cm 0cm 0cm,clip]{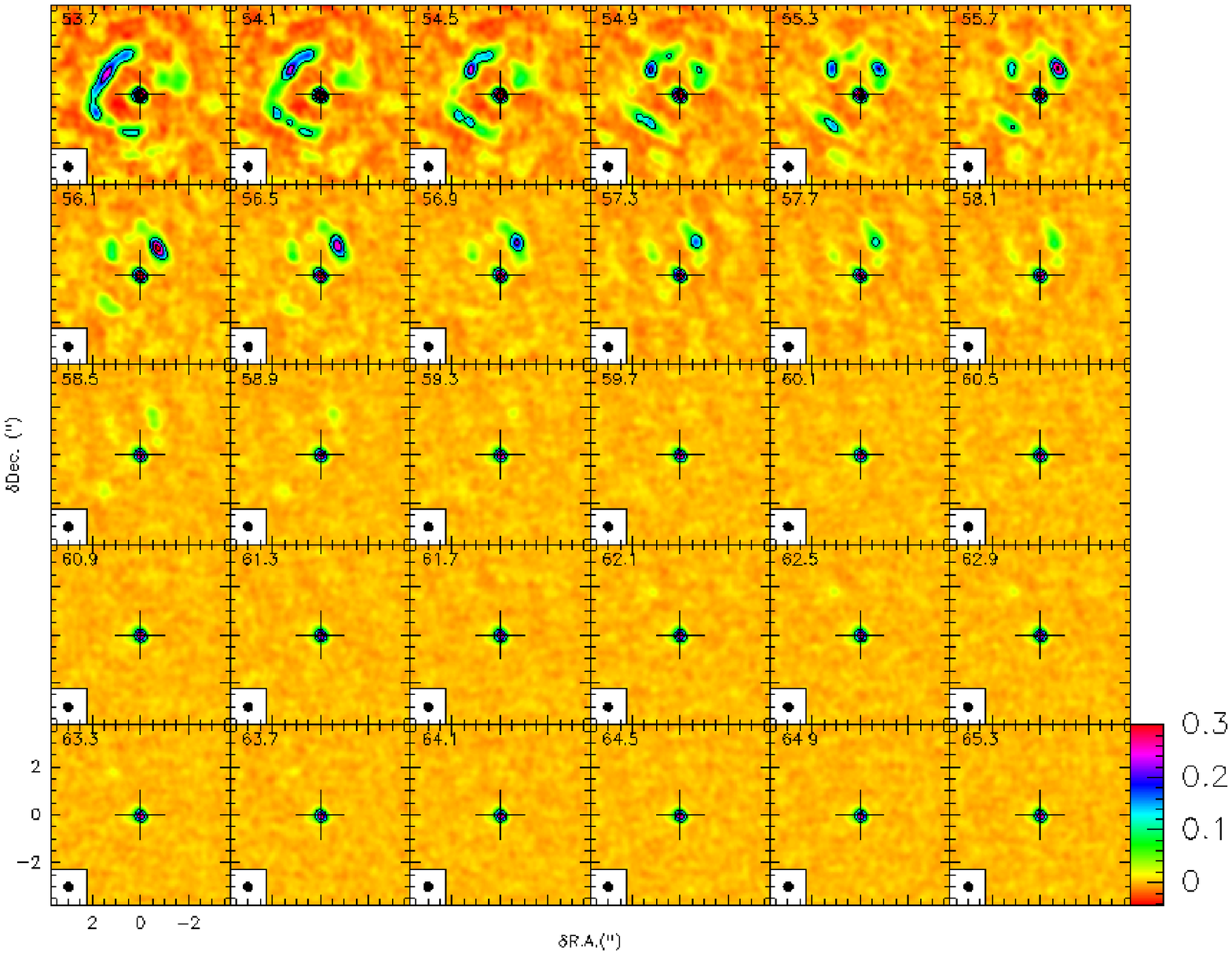}\par
  \textbf{Figure A3.} Channel maps of $^{12}$CO(3-2) emission (cont.).
\end{figure*}

\bsp	
\label{lastpage}
\end{document}